\pdfoutput=1 
\documentclass[prd,twocolumn,superscriptaddress,showpacs,showkeys,floatfix]{revtex4-2}

\usepackage{lineno, blindtext}
\usepackage{graphicx}
\usepackage{amsmath}
\usepackage{enumerate}
\usepackage{placeins}
\usepackage{appendix}
\usepackage[utf8]{inputenc}
\usepackage{chngcntr}
\usepackage{microtype}
\usepackage{color}
\usepackage{epstopdf}
\usepackage{colortbl}
\definecolor{darkred}{rgb}{0.5,0,0}
\definecolor{darkblue}{rgb}{0,0,0.5}
\definecolor{firebrick}{rgb}{0.75,0.125,0.125}
\definecolor{darkgreen}{rgb}{0,0.5,0}

\usepackage{booktabs}
\usepackage{tabularx}
\usepackage{dcolumn}
\usepackage{bm}
\usepackage{xspace} 
\usepackage{caption} 
\usepackage{url}
\usepackage{lmodern} 
\usepackage[T1]{fontenc} 
\usepackage[utf8]{inputenc} 
\usepackage[colorinlistoftodos]{todonotes}

\usepackage{hyperref}

\newcommand{\nova}{NOvA}
\newcommand{\minerva}{MINERvA}

\newcommand{\gevc}{\mbox{GeV/$c$}\xspace}

\newcommand{\mevcc}{\mbox{MeV/$c^2$}\xspace}

\newcommand{\piBe}{\ensuremath{\pi^+ + \textup{Be at 60}\:\gevc}\xspace}
\newcommand{\piCnoE}{\ensuremath{\pi^+ + \textup{C}}\xspace}

\newcommand{\pCnoE}{\ensuremath{p +\textup{C}}\xspace}

\newcommand{\pip}{\ensuremath{\pi^+}\xspace}
\newcommand{\pim}{\ensuremath{\pi^-}\xspace}
\newcommand{\pipm}{\ensuremath{\pi^\pm}\xspace}
\newcommand{\Kp}{\ensuremath{K^+}\xspace}
\newcommand{\Km}{\ensuremath{K^-}\xspace}
\newcommand{\Kpm}{\ensuremath{K^\pm}\xspace}

\newcommand{\kos}{\ensuremath{K^0_{\textup{S}}}\xspace}

\newcommand{\p}{\ensuremath{p}\xspace}
\newcommand{\antip}{\ensuremath{\bar{p}}\xspace}
\newcommand{\lam}{\ensuremath{\Lambda}\xspace}
\newcommand{\alam}{\ensuremath{\bar{\Lambda}}\xspace}

\newcommand{\GeantFour}{{\scshape Geant4}\xspace}

\begin{document}
\title{Measurements of \pip, \pim, $p$, $\bar{p}$, \Kp and \Km production in 120 GeV/$c$ p + C interactions}




\affiliation{National Nuclear Research Center, Baku, Azerbaijan}
\affiliation{Faculty of Physics, University of Sofia, Sofia, Bulgaria}
\affiliation{Ru{\dj}er Bo\v{s}kovi\'c Institute, Zagreb, Croatia}
\affiliation{LPNHE, Sorbonne University, CNRS/IN2P3, Paris, France}
\affiliation{Karlsruhe Institute of Technology, Karlsruhe, Germany}
\affiliation{University of Frankfurt, Frankfurt, Germany}
\affiliation{Wigner Research Centre for Physics, Budapest, Hungary}
\affiliation{E\"otv\"os Lor\'and University, Budapest, Hungary}
\affiliation{Institute for Particle and Nuclear Studies, Tsukuba, Japan}
\affiliation{Okayama University, Japan}
\affiliation{University of Bergen, Bergen, Norway}
\affiliation{University of Oslo, Oslo, Norway}
\affiliation{Jan Kochanowski University, Kielce, Poland}
\affiliation{Institute of Nuclear Physics, Polish Academy of Sciences, Cracow, Poland}
\affiliation{National Centre for Nuclear Research, Warsaw, Poland}
\affiliation{Jagiellonian University, Cracow, Poland}
\affiliation{AGH - University of Science and Technology, Cracow, Poland}
\affiliation{University of Silesia, Katowice, Poland}
\affiliation{University of Warsaw, Warsaw, Poland}
\affiliation{University of Wroc{\l}aw,  Wroc{\l}aw, Poland}
\affiliation{Warsaw University of Technology, Warsaw, Poland}
\affiliation{Affiliated with an institution covered by a cooperation agreement with CERN}
\affiliation{University of Belgrade, Belgrade, Serbia}
\affiliation{Fermilab, Batavia, USA}
\affiliation{Los Alamos National Laboratory, USA}
\affiliation{University of Notre Dame, Notre Dame, USA}
\affiliation{University of Colorado, Boulder, USA}
\affiliation{University of Hawaii at Manoa, Honolulu, USA}
\affiliation{University of Pittsburgh, Pittsburgh, USA}

\author{H.~\surname{Adhikary}}
\affiliation{Jan Kochanowski University, Kielce, Poland}
\author{P.~\surname{Adrich}}
\affiliation{National Centre for Nuclear Research, Warsaw, Poland}
\author{K.K.~\surname{Allison}}
\affiliation{University of Colorado, Boulder, USA}
\author{N.~\surname{Amin}}
\affiliation{Karlsruhe Institute of Technology, Karlsruhe, Germany}
\author{E.V.~\surname{Andronov}}
\affiliation{Affiliated with an institution covered by a cooperation agreement with CERN}
\author{T.~\surname{Anti\'ci\'c}}
\affiliation{Ru{\dj}er Bo\v{s}kovi\'c Institute, Zagreb, Croatia}
\author{I.-C.~\surname{Arsene}}
\affiliation{University of Oslo, Oslo, Norway}
\author{M.~\surname{Bajda}}
\affiliation{Jagiellonian University, Cracow, Poland}
\author{Y.~\surname{Balkova}}
\affiliation{University of Silesia, Katowice, Poland}
\author{M.~\surname{Baszczyk}}
\affiliation{AGH - University of Science and Technology, Cracow, Poland}
% ...

\collaboration{NA61/SHINE Collaboration}
\noaffiliation

\begin{abstract}
    This paper presents multiplicity measurements of charged hadrons produced in 120 \gevc proton-carbon interactions. The measurements were made using data collected at the NA61/SHINE experiment during two different data-taking periods, with increased phase space coverage in the second configuration due to the addition of new subdetectors. Particle identification via $dE/dx$ was employed to obtain double-differential production multiplicities of \pip, \pim, \p, \antip, \Kp, and \Km. These measurements are presented as a function of laboratory momentum in intervals of laboratory polar angle covering the range from 0 to 450 mrad. They provide crucial inputs for current and future long-baseline neutrino experiments, where they are used to estimate the initial neutrino flux.
\end{abstract}

\maketitle


\section{Introduction}

The 120 \gevc proton-carbon interaction is of particular importance for long-baseline neutrino oscillation experiments at Fermilab. The NuMI facility at Fermilab creates its neutrino beam by striking a long carbon target with 120 \gevc protons~\cite{numibeamline}. This neutrino beam has served several experiments over the years, including \nova, \minerva, and MINOS. The Long-Baseline Neutrino Facility (LBNF), which will provide the neutrino beam for the Deep Underground Neutrino Experiment (DUNE), will likely use the same primary interaction to create its beam~\cite{dune_physics}. 

The reaction that initiates neutrino beam creation will produce a variety of charged and neutral hadrons. These hadrons will go on to decay into neutrinos or re-interact and create other neutrino-producing particles. Understanding the initial hadron production in a neutrino beam's primary interaction is crucial for estimating the neutrino beam flux. Varying contributions from decays of different hadron species lead to a neutrino beam with complex flavor content. In a long-baseline neutrino oscillation experiment, the initial neutrino beam flux and flavor content must be well-understood in order to precisely measure neutrino flavor oscillation. 

The NA61/SPS Heavy Ion and Neutrino Experiment (NA61/SHINE) is a fixed-target experiment located at the North Area of the CERN Super Proton Synchrotron (SPS). NA61/SHINE makes dedicated hadron production measurements in reactions relevant to neutrino physics. Hadron production measurements made at NA61/SHINE have been successfully used to improve neutrino flux estimates at existing long-baseline neutrino experiments such as T2K~\cite{Abgrall:2011ae, Abgrall:2011ts, Abgrall:2013wda, Abgrall:2015hmv, Abgrall:2012pp, Abgrall:2016jif, Berns:2018tap, Acharya:2020jtn}. NA61/SHINE has published several papers measuring hadron production processes relevant to Fermilab neutrino experiments~\cite{Aduszkiewicz:2019hhe,Aduszkiewicz:2018uts,Aduszkiewicz:2019xna}.

In 2016 and 2017, NA61/SHINE recorded two complementary datasets measuring hadron production in 120 \gevc protons on a thin carbon target (3.1\% $\lambda$). The measured differential multiplicities include the important $\nu_\mu$- and $\overline{\nu}_\mu$-producing reactions $\pCnoE \to \pipm + X$ and $\pCnoE \to \Kpm + X$ as well as the reactions $\pCnoE \to p + X$ and $\pCnoE \to \bar{p} + X$ where the outgoing (anti)protons can re-interact and lead to additional (anti)neutrino production. Each of these reactions will contribute to the DUNE neutrino beam flux. Previous flux predictions show substantial uncertainty associated with the primary proton beam interaction, and the measurements presented in this publication will be used to reduce these uncertainties~\cite{fields_beyond2020}.

This publication details the charged-hadron analysis methods, including particle identification via $dE/dx$, and reports measured double-differential multiplicities and uncertainties. A separate paper~\cite{neutralHadronArxivPaper} details \kos, \lam and \alam production in the same reaction.

\section{Experimental Setup}\label{sec:Setup}

\begin{figure*}[t]
  \centering
  \includegraphics[width=\textwidth]{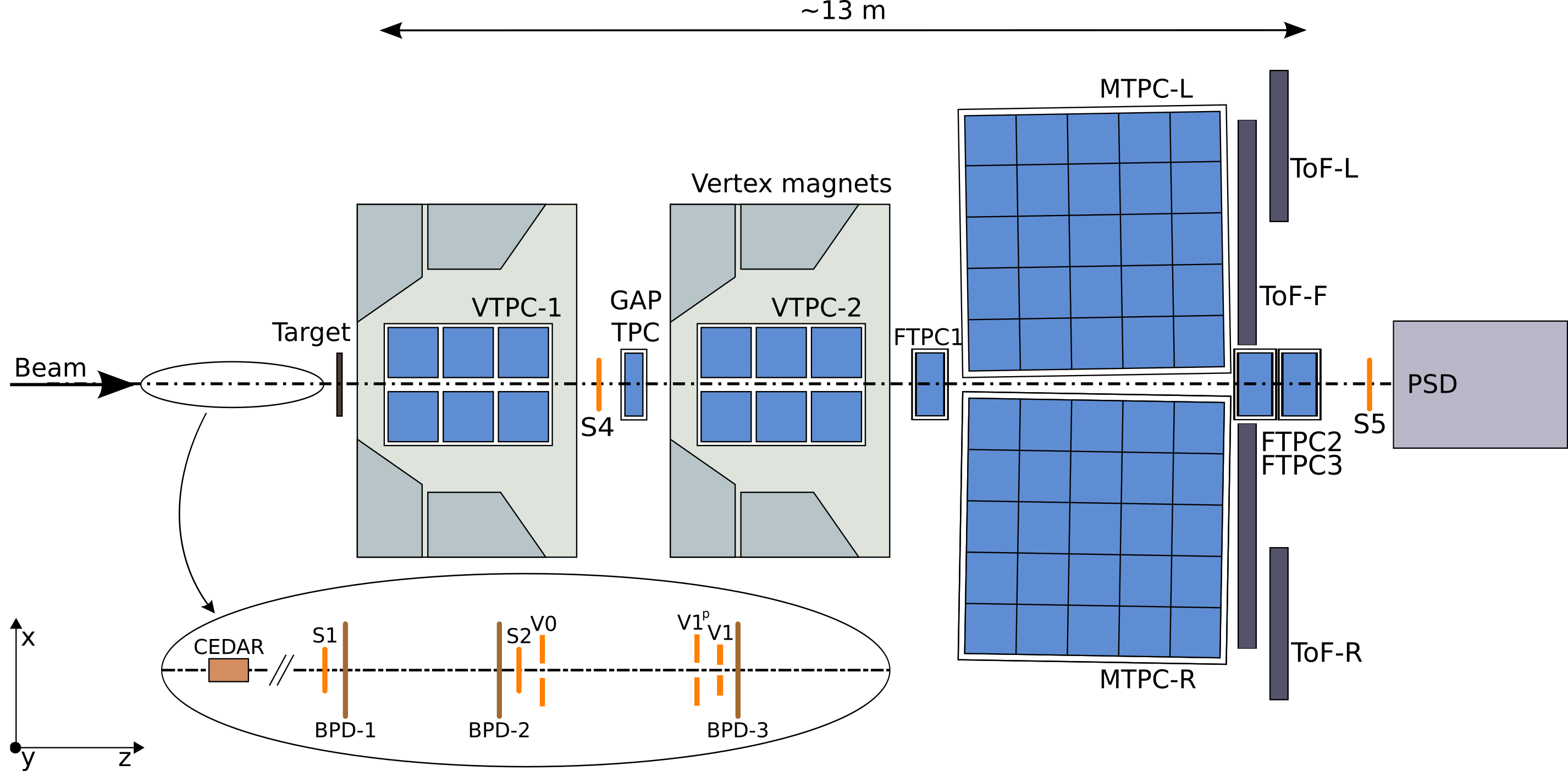}
 \caption{Top view of the NA61/SHINE experiment in the configuration used during the 2017 proton data taking. In 2016 the FTPCs were not present. Adapted from~\cite{na61detector}.} \label{fig:detectorConfiguration}
\end{figure*}

NA61/SHINE is a large-acceptance hadron spectrometer~\cite{na61detector}. Its Time Projection Chamber (TPC)-based tracking detectors are capable of recording charged particle trajectories and identifying particle species via specific ionization energy loss ($dE/dx$).

NA61/SHINE is located on the H2 beamline in Experimental Hall North 1 (EHN1) in CERN's North Area complex. The SPS provides the North Area with beams of primary 400 \gevc protons or ions with momenta in the range [13$A$ - 158$A$] \gevc. The protons can be directed into a production target to provide a beam of secondary hadrons in the range of 13 - 350 \gevc. These secondary beams contain a mixture of hadrons and leptons, and the desired beam particle species must be selected at the event level. Beam particle identification is performed by the Cherenkov Differential Counter with Achromatic Ring Focus (CEDAR)~\cite{cedar, cedar82}, located upstream of the NA61/SHINE spectrometer.

The components of the NA61/SHINE detector used to record these datasets are shown in Fig.~\ref{fig:detectorConfiguration}. Eight TPCs act as the main tracking detectors and provide $dE/dx$ measurements for particle identification. The Vertex TPCs (VTPC-1 and VTPC-2) are located inside two superconducting vertex magnets, which provide up to 9 T$\cdot$m of total bending power and enable track momentum measurement. A Time-of-Flight (ToF) system enables particle identification in select regions of phase space. Three gaseous strip Beam Position Detectors (BPDs) measure incoming beam track trajectories. The BPDs are placed 29.5 m upstream (BPD1), 8.2 m upstream (BPD2), and 0.7 m upstream of the target (BPD3). A straight line is fit to the three $(x,y)$ measurements made by the BPDs to represent the beam track trajectory.

The Gap TPC (GTPC) and three Forward TPCs (FTPCs), collectively referred to as the Beamline TPCs, enable measurement of the most forward-going tracks that pass through the beam gap in the VTPCs and Main TPCs. The FTPCs were constructed specifically to improve the forward acceptance of NA61/SHINE, and were installed in 2017~\cite{Rumberger:FTPCPaper}. The 2016 and 2017 datasets thus have significantly different track acceptance: The forward acceptance was increased for the 2017 dataset, and the 2017 magnetic field strength was reduced by half compared to the field used in 2016.

The beam trigger system, constructed from scintillators S1 and S2, veto scintillators V0 and V1 (scintillators with cylindrical holes centered on the beam), and the CEDAR detector, selects beam particles with acceptable trajectories and of the desired particle type. An interaction scintillator S4, placed downstream of the target, detects beam particles and provides information about whether or not a significant angular scatter has occurred upstream of the scintillator. S4 has a radius of 1 cm.

Interactions of 120 \gevc protons and carbon nuclei were measured in 2016 and 2017 using a thin carbon target with dimensions 25\,mm (W) x 25\,mm (H) x 14.8\,mm (L) and density $\rho$ = 1.80 g/cm$^3$, corresponding to 3.1\% of a proton-nuclear interaction length. Data was collected with the target removed to study interactions outside the carbon target. (see Table~\ref{tab:selectedEventCounts}).

\begin{table*}[htbp]
\centering
\begin{tabular}{cccccccc}
Dataset  & Target-Inserted & Target-Inserted & Target-Removed & Target-Removed \\ 
& (Recorded) & (Selected) & (Recorded) & (Selected) \\
\hline
2016  & 2.5 M & 1.5 M & 0.14 M  & 0.05 M\\
2017  & 1.5 M & 1.1 M & 0.13 M  & 0.07 M
\end{tabular}
\caption[Event Counts]{The number of recorded and selected target-inserted (target-removed) events for the 2016 and 2017 data samples.}
\label{tab:selectedEventCounts}
\end{table*}

\begin{figure*}[!ht]
  \centering
  \includegraphics[width=0.49\textwidth]{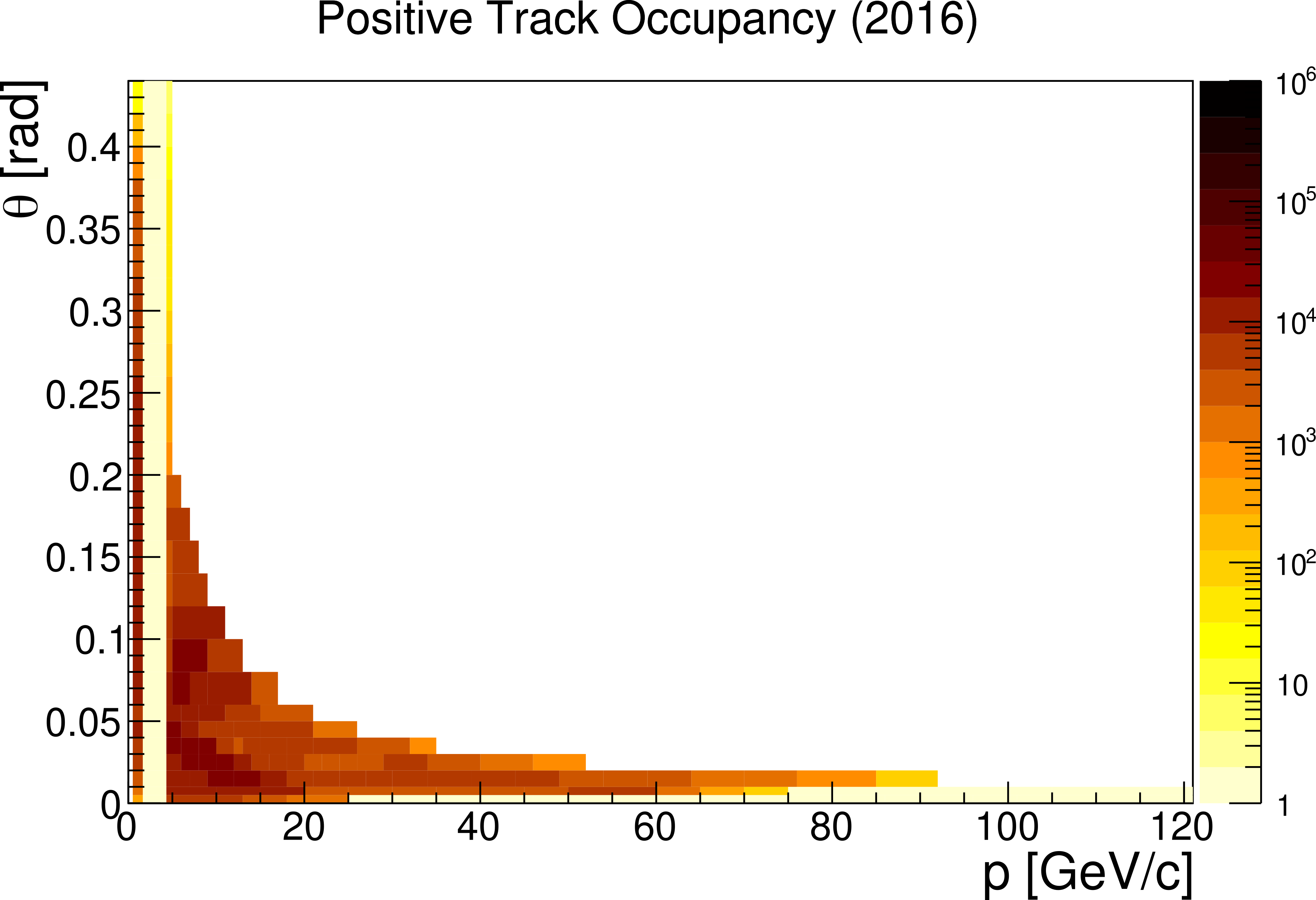}
  \includegraphics[width=0.49\textwidth]{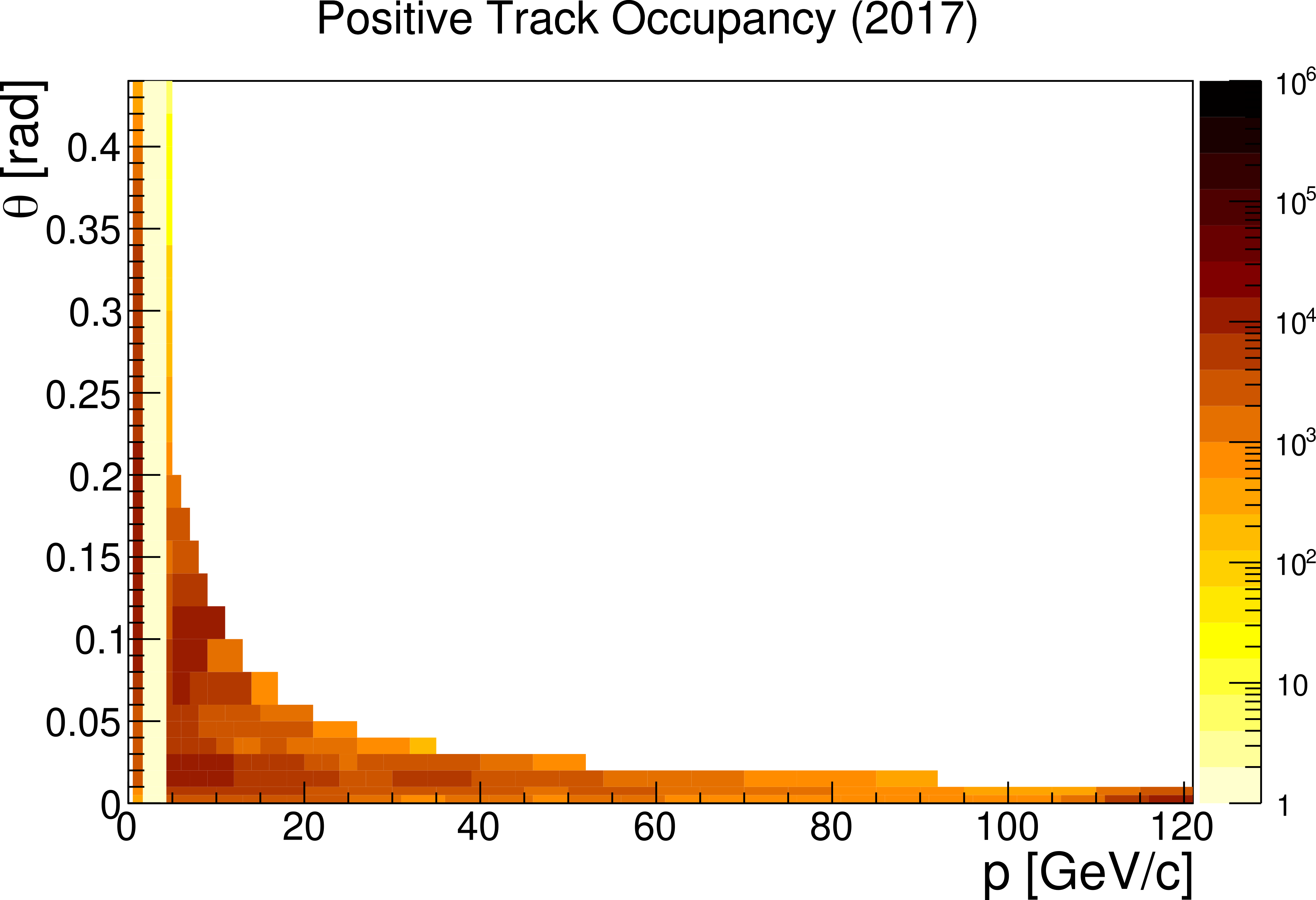}
\caption{Binning scheme and track occupancy comparison for positive tracks between 2016 dataset (\textit{left}) and 2017 dataset (\textit{right}) for the proton analysis. Note the significantly increased phase space occupancy in the forward region for the 2017 analysis. This is the result of adding the FTPCs to the NA61/SHINE detector. The empty region at low momenta corresponds to the omitted Bethe-Bloch crossing region for protons and pions. }\label{fig:occupancyComparison}
\end{figure*}

Differences in detector configuration between 2016 and 2017 lead to significantly different acceptance between the two analyses. In 2016, the magnetic field was set to the maximum possible bending strength in order to deflect forward-going charged particles into the MTPCs. This magnetic field setting has the effect of sweeping low-momentum charged particles out of detector acceptance, but decreasing fractional momentum uncertainty. In 2017, the magnetic field was reduced by half since the forward region was fully instrumented. This configuration significantly increases coverage in both the forward and low-momentum regions of phase space but comes with increased fractional momentum uncertainty. A comparison of the 2016 and 2017 charged track occupancy can be seen in Fig.~\ref{fig:occupancyComparison}.

\section{Data Reconstruction \& Simulation}\label{sec:Reconstruction}

New TPC track reconstruction software was developed for the FTPC tracking system. This track reconstruction software, called the SHINE-native Reconstruction Chain, was used to reconstruct charged tracks in all TPCs for both the 2017 dataset (with FTPCs) and the 2016 dataset (without FTPCs)~\cite{Rumberger:Thesis}. This is the first published analysis exclusively using the new reconstruction software framework. The multiplicities given by the new reconstruction framework were cross-checked with the previously-used NA61/SHINE TPC reconstruction software, and results in overlapping regions of phase space were consistent. 

\begin{figure*}[!ht]
  \centering
  \includegraphics[width=0.7\textwidth,angle=270]{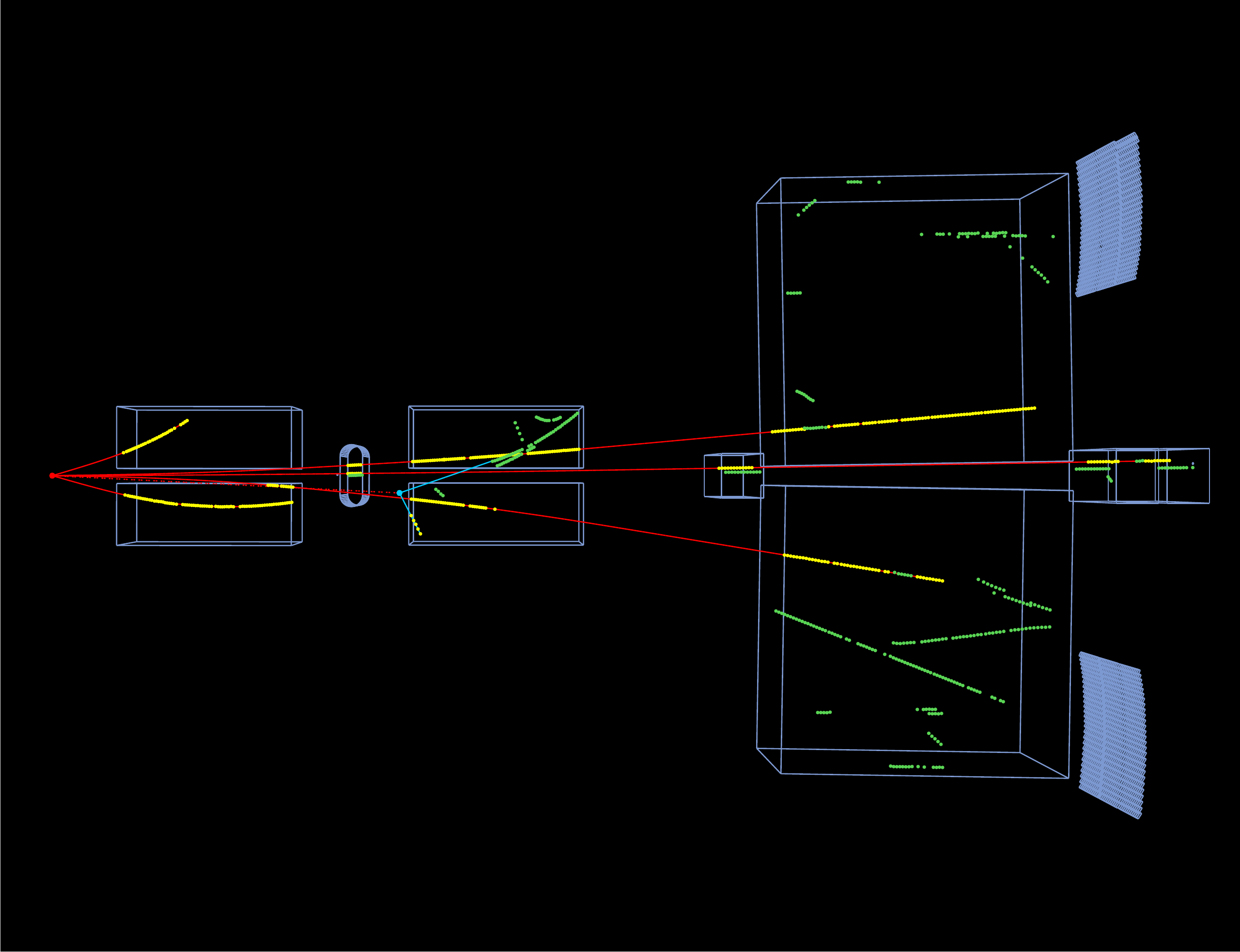}
\caption{Reconstructed event from the 2017 120 \gevc proton-carbon dataset. Event was reconstructed using the SHINE-native Reconstruction Chain. A forward-going track spanning the GTPC and all three FTPCs can be seen. Yellow points and red lines represent TPC point measurements associated with vertex tracks. Green points represent TPC point measurements associated with out-of-time beam particles or tracks produced by out-of-time beam particles.}\label{fig:FTPCTrack}
\end{figure*}

The reconstruction framework uses a Cellular-Automaton-based track seeding algorithm and a Kalman Filter track fitter\cite{Rumberger:Thesis}. Tracks are extrapolated to other TPCs, where compatible track segments are searched for and merged into the extrapolated track. The main interaction vertex is fit for using a least-squares fitter combining all compatible tracks and the BPD trajectory. Each track is re-fit with the main vertex position as an additional measurement point. Tracks originating from the main interaction vertex, called vertex tracks, are the basic input for the charged-hadron multiplicity analysis. A reconstructed vertex track passing through all three FTPCs and the GTPC can be seen in Fig.~\ref{fig:FTPCTrack}.

The SHINE software framework includes a comprehensive \GeantFour~\cite{Agostinelli:2002hh,Allison:2006ve,Allison:2016lfl} detector description called Luminance. This description includes propagation of primary particles through the detector, simulation of secondary interactions in detector components, and digitization of \GeantFour energy deposition events in the TPCs. The digitized simulated events are identical in structure to the detector raw data, and are subsequently processed with the SHINE-native Reconstruction Chain. The reconstructed simulated events form the basis for Monte-Carlo-based corrections, including the acceptance, reconstruction, and selection corrections. This analysis used \GeantFour version 10.7.0.

\section{Charged-Hadron Multiplicity Analysis} \label{sec:ChargedAnalysis}

The charged-hadron analysis includes reconstructing charged tracks associated with an event's main vertex, applying track selection criteria, fitting track $dE/dx$ distributions, and calculating identified multiplicities. The charged hadrons included in this analysis are \pipm, $p / \bar{p}$, and \Kpm.

Event and track selection for the charged-hadron analyses follow a similar methodology to previous NA61/SHINE measurements for \piCnoE and \piBe~\cite{Aduszkiewicz:2019hhe}. Selection criteria used in this analysis are discussed in the following subsections.

\subsection{Event Selection}

Three selection criteria are applied at the event level prior to track selection. The total number of recorded events and the number of events passing event selection criteria are shown in Table~\ref{tab:selectedEventCounts}.

\begin{itemize}
\item{Beam Divergence Cut (BPD Cut)}
\end{itemize}

To mitigate systematic effects related to beam particles with large angles, a cut is applied to each measured beam particle trajectory. Beam tracks with significant angle will miss the S4 scintillator and cause an interaction trigger, even if no significant interaction occurred. The BPD cut ensures that the unscattered trajectory of each beam track points to within 0.95 cm of the center of the S4 scintillator.

\begin{itemize}
\item{Well-Measured Beam Trajectory Cut (BPD Status Cut)}
\end{itemize}

Events with a well-measured beam trajectory are selected using the BPD status. Any one of the three BPDs may report an error during the clusterization and fitting process due to transient noise in the detector or another ionizing particle passing through the detector simultaneously. The BPD status cut ensures that either all three detectors measured the six coordinates of a particle's trajectory and a straight line fit converged, or that two of the detectors reported satisfactory measurements and a straight line fit converged. BPD3 is required to have a single cluster with well-measured $(x,y)$ coordinates, ensuring that no significant scatter occurred upstream of BPD3.

\begin{itemize}
\item{Off-Time Beam Particle Cut (WFA Cut)}
\end{itemize}

Events containing an off-time beam particle within $\pm 0.8$ \textmu{}s of the triggering particle are removed. The Waveform Analyzer (WFA) records signals in the trigger scintillators near the triggered event, including those from beam particles not associated with the interaction trigger. These are known as off-time beam particles. The arrival of a subsequent beam particle closely-spaced in time may hit the S4 scintillator and appear to be a non-interaction. In addition, off-time beam particles may interact in the target. If the off-time particle arrives several hundred nanoseconds after the triggering particle, off-time tracks may be reconstructed to the event main vertex.

For spectra analysis, only interaction trigger events are considered.
After the described event selection cuts, 2.1 M (2016) and 1.5 M (2017) target-inserted and 0.07 M (2016) and 0.08 M (2017) target-removed events were selected. Differences in the target-inserted and target-removed ratios between the two years are due to different amounts of beam time being devoted to target-removed event collection.

\subsection{Selection of Charged Tracks}

\begin{itemize}
\item{Topological Cuts}
\end{itemize}

This analysis classifies charged tracks into two categories, Right-Side Tracks (RST) and Wrong-Side Tracks (WST), according to a track's charge $q$ and the orientation of its momentum vector $\vec{p}$ with respect to the magnetic field, which points in the positive $y$-direction:

\begin{equation}
\begin{cases}
p_x/q > 0 & \text{RST}, \\
p_x/q < 0 & \text{WST}.
\end{cases}.
\end{equation}

Right-side tracks are aligned to the orientation of the TPC readout pads, which are tilted in order to compensate for average track angles. RSTs typically exhibit a narrower $dE/dx$ distribution for a given particle species and momentum range. In this analysis, WSTs are used to cross-check the RSTs for consistency, and RSTs are used to calculate the final identified hadron spectra. The RST/WST designation is only applied to tracks with polar angle $\theta \geq 10$ mrad, as the azimuthal angle $\phi$ becomes difficult to measure at small polar angles.

\begin{itemize}
\item{Track Quality Cuts}
\end{itemize}

In order for a track to have a well-estimated momentum, the track must have a sufficient number of point measurements (referred to as ``clusters'') in a VTPC or the GTPC. Passing through one of the VTPCs alone is enough for a sufficient momentum estimate and $dE/dx$ measurement. For tracks passing through the GTPC and missing the VTPCs, additional measurements in the MTPCs or FTPCs are required for $dE/dx$ measurement. Allowed topologies for $dE/dx$ analysis are either 20 total clusters in VTPC1 + VTPC2, 3 clusters in the GTPC and 20 additional clusters in the MTPCs, or 3 clusters in the GTPC and 6 additional clusters in the FTPCs. In addition to passing the number of cluster cuts, selected tracks must have an impact parameter less than 2 cm in total distance from the main interaction vertex. The reconstructed main interaction vertex must be within $\pm 5$ cm of the target center along the beam axis.

\begin{itemize}
\item{Acceptance Cuts}
\end{itemize}

The detector acceptance as a function of track azimuthal angle $\phi$ varies significantly with polar angle $\theta$ and track topology. Significant acceptance cuts were implemented for each angular bin in $(p,\phi)$ space in order to accept tracks in regions of uniform acceptance as a function of $\phi$. This allows for the extrapolation of track multiplicity into unmeasured regions, as particle production is independent of azimuthal angle. This extrapolation is performed by using a Monte Carlo correction factor, which will be described in Sec.~\ref{sec:chargedMCCorrections}.

\begin{itemize}
\item{$dE/dx$ Cuts}
\end{itemize}

\begin{figure*}[t]
  \centering
  \includegraphics[width=0.45\textwidth]{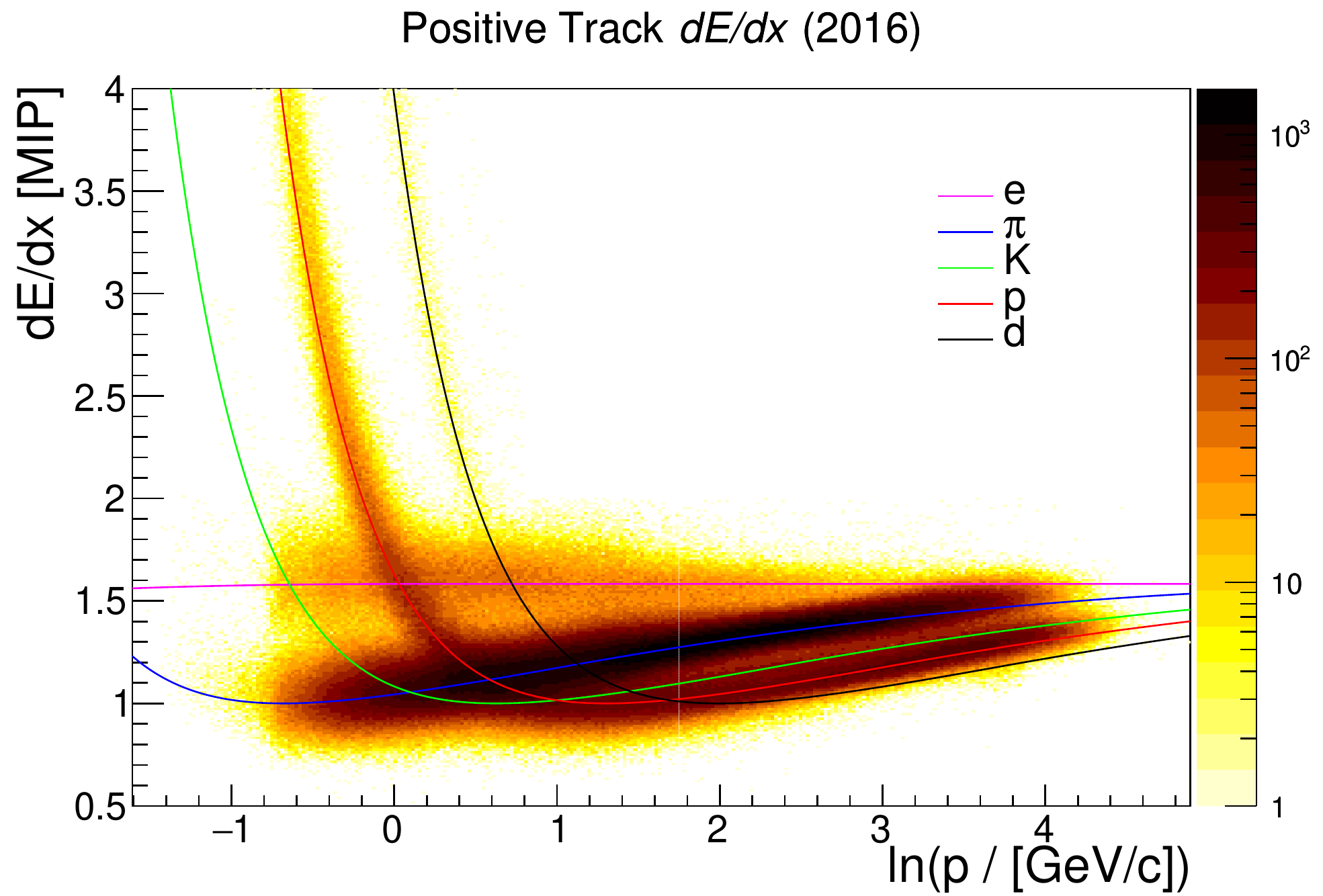}
  \includegraphics[width=0.45\textwidth]{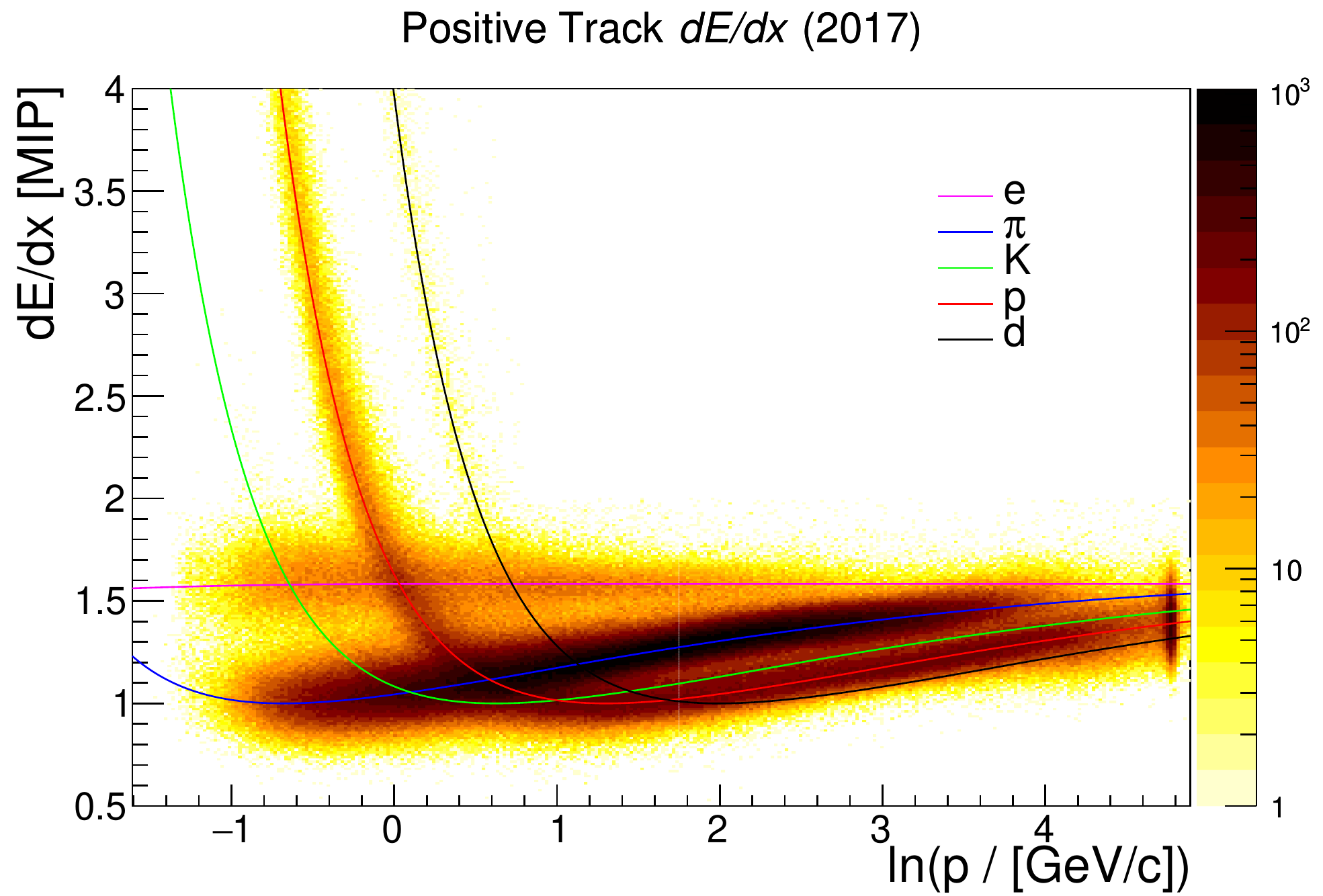}
  \includegraphics[width=0.45\textwidth]{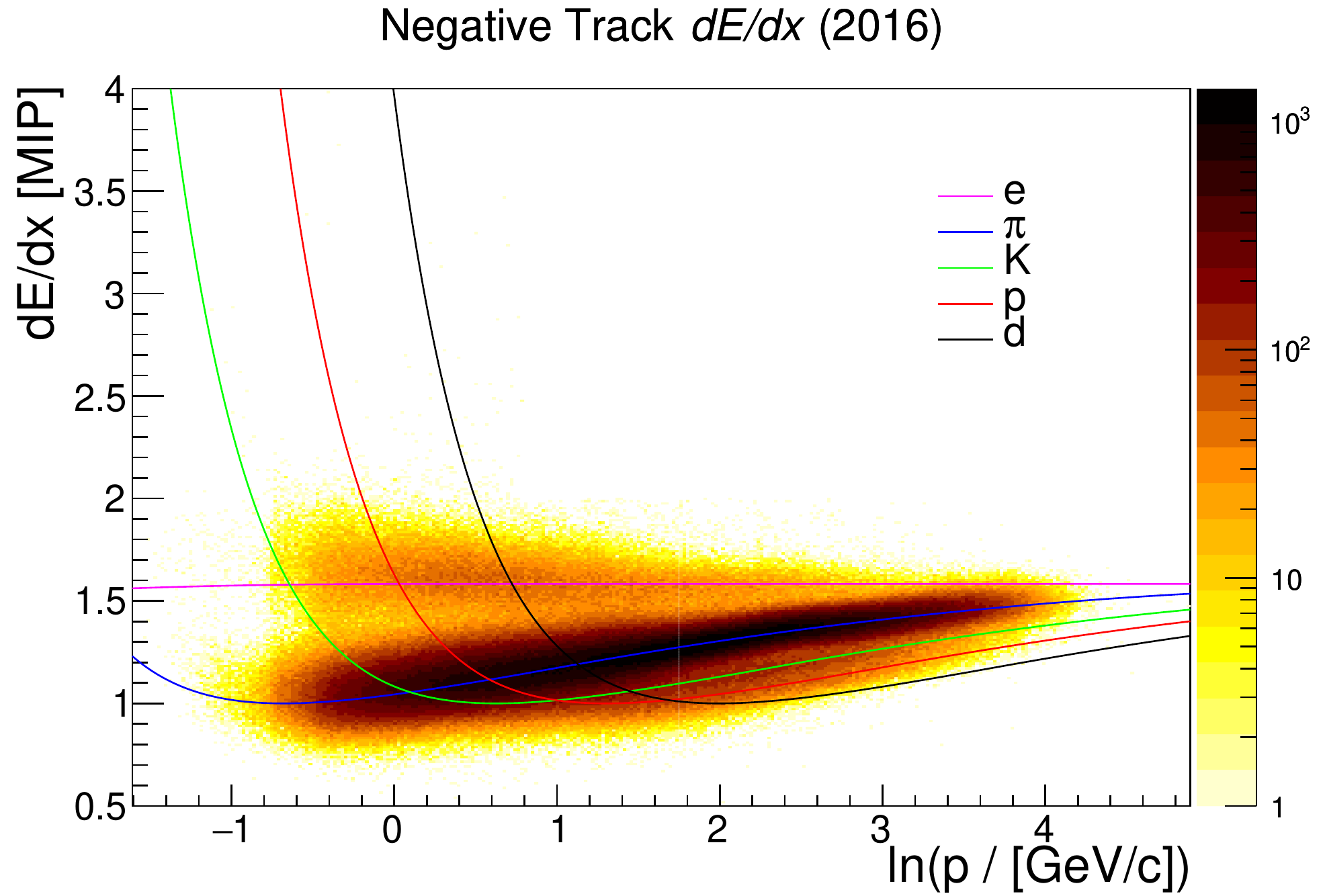}
  \includegraphics[width=0.45\textwidth]{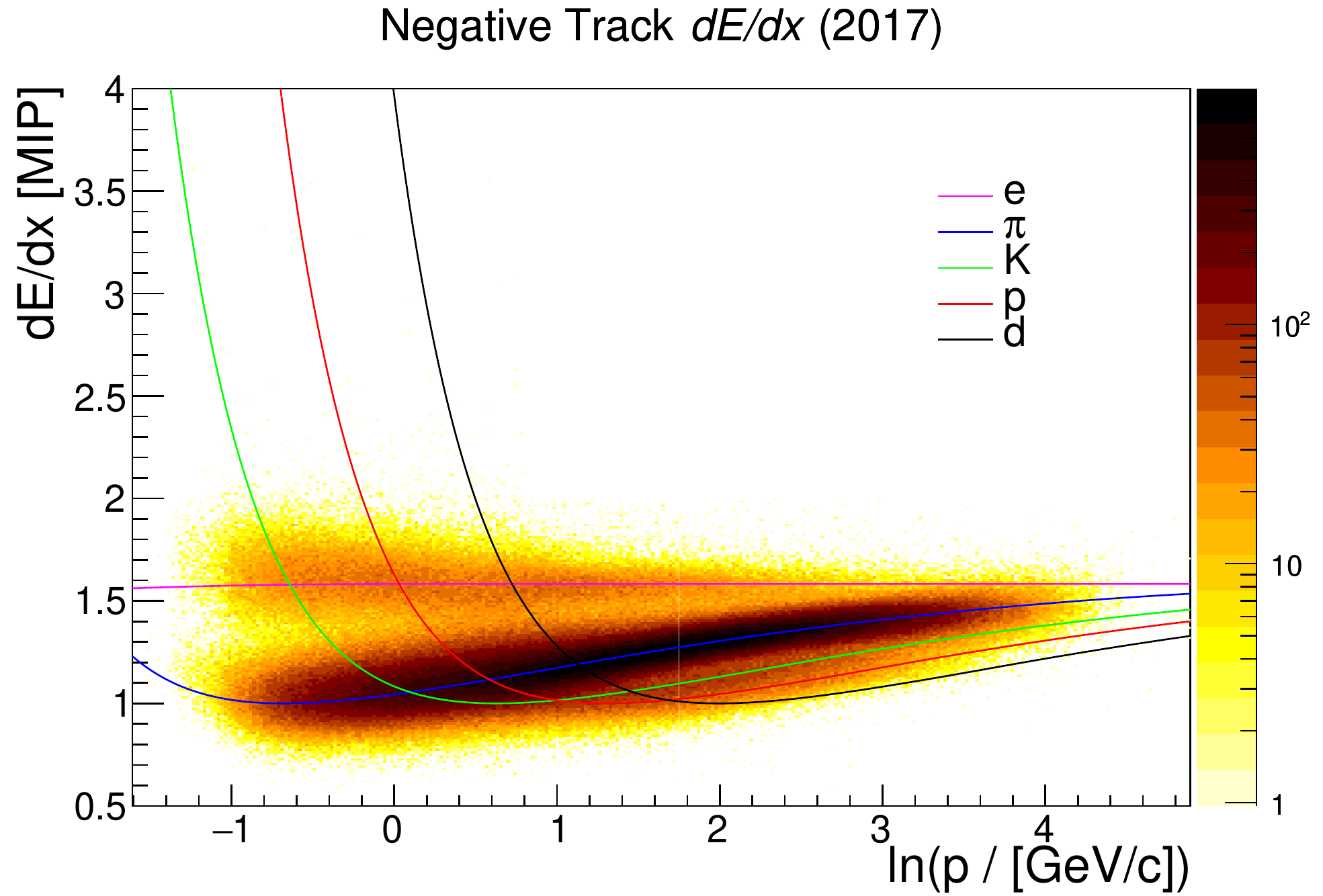}
\caption{Two-dimensional distributions of charged track $dE/dx$ vs. $\ln(p)$  for the 2016 and 2017 datasets after applying track quality cuts. The lines represent Bethe--Bloch predictions for various particle species. Increased acceptance in the 2017 dataset is visible in the extension of the distribution to lower total momenta (due to a lower magnetic field setting) and a prominent peak in positively charged track $dE/dx$ at the beam momentum (ln($p$ / [\gevc]) = 4.78). }\label{fig:dEdxVsLogP}
\end{figure*}

This analysis identifies charged hadrons using track $dE/dx$, as shown vs vs. $\ln(p)$ in Fig.~\ref{fig:dEdxVsLogP}, and therefore cannot report results in the vicinity of Bethe--Bloch crossings. Bethe--Bloch crossings are defined as momentum regions in which two species' Bethe--Bloch expectations are within 5\% of one another. For the \pipm analysis, the proton Bethe--Bloch crossing region $p \in [1.64,2.02]$ \gevc is omitted. For the $p/\bar{p}$ analysis, both the \pipm and \Kpm Bethe--Bloch crossings are omitted, as is the small momentum region between the two crossing ranges, giving a total omitted region of $p \in [1.64,4.32]$ \gevc. For the \Kpm analysis, the pion and proton crossing regions are omitted, giving a total omitted region of $p \in [0.95,2.02]$ \gevc. A final cut on $dE/dx$ quality was imposed in order to exclude doubly charged tracks and tracks with large $dE/dx$ distortions. This cut omits tracks with $p > 2.2$ \gevc and $dE/dx$ $> 2.0$ times that of a mimimum-ionizing particle (MIP).

The number of remaining charged tracks for each charged-hadron analysis can be seen in Table~\ref{tab:selectedTracks}.

\begin{table*}[htbp]
\centering
\begin{tabular}{cccccccc}
Dataset & $\pi^\pm$ Analysis & 
$p$ / $\bar{p}$ Analysis &
$K^\pm$  Analysis\\
\hline
2016  & 2.1 M (9 K) & 1.5 M (8 K) & 1.2 M (7 K) \\
2017  & 1.3 M (15 K) & 0.9 M (13 K) & 0.7 M (12 K) \\
\end{tabular}
\caption[Charged Tracks Passing Selection Cuts]{The number of target-inserted (target-removed) charged tracks passing selection cuts for the 2016 and 2017 data samples.}
\label{tab:selectedTracks}
\end{table*}

\subsection{$dE/dx$ Distribution Fits}

Charged tracks passing selection criteria are separated by charge, sorted into kinematic analysis bins, and fit for yield fractions corresponding to five charged particles: $e$, $\pi$, $K$, $p$, and $d$. The mean value of $dE/dx$ while traversing a specific medium, $\langle\epsilon\rangle$, depends on particle velocity $\beta$. This enables the separation of particles with different masses for a given range of momentum. A likelihood-based fit is performed in each analysis bin to estimate the fractional content of each particle species. 

\subsection{$dE/dx$ Fit Function}

This section details the fit function used to obtain the fractional particle species content. The fit function is identical to the one used in the analysis of 2016 \pip + C and \pip + Be data~\cite{Aduszkiewicz:2019hhe}, with additional support for the inclusion of tracks with clusters in the FTPCs. 

A projection of the $dE/dx$ distribution for a given momentum range and particle species will resemble a straggling function~\cite{StragglingFunctions}, exhibiting a long tail toward large energy deposit. When this distribution is truncated at the $[0,50]$ percentiles, i.e. the largest 50\% of the samples are removed, the remaining $dE/dx$ samples are well-described by an asymmetric Gaussian function: 

\begin{equation}
  f(\epsilon,\sigma) =
  \frac{1}{\sqrt{2\pi\sigma}}e^{-\frac{1}{2}\left(\frac{\epsilon-\mu}{\delta\sigma}\right)^2}
  \qquad \delta =
  \begin{cases}
    1-d, & \epsilon \leq \mu \\
    1+d, & \epsilon > \mu
  \end{cases}.
\end{equation}

Here $\epsilon$ is the mean $dE/dx$ given by the Bethe--Bloch equation, $d$ is a parameter describing the asymmetry of the distribution, $\sigma$ is the base distribution width, and $\mu$ is the distribution peak location, given by $\langle\epsilon\rangle - \frac{4d\sigma}{\sqrt{2\pi}}$.

The width of this distribution depends on the number of $dE/dx$ samples recorded in each detector, denoted by $N_{cl\;A}$ where $A$ indicates one of the TPCs, the mean $dE/dx$ itself, $\langle\epsilon\rangle$, a scaling parameter, $\alpha$:

\begin{equation} \label{eq:DEDXWidth}
  \sigma = \frac{ \langle\epsilon\rangle^\alpha }{ \sqrt{ \frac{N_\textrm{cl\;Up}}{\sigma^2_\textrm{0\;Up}} +
      \frac{N_\textrm{cl\;V}}{\sigma^2_\textrm{0\;V}} +
      \frac{N_\textrm{cl\;M}}{\sigma^2_\textrm{0\;M}} +
      \frac{N_\textrm{cl\;F}}{\sigma^2_\textrm{0\;F}}} }
\end{equation}

where $N_\textrm{cl\;A}$ denotes the number of $dE/dx$ samples measured in detector $A$ and $\sigma_\textrm{0\;A}$ denotes the base $dE/dx$ width corresponding to detector $A$. $Up$ denotes the upstream two sectors of VTPC1, $V$ denotes the remainder of the VTPCs, $M$ denotes the MTPCs, and $F$ denotes the FTPCs. The difference in base $dE/dx$ widths $\sigma_\textrm{0\;A}$ originates in the differing pad geometries in the detectors.

The likelihood function $LL$ is a sum over all tracks, separated by charge:

\begin{widetext}
\begin{align*}
  LL(\epsilon,p,N_{cl\;A};Y_e^\pm,Y_\pi,Y_K^\pm,Y_P^\pm,Y_d^\pm) = 
  \sum_{i}^{i \; \in                                                                                        
    \; + \; \textrm{tracks}} \left( \sum_j
  \frac{Y_j}{\sqrt{2\pi\sigma_i}}e^{-\frac{1}{2}\left(\frac{\epsilon_i-\mu_j}{\delta\sigma_i}\right)^2}
  \right) + \\
  \sum_{k}^{k \; \in                                                                                        
    \; - \; \textrm{tracks}} \left( \sum_l
  \frac{Y_l}{\sqrt{2\pi\sigma_k}}e^{-\frac{1}{2}\left(\frac{\epsilon_k-\mu_l}{\delta\sigma_k}\right)^2}
  \right),
  \qquad
  \begin{cases}
    j \in e^+, \pi^+, K^+, p^+, d^+ \\
    l \in e^-, \pi^-, K^-, p^-, d^-
  \end{cases}.
\end{align*}
\end{widetext}

Here $Y_j$ is the fractional yield corresponding to particle species $j$. 

Several constraints are imposed when fitting for the species yields. The fractional yields for each charge are constrained to sum to unity. Soft constraints are employed to enforce physical limits, such as the ordering of particle species $dE/dx$ for a given momentum. The raw yield for a given species is obtained by multiplying the fractional yield of species $j$ in kinematic bin $k$ by the total number of tracks in the bin $N_k$ for a given charge:

\begin{equation}
    y_{j,\;k}^\textrm{raw} = N_k Y_{j,\;k}.
\end{equation}

\begin{figure*}[t]
  \centering
  \includegraphics[width=0.495\textwidth]{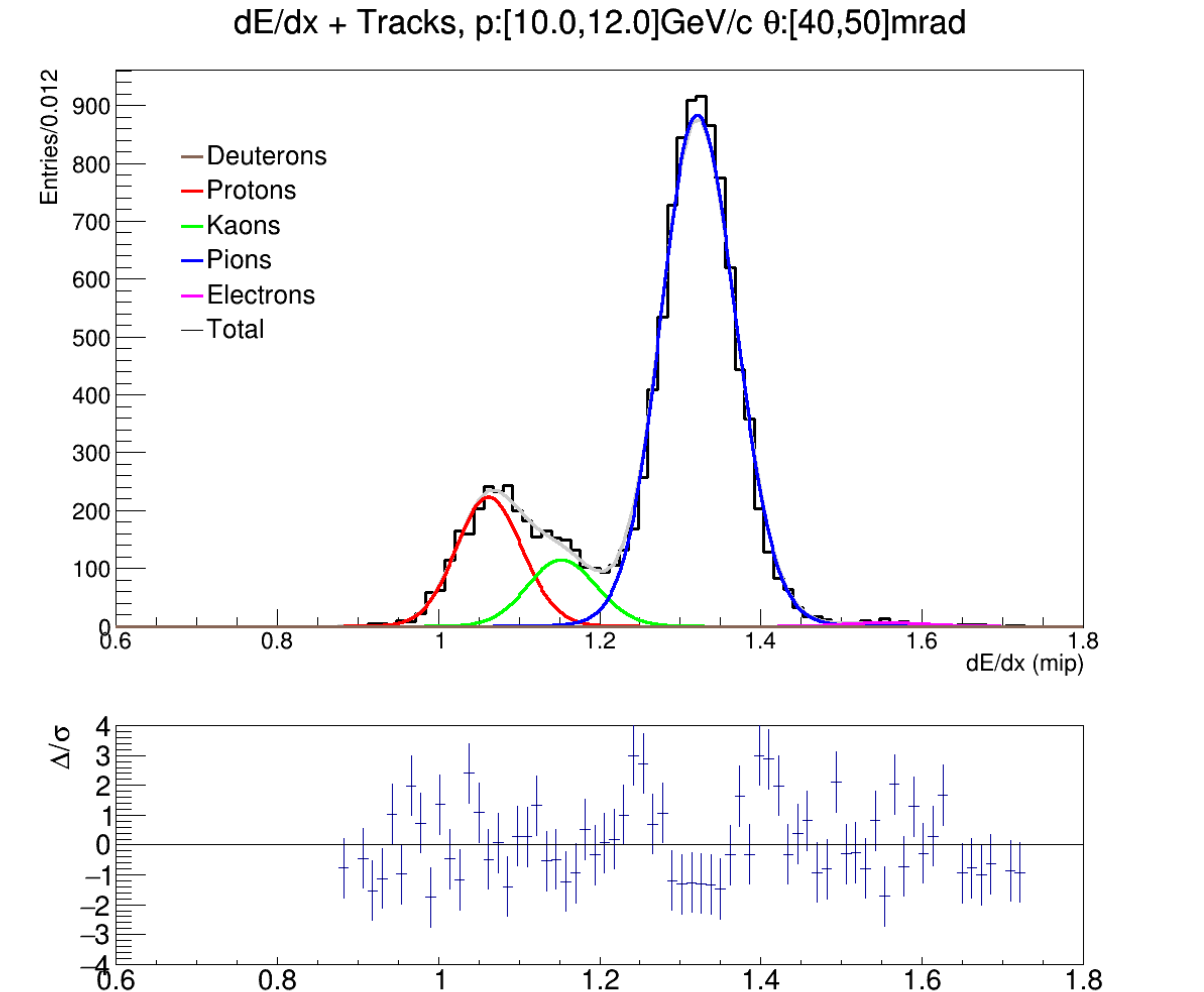}
  \includegraphics[width=0.495\textwidth]{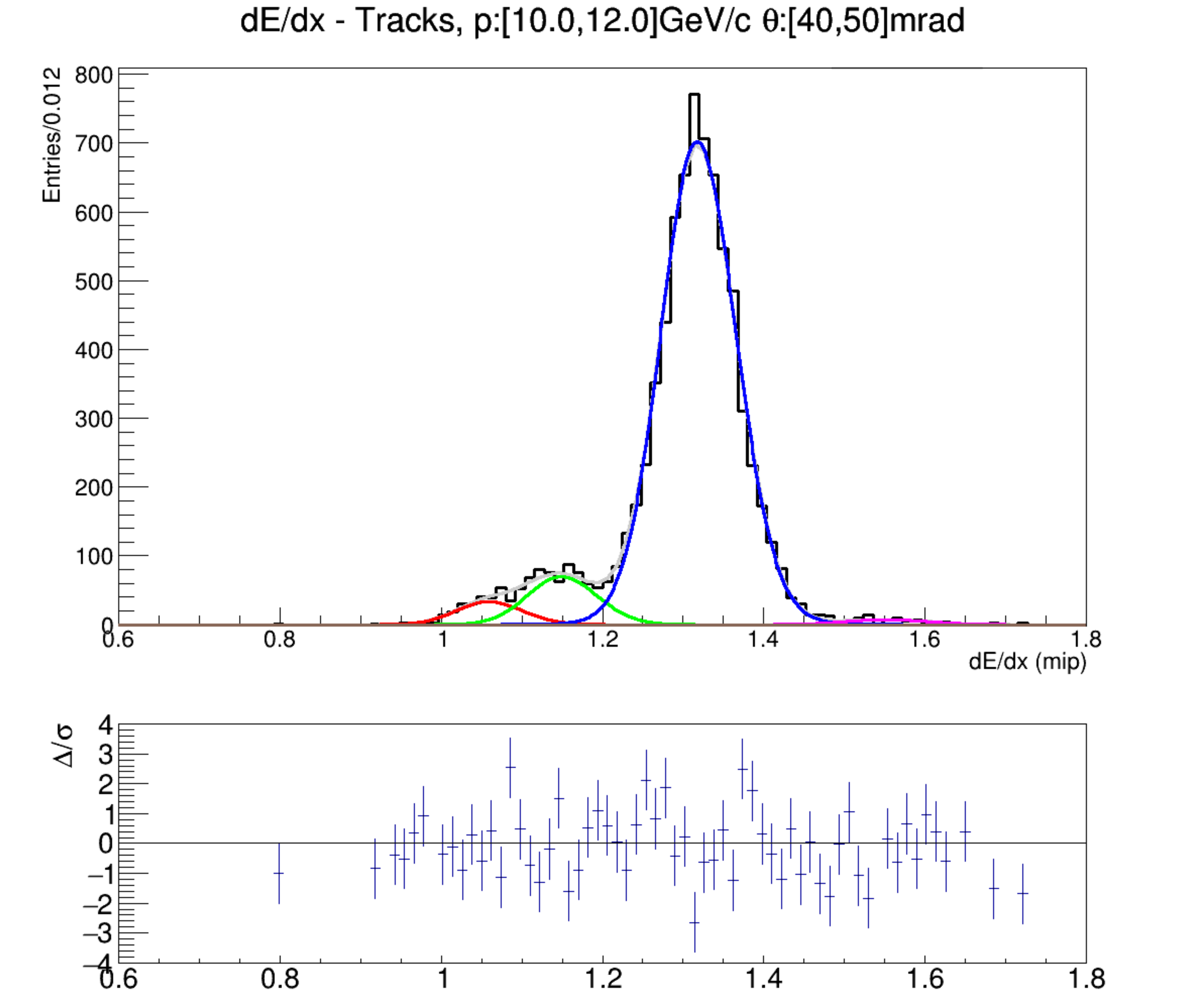}
\caption{Example $dE/dx$ distribution fit for one kinematic bin showing positively (left) and negatively (right) charged track distributions. This kinematic bin shows an abundance of pions in both the positively and negatively charged track distributions, a significant fraction of protons, and a lack of antiprotons. This is characteristic of the 120 \gevc proton-carbon reaction.}\label{fig:dedxFit}
\end{figure*}

Raw yields are obtained for both target-inserted and target-removed data samples. An example $dE/dx$ distribution fit for one kinematic bin can be seen in Fig.~\ref{fig:dedxFit}.

\subsection{Monte Carlo Corrections}\label{sec:chargedMCCorrections}

Monte Carlo corrections are used to restore tracks removed by various cuts, and correct for detector acceptance, background contributions, and reconstruction inefficiencies. The total correction factor for a given analysis bin $k$ may be broken down into its constituent parts:

\begin{widetext}
\begin{equation} 
  c_k = \frac{N(\textrm{Simulated charged tracks from production events})_k}
  {N(\textrm{Selected reconstructed charged tracks})_k} = 
  c_{\textrm{acc}} \times c_{\textrm{sel}} \times c_{\textrm{rec eff}} \times c_{\textrm{fd}}.
\end{equation}
\end{widetext}

These corrections are calculated by counting the number of simulated charged tracks in each analysis bin and dividing by the number of accepted reconstructed simulated charged tracks in each bin. $c_{\textrm{acc}}$ is the correction associated with acceptance cuts, $c_{\textrm{sel}}$ is the correction associated with track quality cuts, $c_{\textrm{rec eff}}$ is the correction associated with reconstruction efficiency, and $c_{\textrm{fd}}$ is the correction associated with feed-down tracks, or tracks originating from weakly decaying \kos, \lam, or \alam. 

\subsection{$dE/dx$ Fit Bias Corrections}\label{sec:dedxFitBiasCorrections}

An additional correction was calculated to remove biases introduced during the $dE/dx$ fitting procedure. To estimate these biases, a dedicated $dE/dx$ Monte Carlo was made. Using the full fit results from each analysis bin, the fit parameters were varied, and the individual track $dE/dx$ in each bin was re-simulated with the varied parameters. The resulting $dE/dx$ distributions were then re-fit and studied. The difference between simulated and fit particle yields was recorded and the mean of the values was taken as the fit bias. The explicit correction is given by

\begin{equation}
  c^\textrm{Fit}_i = \frac{1}{N_\textrm{trials}}
  \sum^{N_\textrm{trials}}_{n = 1}\left( \frac{y_n^\textrm{fit} -
    y_n^\textrm{true}}{y_n^\textrm{true}} \right).
\end{equation}

A trial represents a re-simulation of track $dE/dx$ conducted with an independent set of varied $dE/dx$ fit parameters. The particle species yields were kept fixed and the fit parameters ($\sigma_A$, $\alpha$, $d$) were varied according to their variations observed across kinematic bins. Fifty total trials were created. The standard deviation of the differences was also recorded and taken as the $dE/dx$ fit uncertainty. Typical fit bias corrections for the charged pion analysis are less than 2\%, and for the proton and charged kaon analyses are less than 4\%.

\subsection{Feed-Down Re-Weighting}

\begin{figure*}[!ht]
  \centering
  \includegraphics[width=0.3\textwidth]{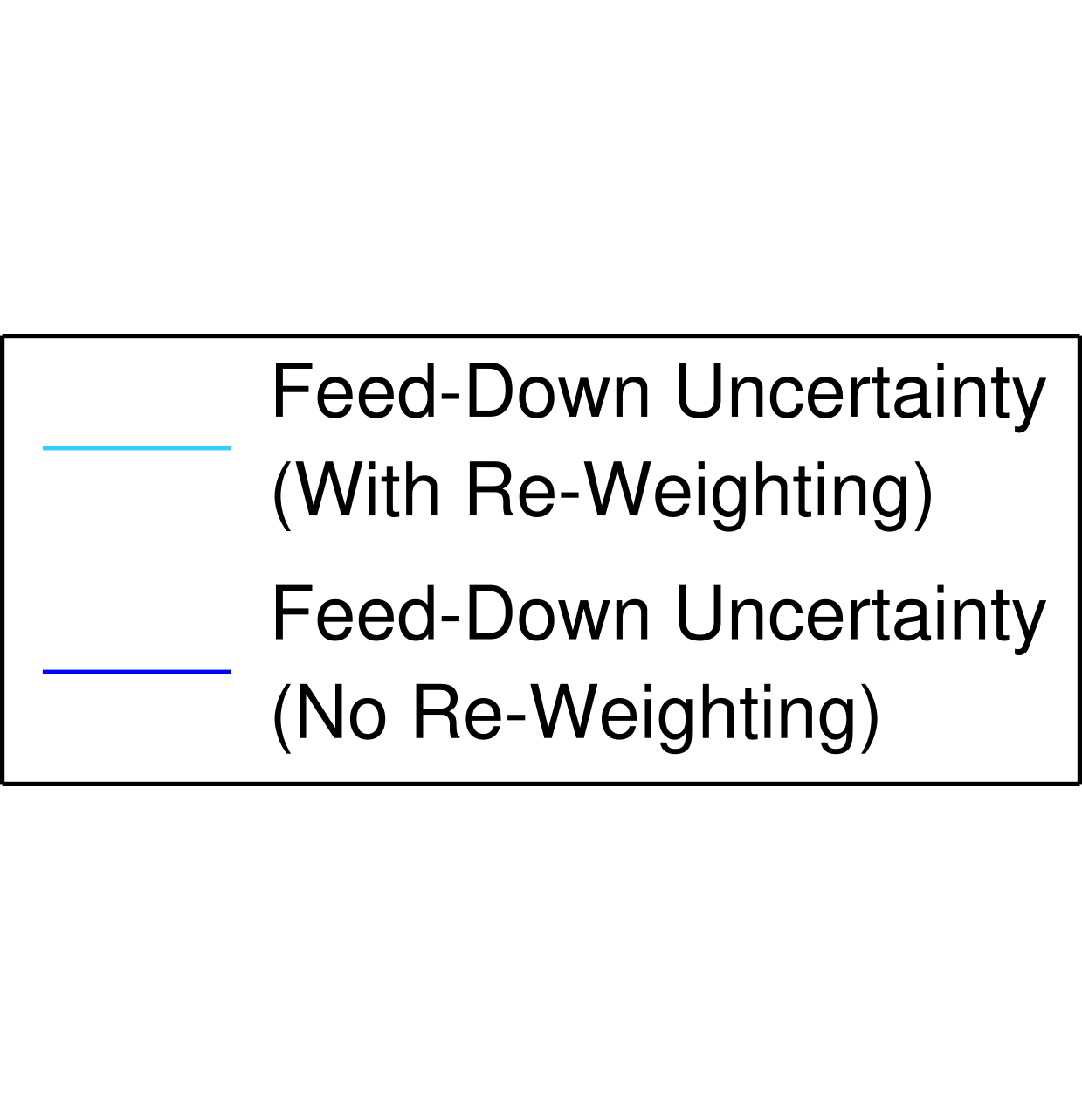}
  \includegraphics[width=0.33\textwidth]{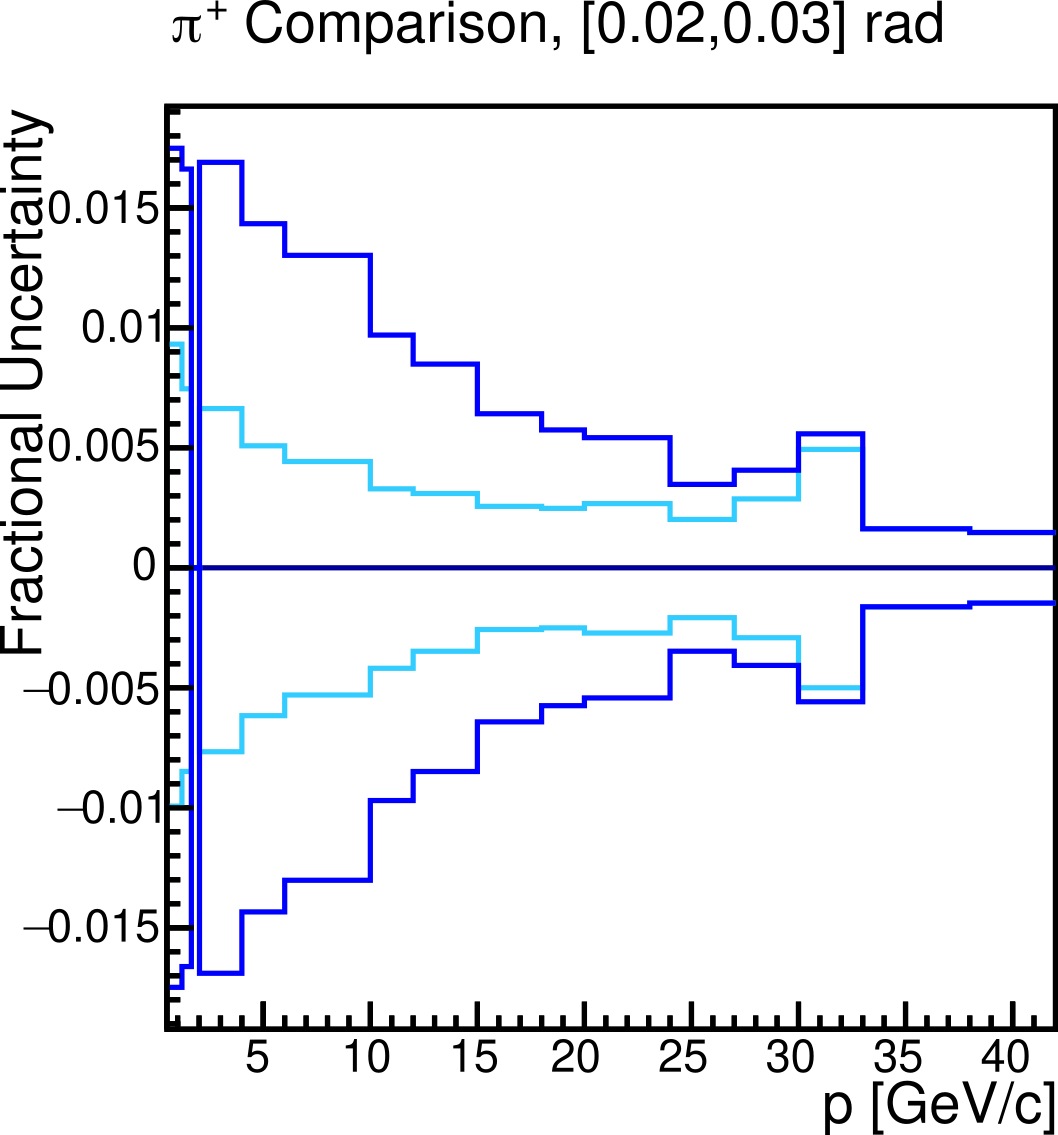}
  \includegraphics[width=0.33\textwidth]{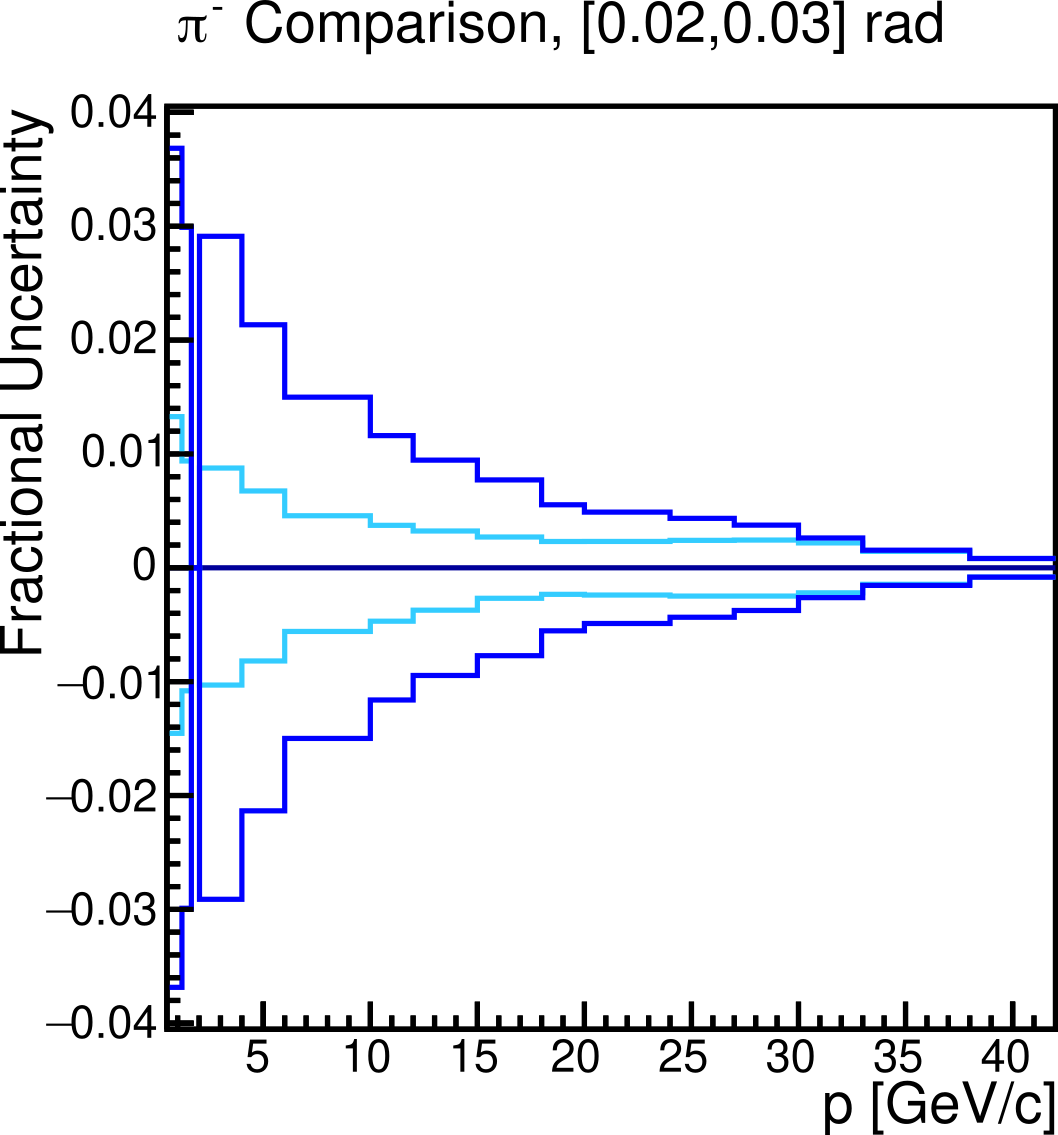}
\caption{Comparison of uncertainties associated with feed-down correction with and without the inclusion of neutral-hadron multiplicity measurements as constraints~\cite{neutralHadronArxivPaper}. Uncertainties are reduced from more than 1.5\% to less than 0.5\% for $\pi^+$ (\emph{left}) and from more than 3\% to less than 1\% for $\pi^-$ (\emph{right}). Only one representative angular bin is shown.} \label{fig:feedDownComparisonPi}
\end{figure*}

The feed-down correction $c_{\textrm{fd}}$ is estimated using Monte Carlo models. However, these models do not accurately predict weakly decaying neutral-hadron multiplicities, and large variation among the model predictions is common. The feed-down corrections can be constrained and improved using NA61/SHINE measurements of \kos, \lam, and \alam production in 120 \gevc proton-carbon interactions~\cite{neutralHadronArxivPaper}, significantly reducing systematic uncertainties associated with Monte Carlo model variations. The re-weighting factor for a kinematic bin $w_i$ is given by

\begin{equation}
    w_i = \frac{m^\textrm{Data}_i}{m^\textrm{MC}_i}
\end{equation}

where ${m^\textrm{Data}_i}$ is the measured multiplicity of a particular neutral hadron in bin $i$, and ${m^\textrm{MC}_i}$ is the Monte Carlo multiplicity in the same kinematic bin. This factor is applied to \pipm, $p$, and $\bar{p}$ originating from decays of simulated \kos, \lam, or \alam in regions covered by NA61/SHINE measurements. In regions not covered by existing measurements, the Monte Carlo predictions are not re-weighted.

\begin{figure*}[!ht]
  \centering
  \includegraphics[width=0.3\textwidth]{FeedDownUncertaintiesLegend.png}
  \includegraphics[width=0.33\textwidth]{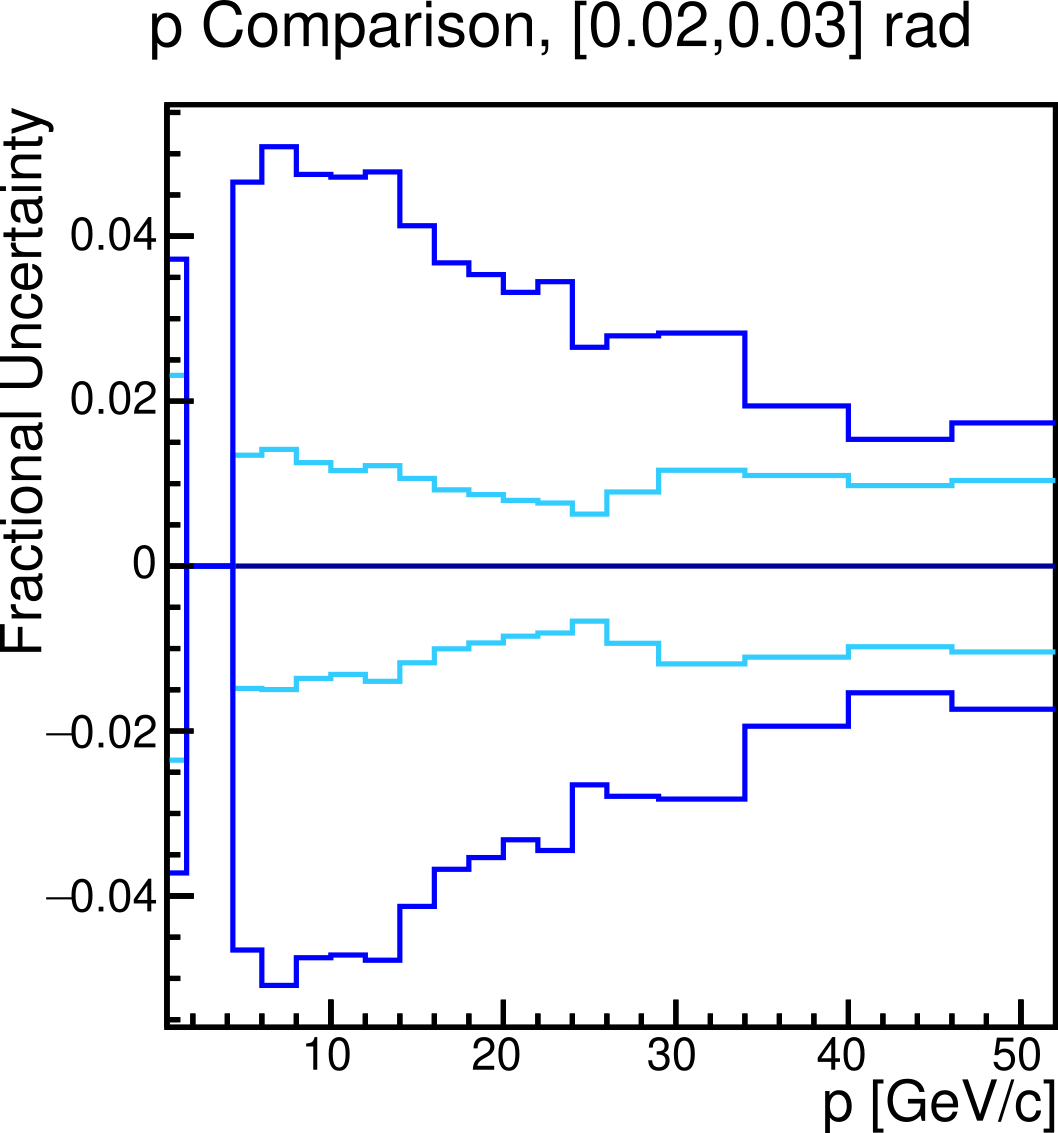}
  \includegraphics[width=0.33\textwidth]{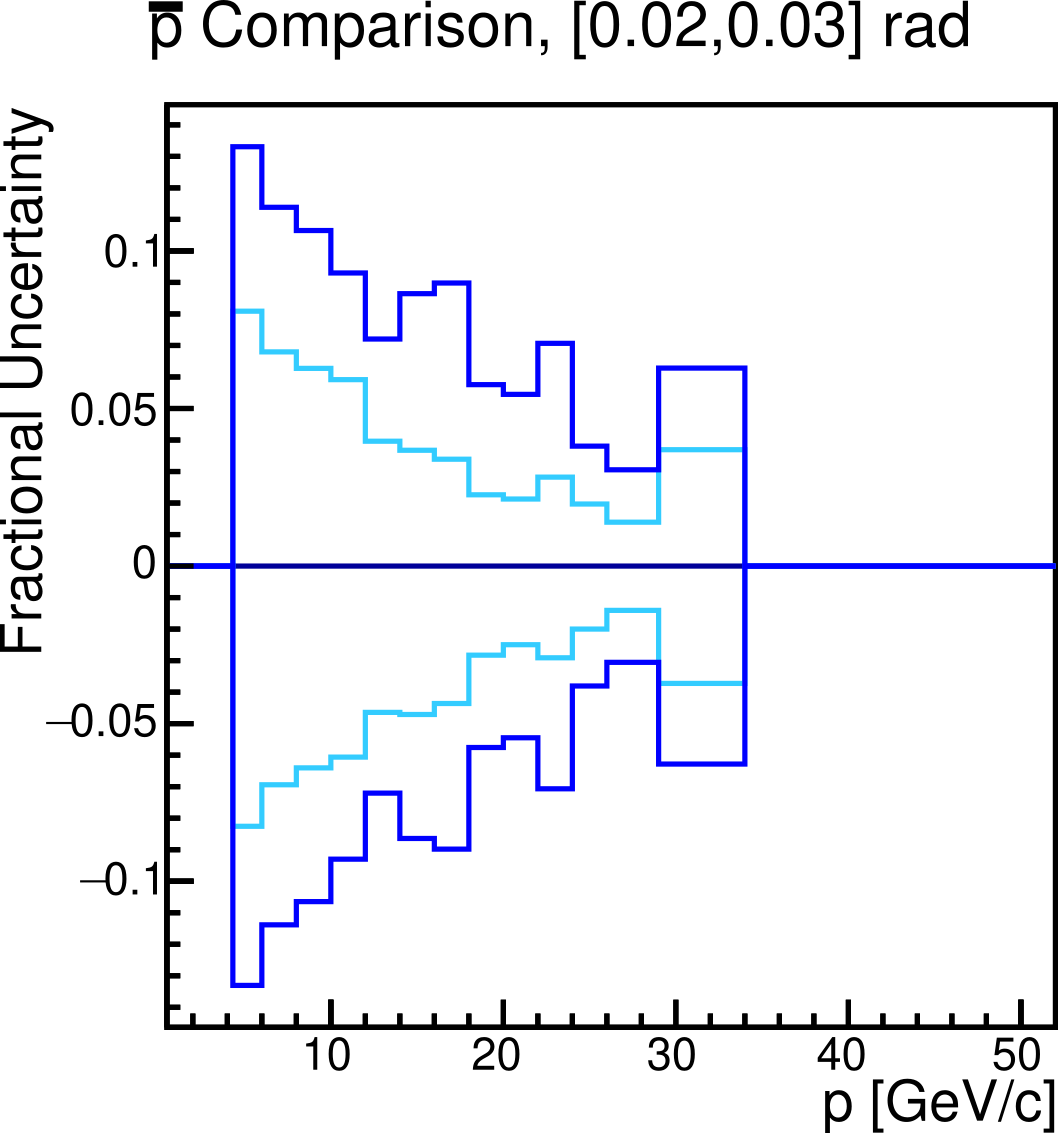}
\caption{Comparison of uncertainties associated with feed-down correction with and without the inclusion of neutral-hadron multiplicity measurements as constraints~\cite{neutralHadronArxivPaper}. Uncertainties are reduced from up to 5\% to less than 2\% for $p$ (\emph{left}) and from more than 10\% to around 6\% for $\bar{p}$ (\emph{right}). Only one representative angular bin is shown.} \label{fig:feedDownComparisonP}
\end{figure*}

Comparisons of the uncertainties associated with the feed-down corrections with and without neutral-hadron re-weighting are shown in Figs.~\ref{fig:feedDownComparisonPi} and~\ref{fig:feedDownComparisonP}. The inclusion of the neutral-hadron measurements significantly decreases these uncertainties, as the multiplicity measurement uncertainties are significantly smaller than the variations in multiplicity predictions by different Monte Carlo models. 

\subsection{Charged-Hadron Multiplicity Measurements} 

The raw yields for \pipm, $p/\bar{p}$, and \Kpm are used to calculate differential production multiplicities, defined as the number of produced hadrons per production interaction. A production interaction is defined as an interaction resulting in the production of new hadrons and excluding quasielastic interactions. The double-differential production multiplicities are given by 

\begin{equation} \label{eq:multiplicity}
  \frac{d^2 n_i}{dp d\theta} = \frac{ c^\textrm{total}_i
    \sigma_\textrm{trig}}{(1-\epsilon) \sigma_\textrm{prod} \Delta p
    \Delta \theta} \left(
  \frac{y^{\textrm{I}}_i}{N_{\textrm{trig}}^{\textrm{I}}} -
  \frac{\epsilon y^{\textrm{R}}_i}{N_{\textrm{trig}}^{\textrm{R}}}
  \right).
\end{equation}

Here $n_i$ is the number of produced hadrons in kinematic bin $i$ with production angle $\theta$ and production momentum $p$, the raw yield given by $y^\textrm{I}_i$ ($y^\textrm{R}_i$) corresponds to the yield in kinematic bin $i$ with the target inserted (removed), $N_{\textrm{trig}}^{\textrm{I}}$ ($N_{\textrm{trig}}^{\textrm{R}}$) is the number of recorded triggers with the target inserted (removed), $c^\textrm{total}_i$ is the total correction (combined Monte Carlo and $dE/dx$ fit) for kinematic bin $i$, $\epsilon$ is the inserted-to-removed trigger probability ratio $P_\textrm{trig}^\textrm{R}/P_\textrm{trig}^\textrm{I}$, $\sigma_\textrm{trig}$ and $\sigma_\textrm{prod}$ are the trigger and production cross sections, respectively, and $\Delta p \Delta \theta$ is the size of kinematic bin i.

\begin{figure*}[!ht]
  \centering
  \hspace*{-1cm}\includegraphics[width=0.35\textwidth]{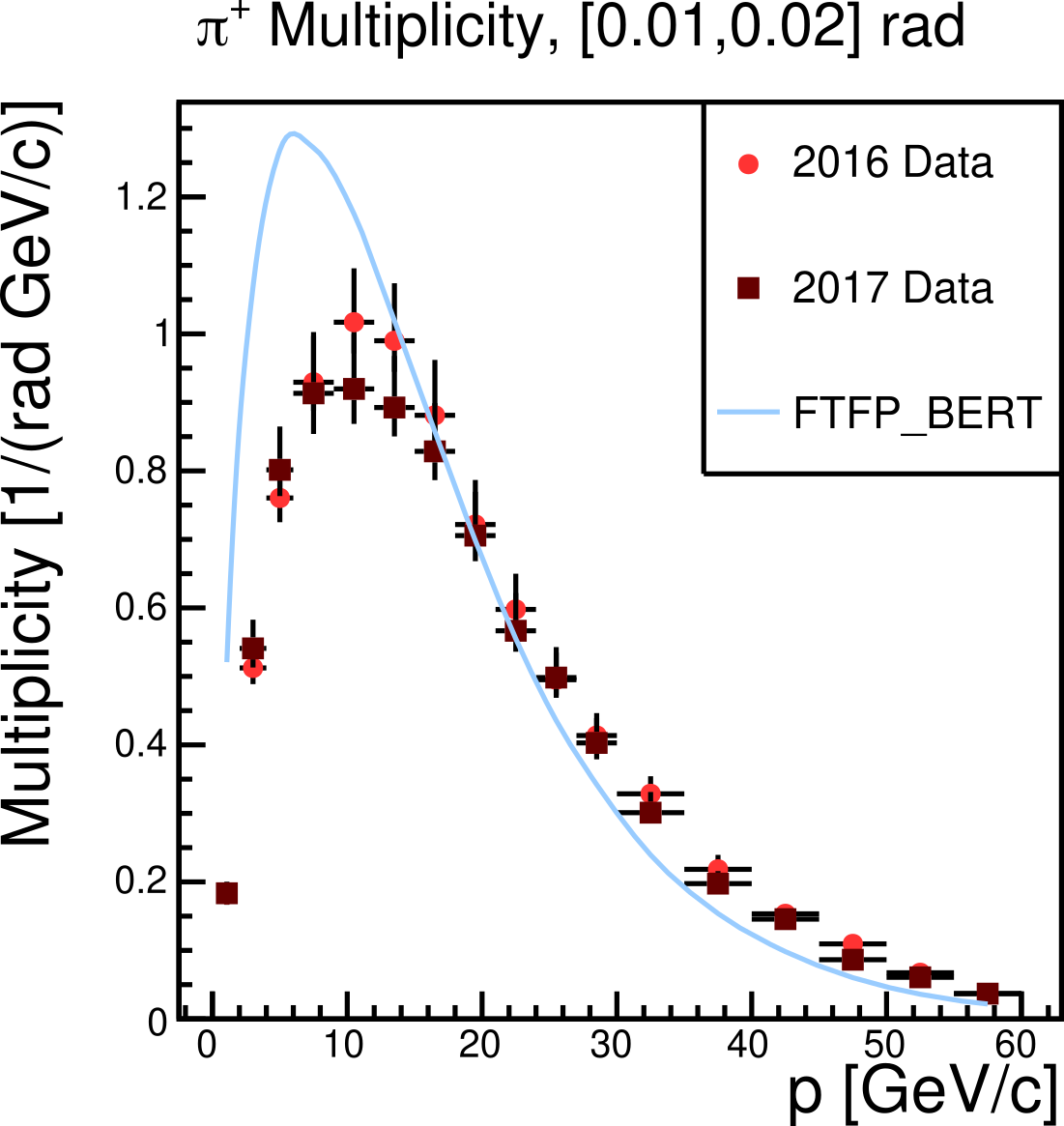}
  \hspace*{1cm}\includegraphics[width=0.35\textwidth]{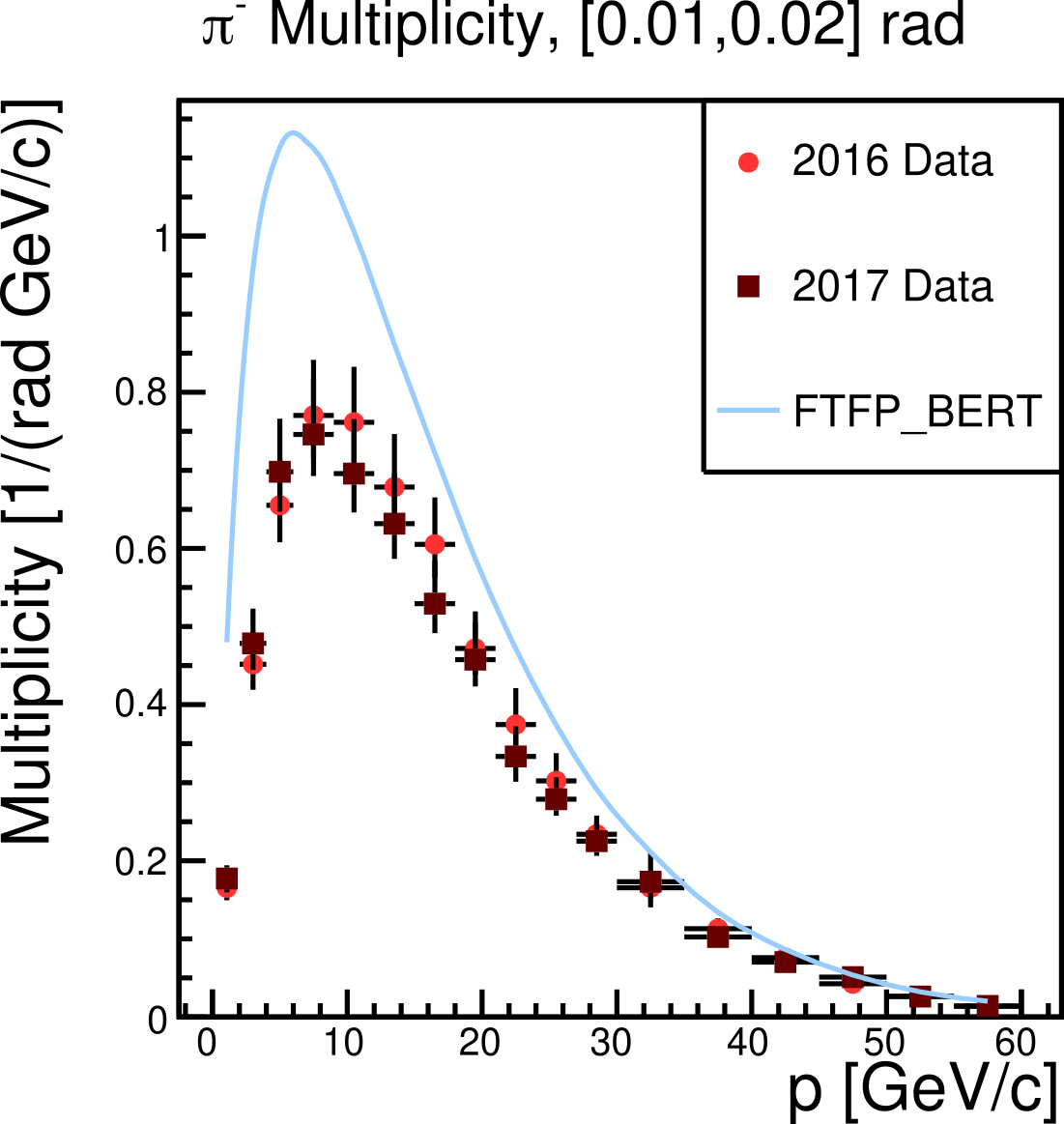}
\caption{Example \pipm multiplicity measurements comparing the 2016 and 2017 analysis results. Uncertainties reflect total uncertainty (statistical and systematic) for the independent analyses.}\label{fig:chargedComparisonPi}
\end{figure*}

\begin{figure*}[!ht]
  \centering
  \hspace*{-1cm}\includegraphics[width=0.35\textwidth]{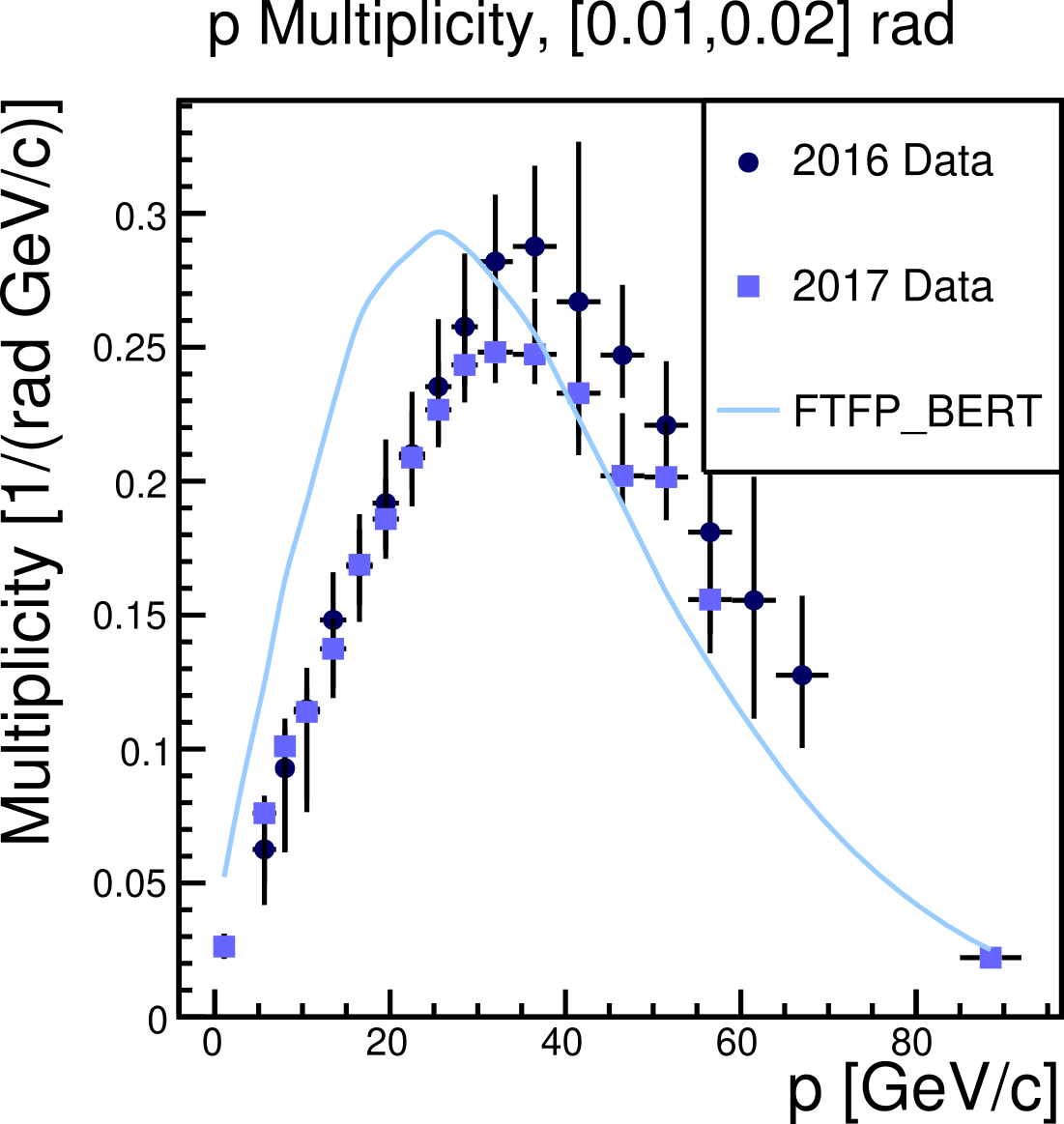}
  \hspace*{1cm}\includegraphics[width=0.35\textwidth]{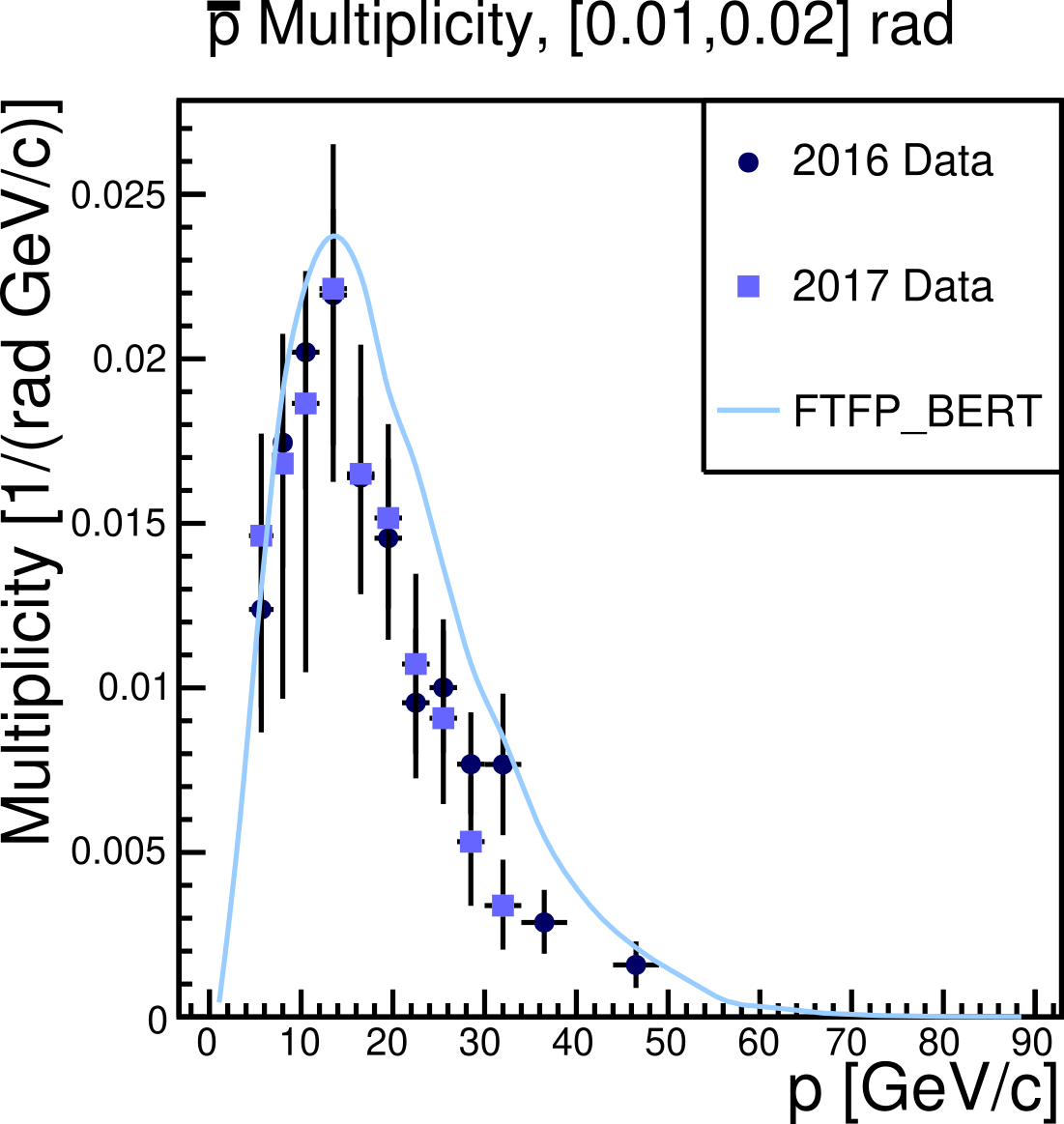}
\caption{Example $p/\bar{p}$ multiplicity measurements comparing the 2016 and 2017 analysis results. Uncertainties reflect total uncertainty (statistical and systematic) for the independent analyses.}\label{fig:chargedComparisonP}
\end{figure*}

\begin{figure*}[!ht]
  \centering
  \hspace*{-1cm}\includegraphics[width=0.35\textwidth]{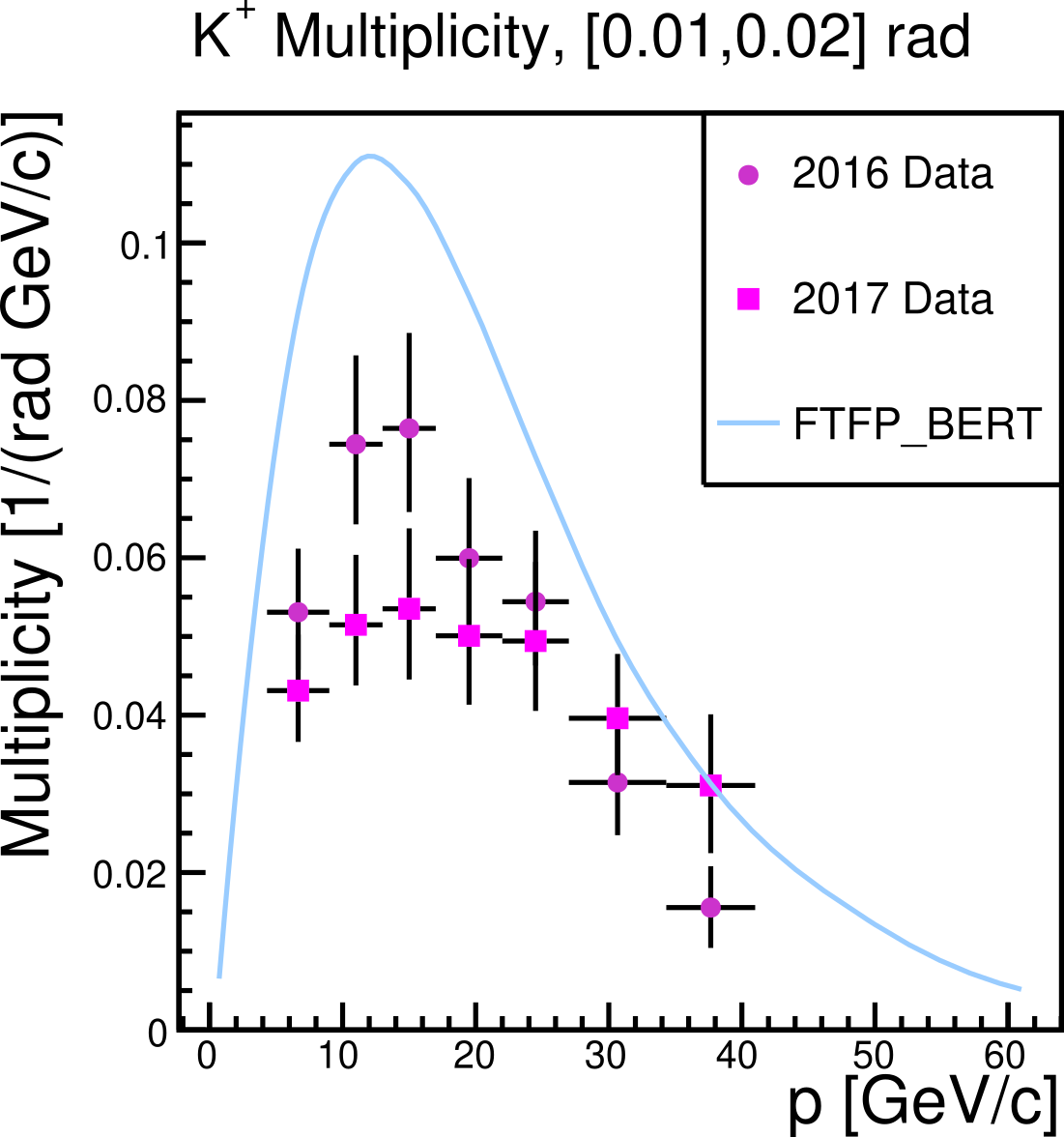}
  \hspace*{1cm}\includegraphics[width=0.35\textwidth]{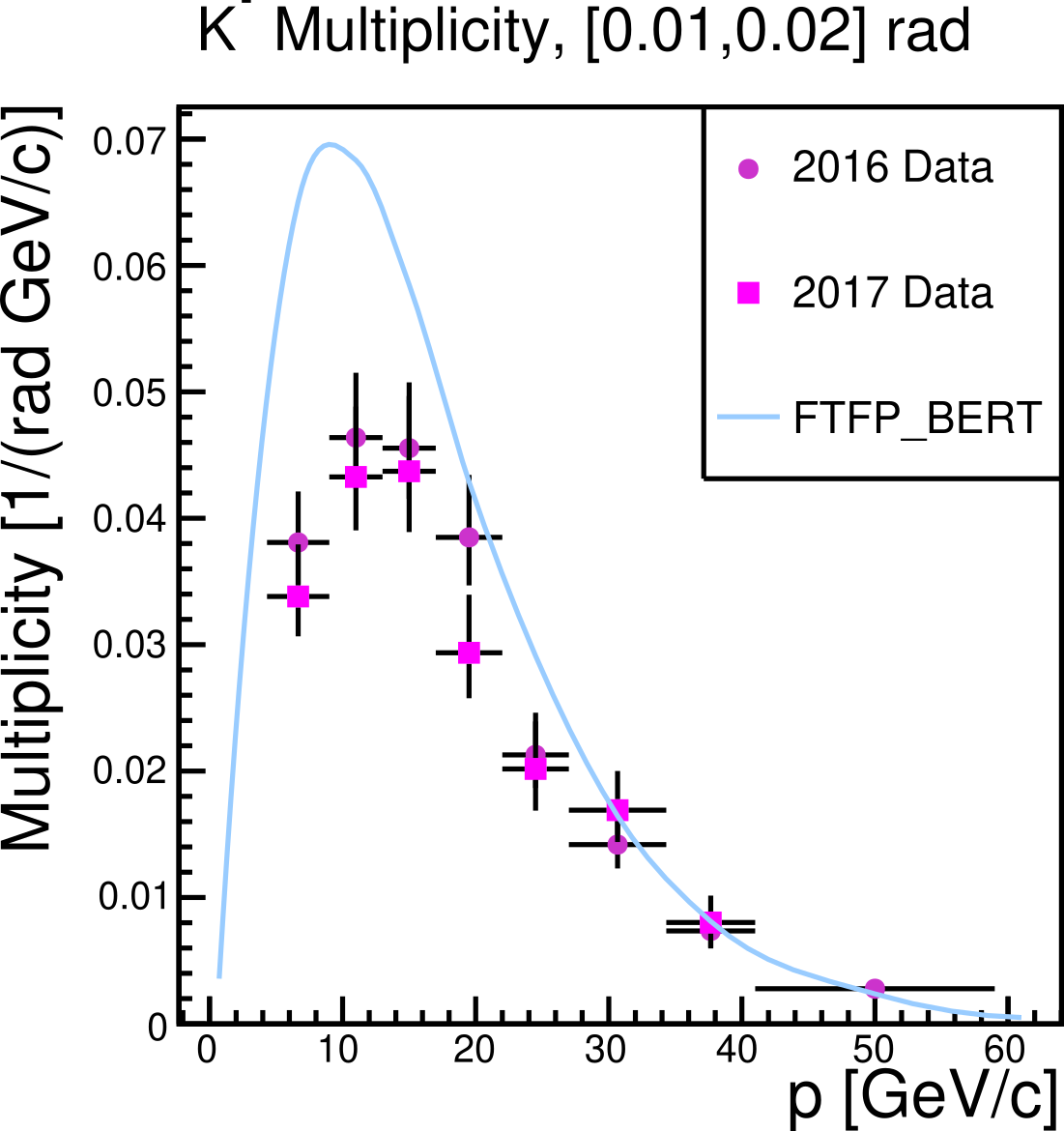}
\caption{Example \Kpm multiplicity measurements comparing the 2016 and 2017 analysis results. Uncertainties reflect total uncertainty (statistical and systematic) for the independent analyses.}\label{fig:chargedComparisonK}
\end{figure*}

Production multiplicities in selected regions of phase space for \pipm, $p/\bar{p}$, and \Kpm are presented in Figs.~\ref{fig:chargedComparisonPi}--\ref{fig:chargedComparisonK}. Comparisons of the 2016 and 2017 measurements show agreement of most measurements within $1\sigma$ (statistical + systematic). A combined measurement, taking into account correlated and uncorrelated systematic uncertainties, will be presented in Sec.~\ref{sec:combinedMultiplicities}.
 

\section{Systematic Uncertainties of 2016 and 2017 Analyses}

Systematic uncertainties from several effects were considered and their effects were evaluated independently for the 2016 and 2017 analyses. This section will detail sources of uncertainty considered and show the individual contributions to total systematic uncertainty.  

\begin{figure*}[!ht]
  \centering
  \includegraphics[width=0.3\textwidth]{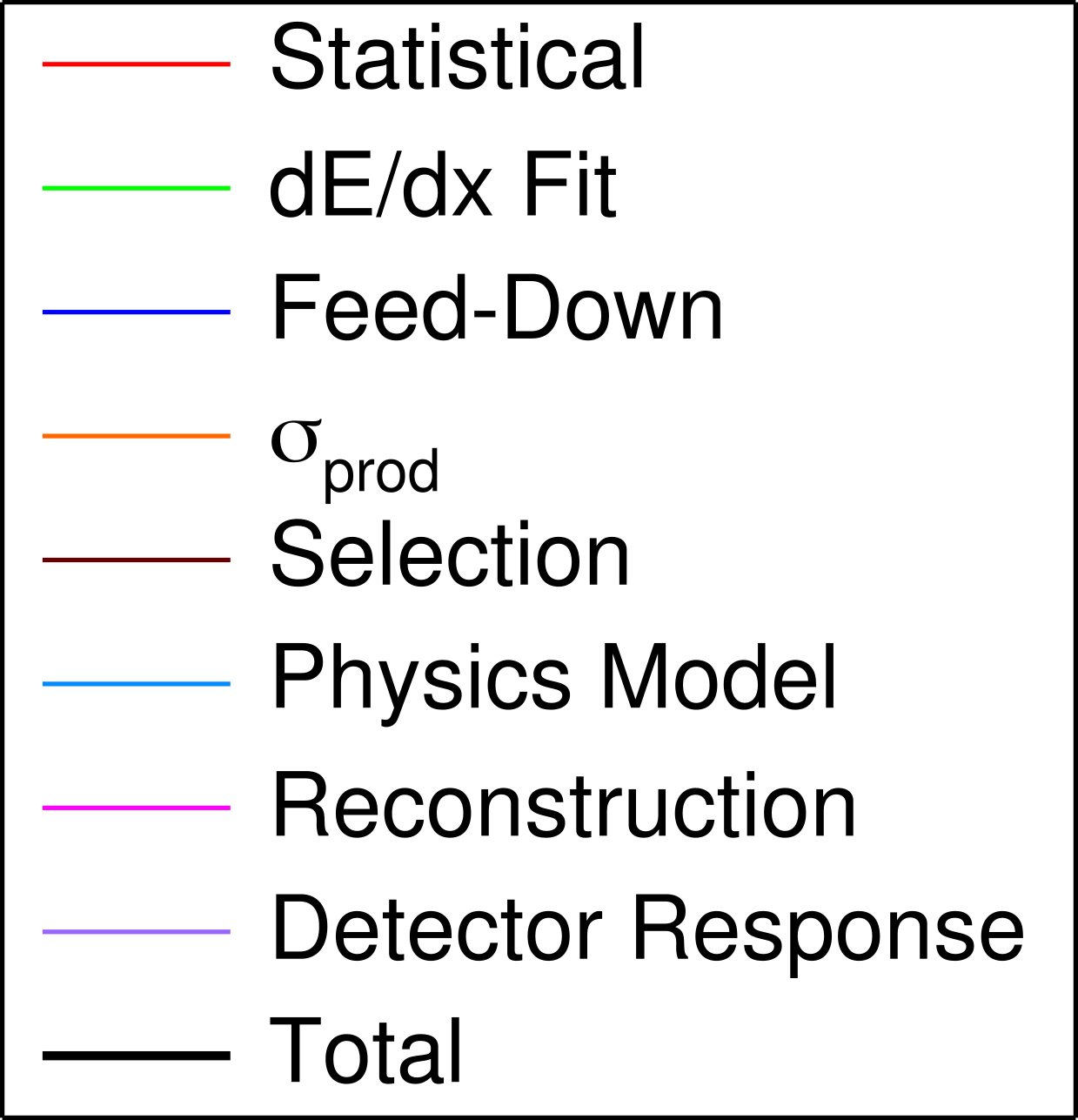}
  \includegraphics[width=0.33\textwidth]{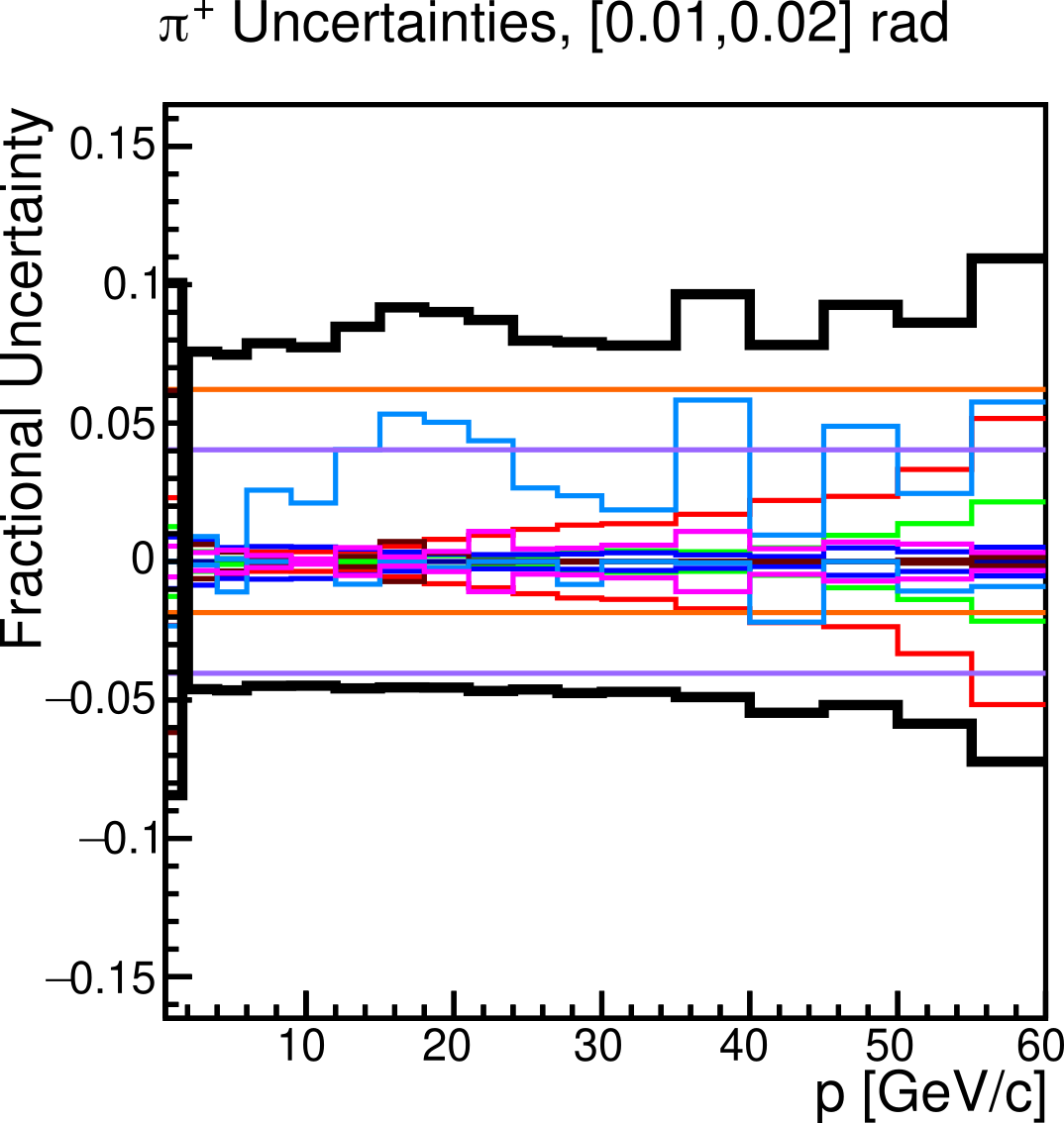}
  \includegraphics[width=0.33\textwidth]{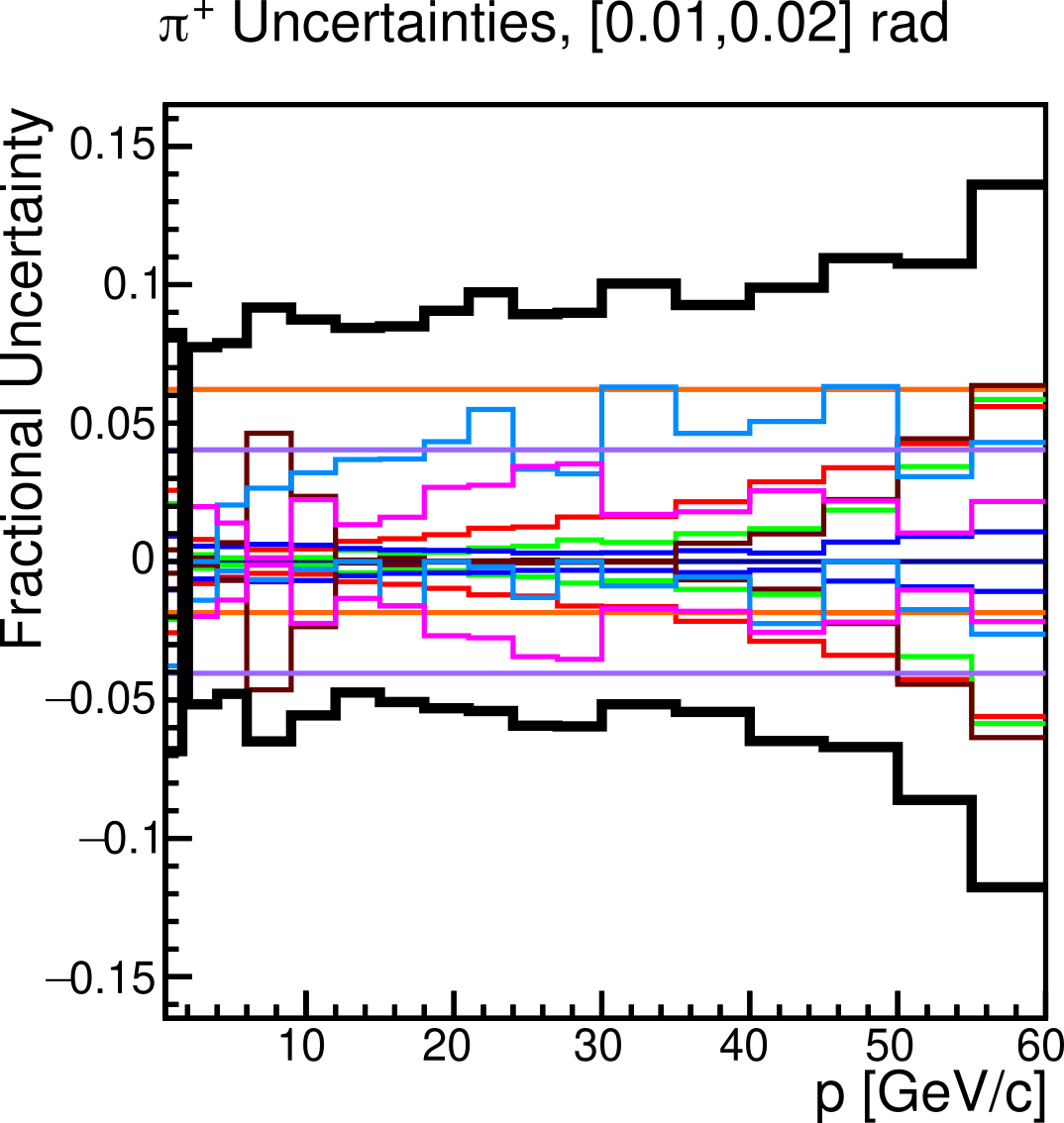}
\caption{Systematic uncertainty breakdown for 2016 and 2017 \pip analyses. One representative angular bin is shown.}\label{fig:systematicsPi}
\end{figure*}

\begin{figure*}[!ht]
  \centering
  \includegraphics[width=0.3\textwidth]{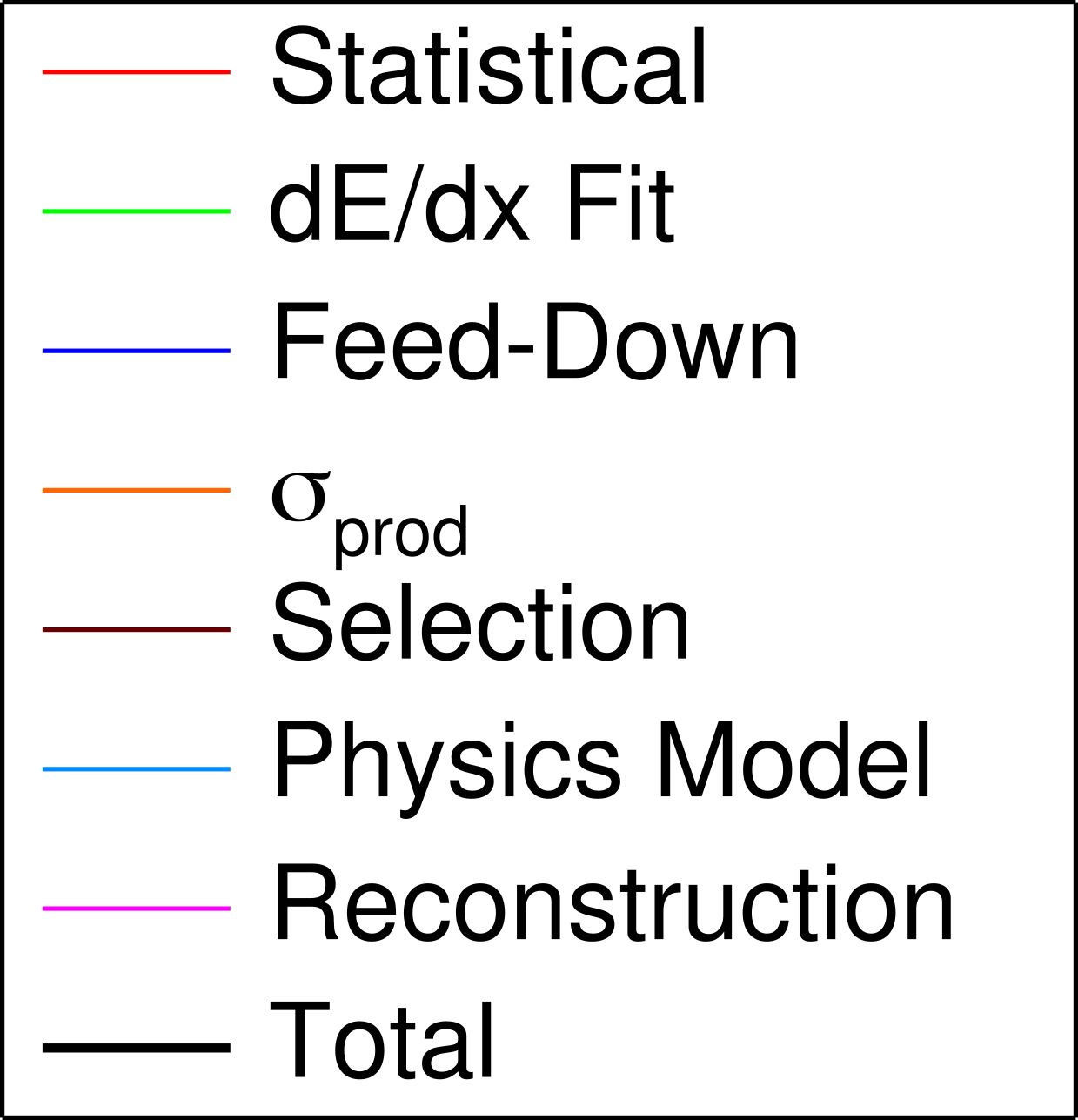}
  \includegraphics[width=0.33\textwidth]{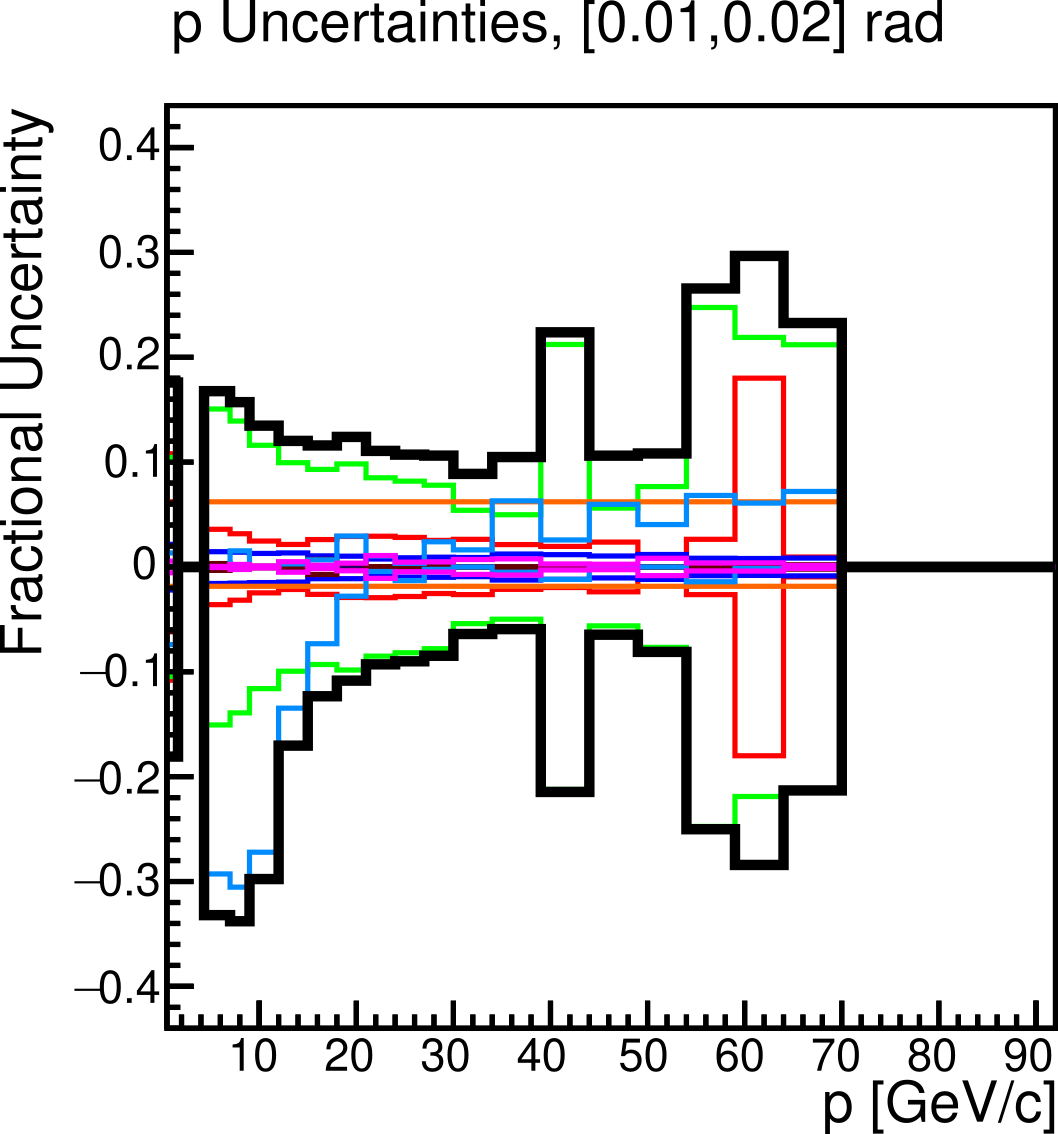}
  \includegraphics[width=0.33\textwidth]{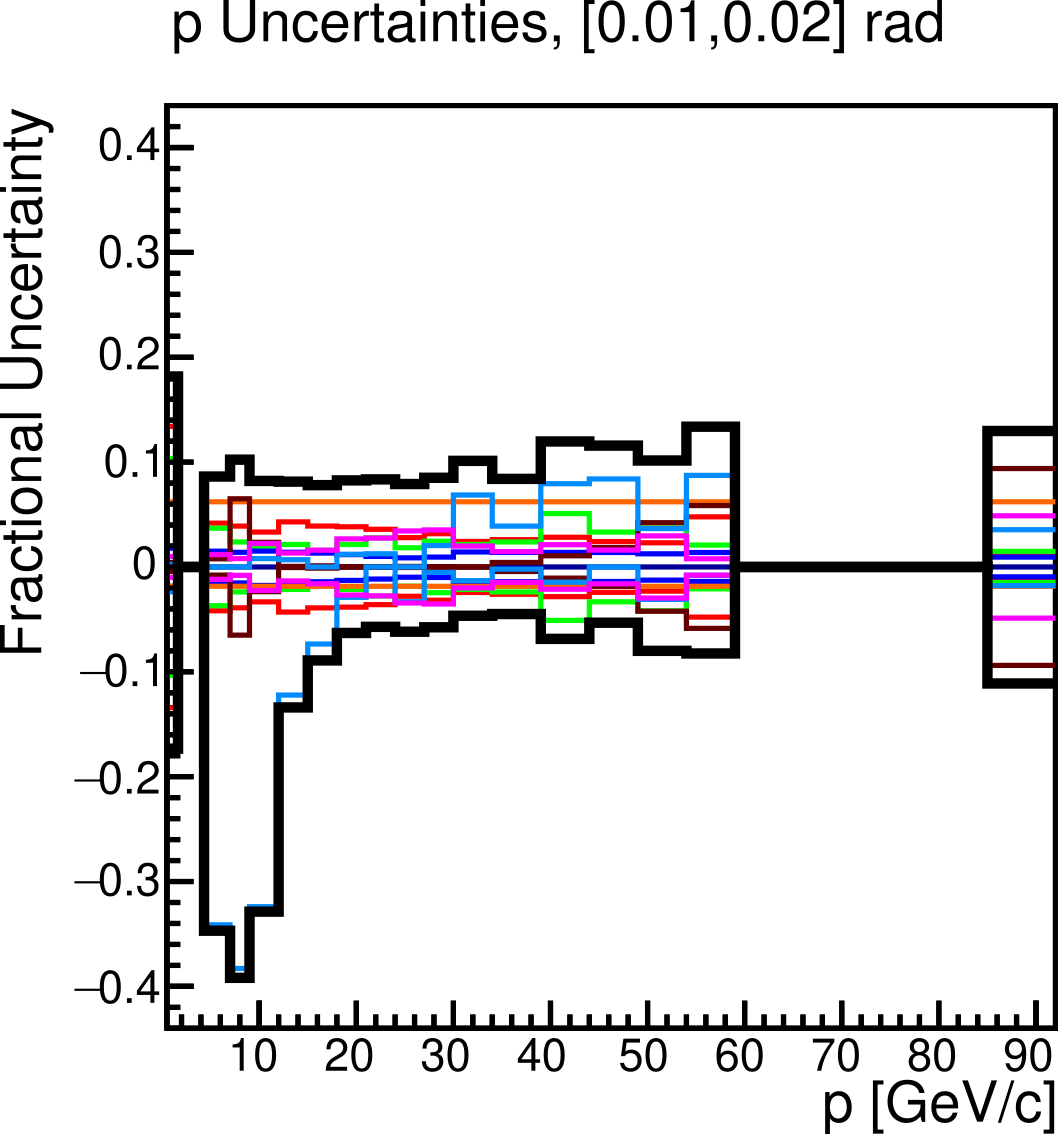}
\caption{Systematic uncertainty breakdown for 2016 and 2017 $p$ analyses. One representative angular bin is shown.}\label{fig:systematicsP}
\end{figure*}

\begin{figure*}[!ht]
  \centering
  \includegraphics[width=0.3\textwidth]{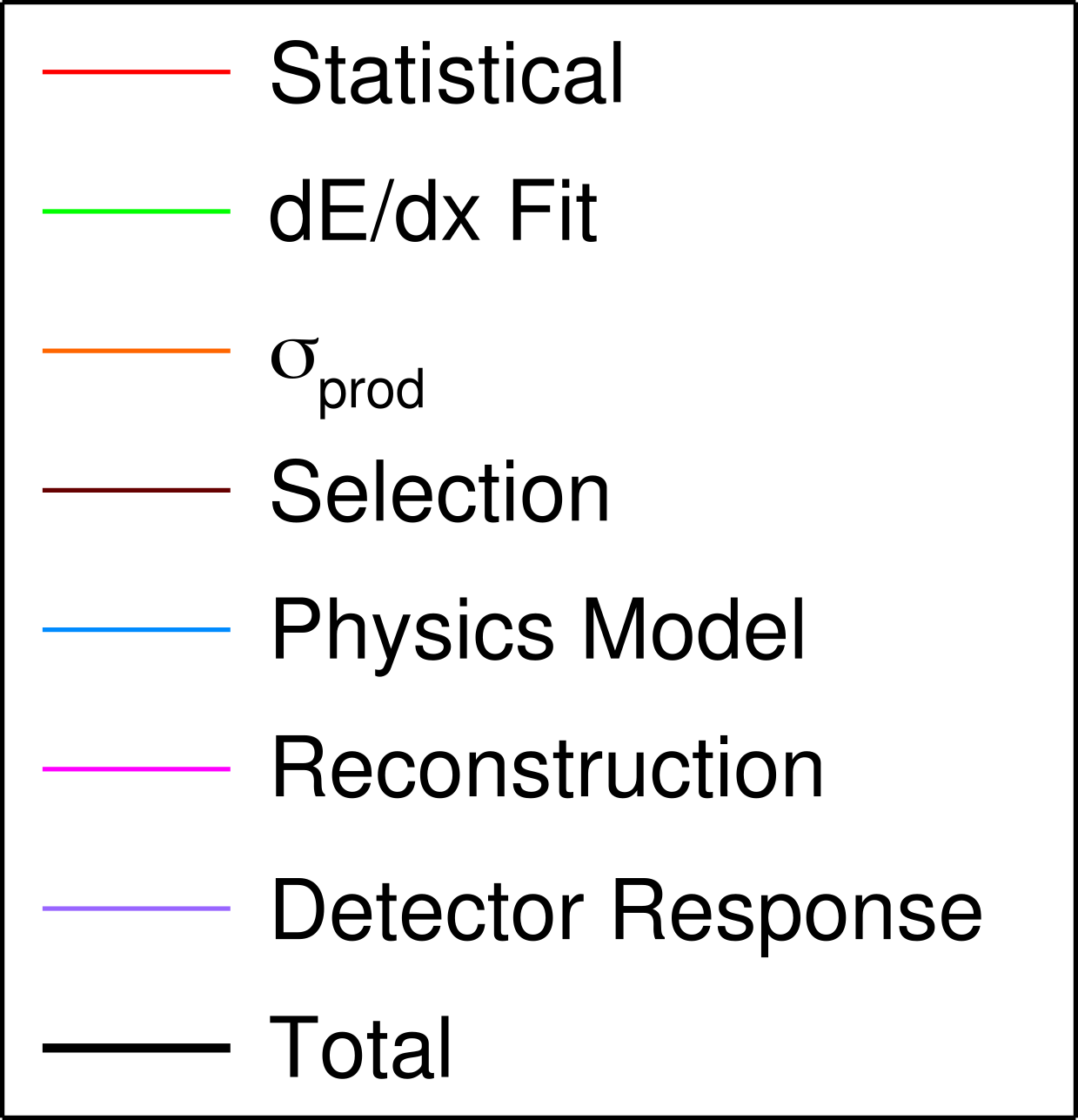}
  \includegraphics[width=0.33\textwidth]{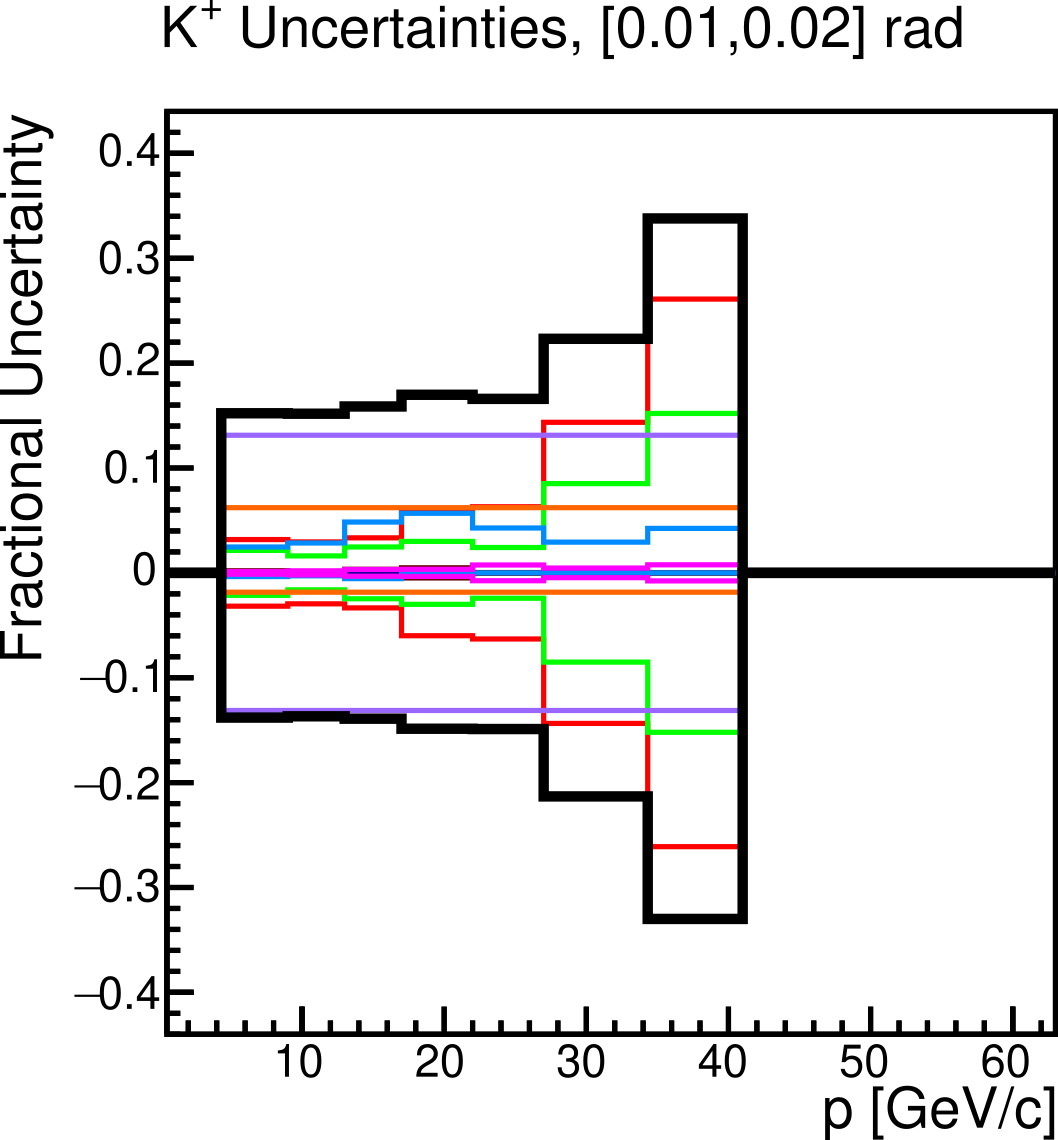}
  \includegraphics[width=0.33\textwidth]{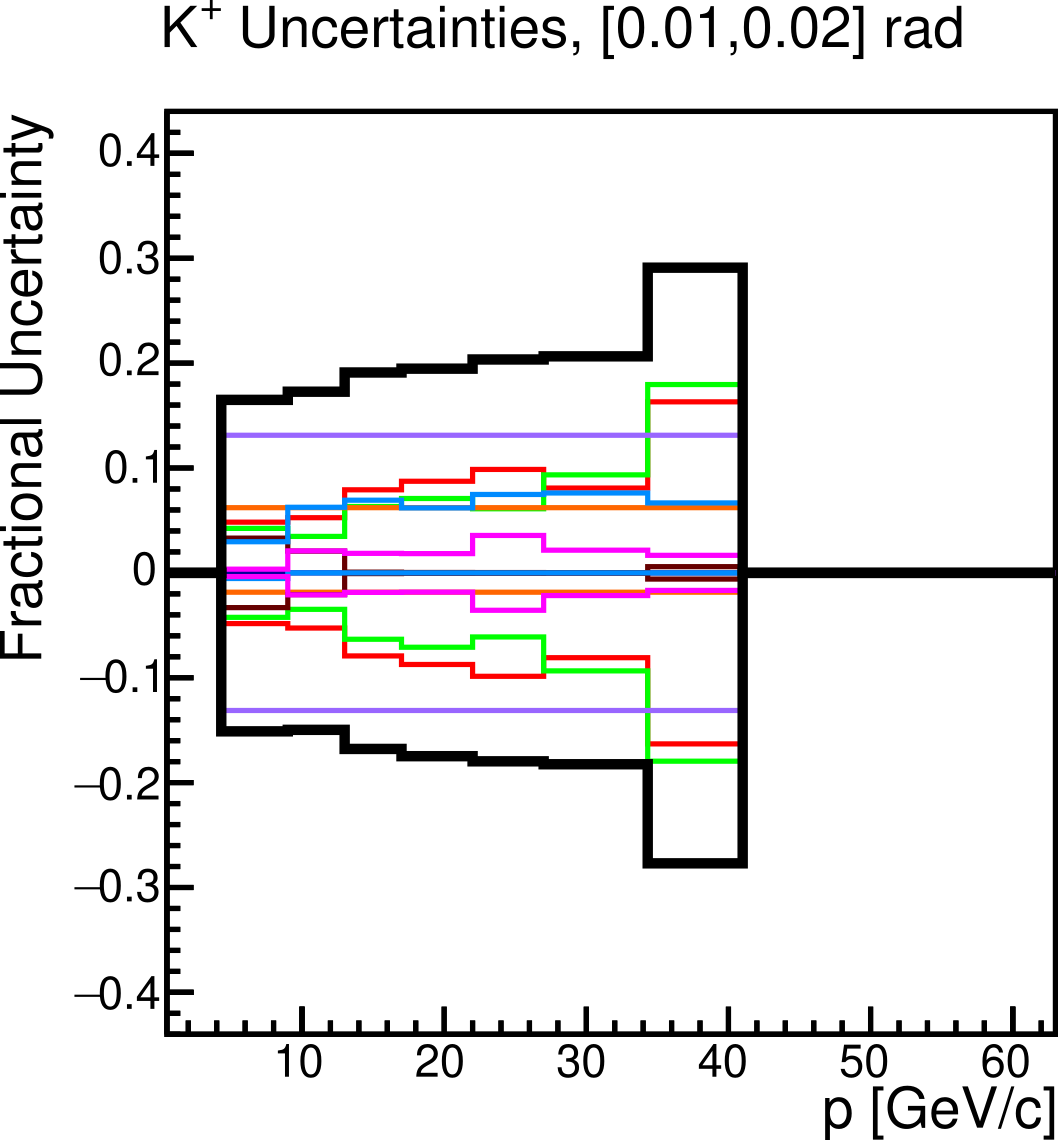}
\caption{Systematic uncertainty breakdown for 2016 and 2017 \Kp analyses. One representative angular bin is shown.}\label{fig:systematicsK}
\end{figure*}

 A breakdown of the individual systematic uncertainties for each analysis can be seen in Figs.~\ref{fig:systematicsPi}--\ref{fig:systematicsK}.

\subsection{Reconstruction}

Differences between true detector positions and those used in the Monte Carlo simulation affect final multiplicity measurements. Residual distributions describing track and point measurement mismatch were used to estimate potential detector misalignment. To estimate the reconstruction uncertainty, the detector central positions were displaced by varying amounts and the change in multiplicity was studied. The VTPCs and GTPC were independently shifted by 100 \textmu{}m in the $x$-dimension, and the FTPCs were independently shifted by 50 \textmu{}m in the $x$-dimension. These distances correspond to the widths of the track residual measurements for each detector. The resulting changes in the multiplicity measurements were added in quadrature to obtain the final reconstruction uncertainty.

\subsection{Selection}

Upon comparing track characteristics between reconstructed Monte Carlo and recorded data, a discrepancy was found in the average number of clusters per track. The simulated tracks contain 5 -- 10 \% more clusters than tracks from data. This is likely due to unsimulated faulty front-end electronics channels and periodic detector noise. These two effects often lead to cluster loss, as the cluster structures become difficult to distinguish from background noise. In order to compensate for this effect, the Monte Carlo corrections were re-calculated after artificially reducing the number of clusters on the simulated track by 15\% for a conservative estimate. The resulting Monte Carlo corrections were used to re-calculate the multiplicity measurements, and the difference was taken as a systematic uncertainty.

\subsection{Physics Model}

The Monte Carlo correction factors are calculated using a given physics model, and varying the underlying physics model will lead to different correction factors. The central values for the Monte Carlo corrections were determined using the FTFP\_BERT physics list, which appears to be more consistent with NA61/SHINE data than other physics models for the used version of \GeantFour. Two other physics models, FTF\_BIC and QBBC, were substituted in independent Monte Carlo samples, and the multiplicities were re-calculated using these correction factors. The largest difference from the nominal multiplicities in each kinematic bin was taken as a systematic uncertainty. The QGSP\_BERT physics model was not used for this uncertainty calculation due to large differences between the model predictions and these measurements (see Figs.~\ref{fig:combinedResultPiPlus}--\ref{fig:combinedResultKMinus}). This systematic uncertainty is naturally asymmetric, as the different model corrections often yield large non-uniform increases or decreases in multiplicities.

\subsection{Production Cross-Section Uncertainty}

The 120 \gevc proton-carbon production cross-section measurement was reported with a highly asymmetric systematic uncertainty~\cite{Aduszkiewicz:2019xna}. The upper and lower uncertainty values were propagated through the multiplicity analysis in order to obtain the associated uncertainty on the multiplicity spectra. The result is a uniform fractional uncertainty on each measurement of (+5.8,-1.8)\%. This uncertainty can be significantly reduced in the future when a more precise measurement of the 120 \gevc proton-carbon quasielastic cross section is made.  

\subsection{Momentum}

Uncertainty on the momentum reconstruction scale was estimated by studying the \kos invariant mass spectrum while performing the neutral-hadron analysis. An aggregate invariant mass sample was created by merging the kinematic analysis bins, and the \kos mass was fit for using a Breit--Wigner signal model and a 3rd order polynomial background model. The fractional difference between the current accepted value for the \kos mass~\cite{ParticleDataGroupWorkman:2022ynf} and the aggregate fit mass was taken as an uncertainty on reconstructed track momentum. The momenta of all tracks were then shifted by this amount and the resulting change in multiplicities was taken as a systematic uncertainty. For the 2016 analysis the measured mass shift was $\Delta m = -0.1$ \mevcc (-0.02\%) and for the 2017 analysis the measured mass shift was $\Delta m = 1.1$ \mevcc (0.22\%). This uncertainty source was significantly less than the other systematic uncertainties, and thus was not included in the uncertainty evaluation. 

\subsection{Feed-down}

The feed-down uncertainty for the charged-hadron analysis is derived from the neutral-hadron multiplicity uncertainties, as the measurements of \kos, \lam and \alam are used to constrain the charged feed-down corrections~\cite{neutralHadronArxivPaper}. For a given neutral particle decaying into a \pipm, $p$ or $\bar{p}$, if the parent particle kinematics are covered by the neutral-hadron multiplicity measurements, the multiplicity uncertainty associated with that kinematic bin is recorded. If the kinematics are not covered by the measurement, an uncertainty of 50\% is used. The collected uncertainties are averaged in the charged analysis bins in order to assign a total feed-down uncertainty for each bin. For regions covered by the neutral-hadron analysis, the uncertainty is typically much smaller than 50\%. Finally, the number of tracks originating from weak neutral hadron decay is varied by the calculated fractional uncertainties and a new feed-down correction is computed. The resulting changes in multiplicities are taken as a systematic uncertainty.

\subsection{$dE/dx$ Fit}

In Section~\ref{sec:dedxFitBiasCorrections}, the procedure for determining $dE/dx$ fit bias in each analysis bin was discussed. The systematic uncertainty associated with the fitting routine was evaluated using a similar procedure: calibration parameters were independently varied according to a Gaussian distribution whose width corresponds to the RMS of each parameter from fits to the data. As the $dE/dx$ simulation parameters were varied and the simulated track $dE/dx$ distributions re-fit, the standard deviation of fit biases in each bin was calculated:

\begin{equation}
  \sigma^\textrm{Fit}_i =  \sqrt{\frac{1}{N_\textrm{trials}} \sum^{N_\textrm{trials}}_{n = 1}\left( \frac{y_n^\textrm{fit} -
      y_n^\textrm{true}}{y_n^\textrm{true}} - \Big \langle
    \frac{y^\textrm{fit} - y^\textrm{true}}{y^\textrm{true}} \Big
    \rangle \right)^2}.
\end{equation}

This standard deviation of fractional multiplicity given by 50 Monte Carlo simulations was taken as the uncertainty associated with the fitting routine and was propagated to the measured multiplicities.

\subsection{Detector Response}

An additional uncertainty arising from detector calibration and acceptance differences between the 2016 and 2017 configurations was applied to the \pipm and \Kpm measurements. During the combination of the independent measurements and uncertainties from the 2016 and 2017 analyses, (see Sec.~\ref{sec:combinedMultiplicities}) some measurements showed disagreement. A uniform uncertainty was added to the \pipm and \Kpm measurements such that the reduced $\chi^2$ corresponding to the combination of the two measurement sets was unity. In order to be conservative, this uncertainty was applied uniformly to each measurement in both the 2016 and 2017 analyses.

\section{Combined Multiplicity Measurements}\label{sec:combinedMultiplicities}

In regions of phase space where detector acceptance overlapped in 2016 and 2017, multiplicity measurements can be combined. The measurements must be weighted by the uncertainty unique to each analysis, referred to here as the uncorrelated uncertainty. This uncertainty includes statistical, reconstruction, selection, momentum, and fit uncertainties, added in quadrature. Correlated uncertainties, consisting of feed-down, production cross-section, and physics model uncertainties, apply to both analyses and are not included in measurement weights during combination.

For the combined multiplicity measurement, a simple weighted mean is calculated using the uncorrelated uncertainties:

\begin{equation}
m_\textrm{combined} = 
\frac{\frac{m_1}{\sigma_1^2} + \frac{m_2}{\sigma_2^2}}{\frac{1}{\sigma_1^2} + \frac{1}{\sigma_2^2}},
\end{equation}

where $m_1$ and $\sigma_1$ are the multiplicity measurement and uncorrelated uncertainty from the 2016 analysis and $m_2$ and $\sigma_2$ are the multiplicity measurement and uncorrelated uncertainty from the 2017 analysis.

\begin{figure*}[!ht]
  \centering
  \includegraphics[width=0.49\textwidth]{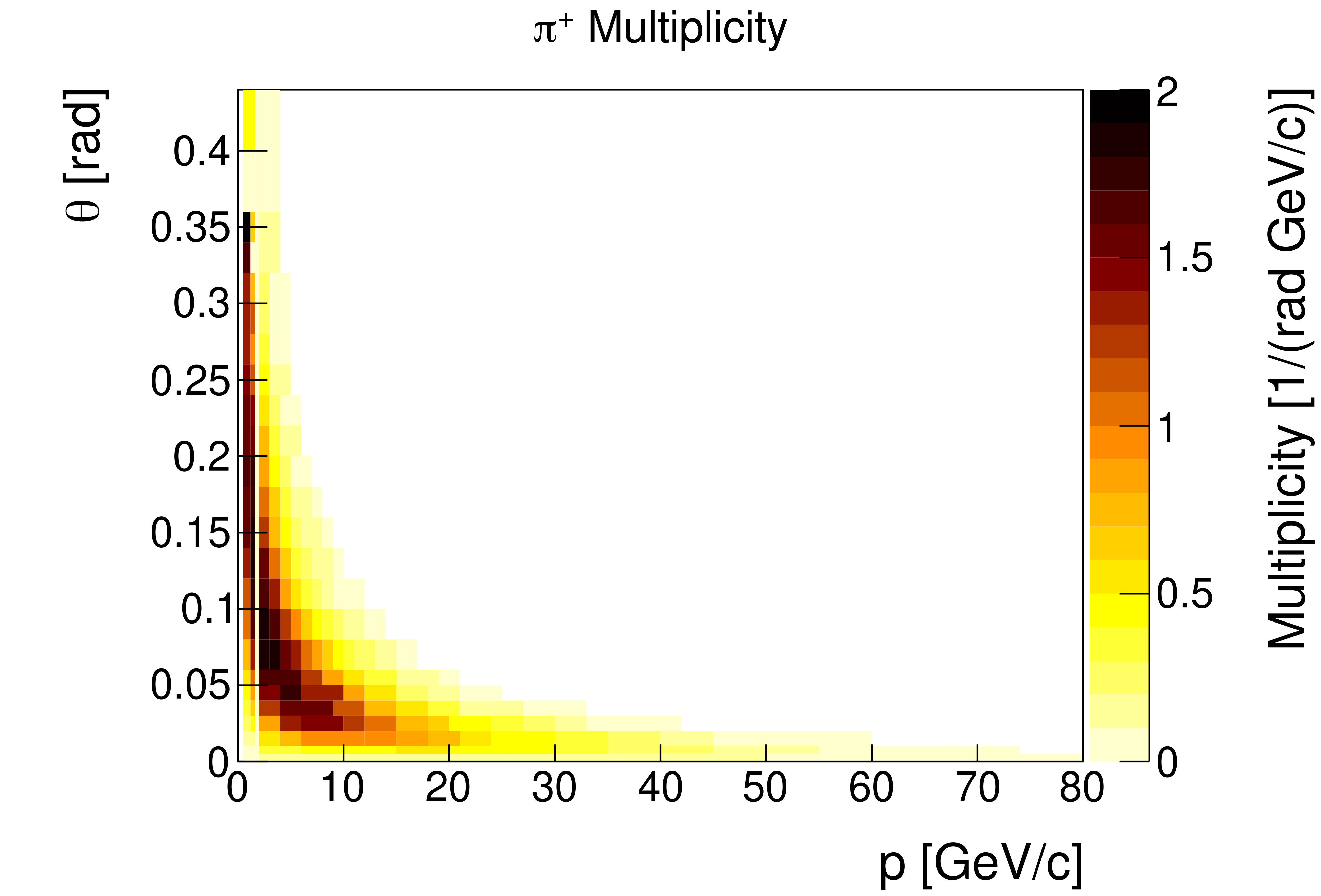}
  \includegraphics[width=0.49\textwidth]{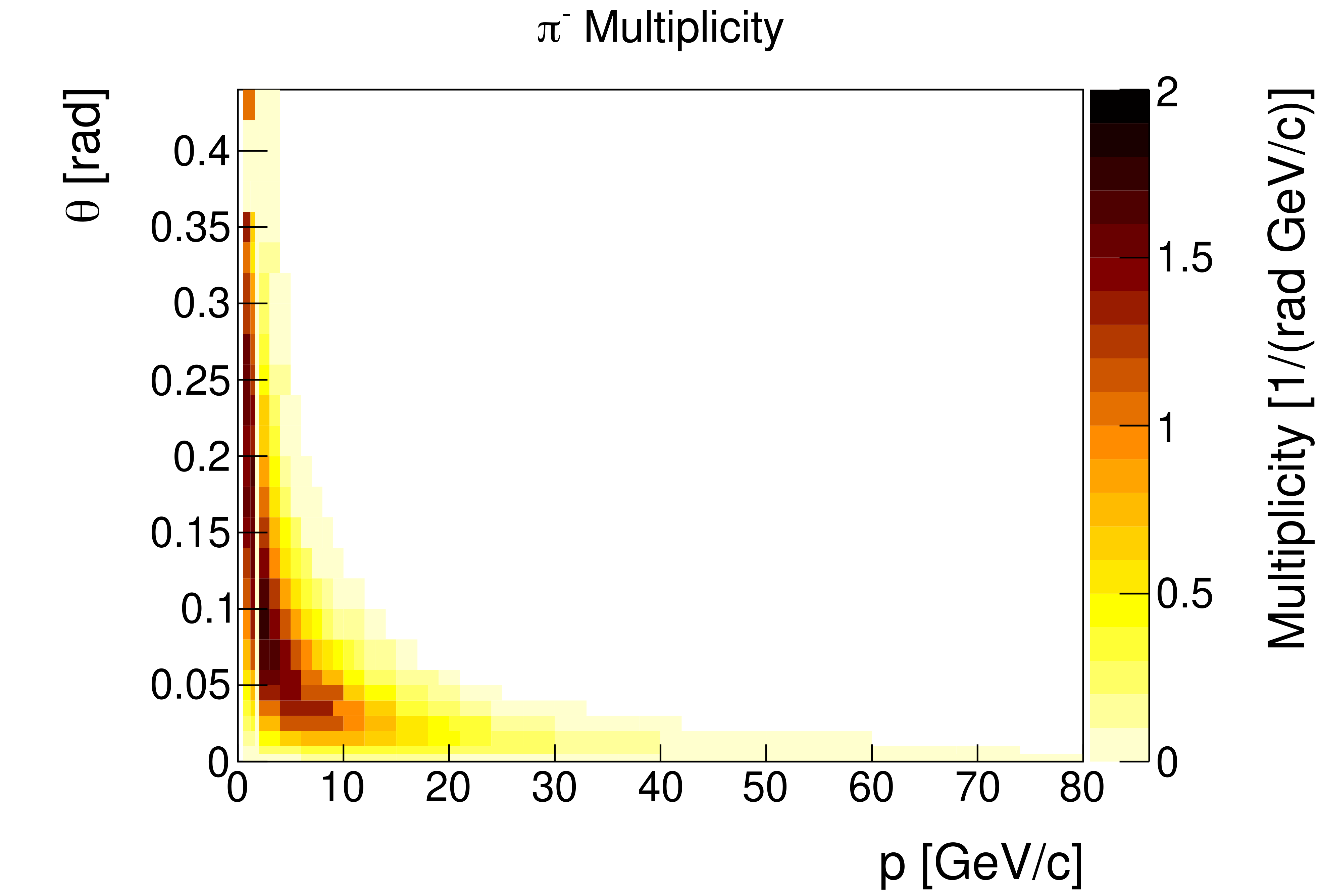}
\caption{Combined multiplicity measurements for \pip and \pim analyses. Numerical values can be found at~\cite{pC120EDMS}.}\label{fig:combinedResultPi2D}
\end{figure*}

\begin{figure*}[!ht]
  \centering
  \includegraphics[width=0.49\textwidth]{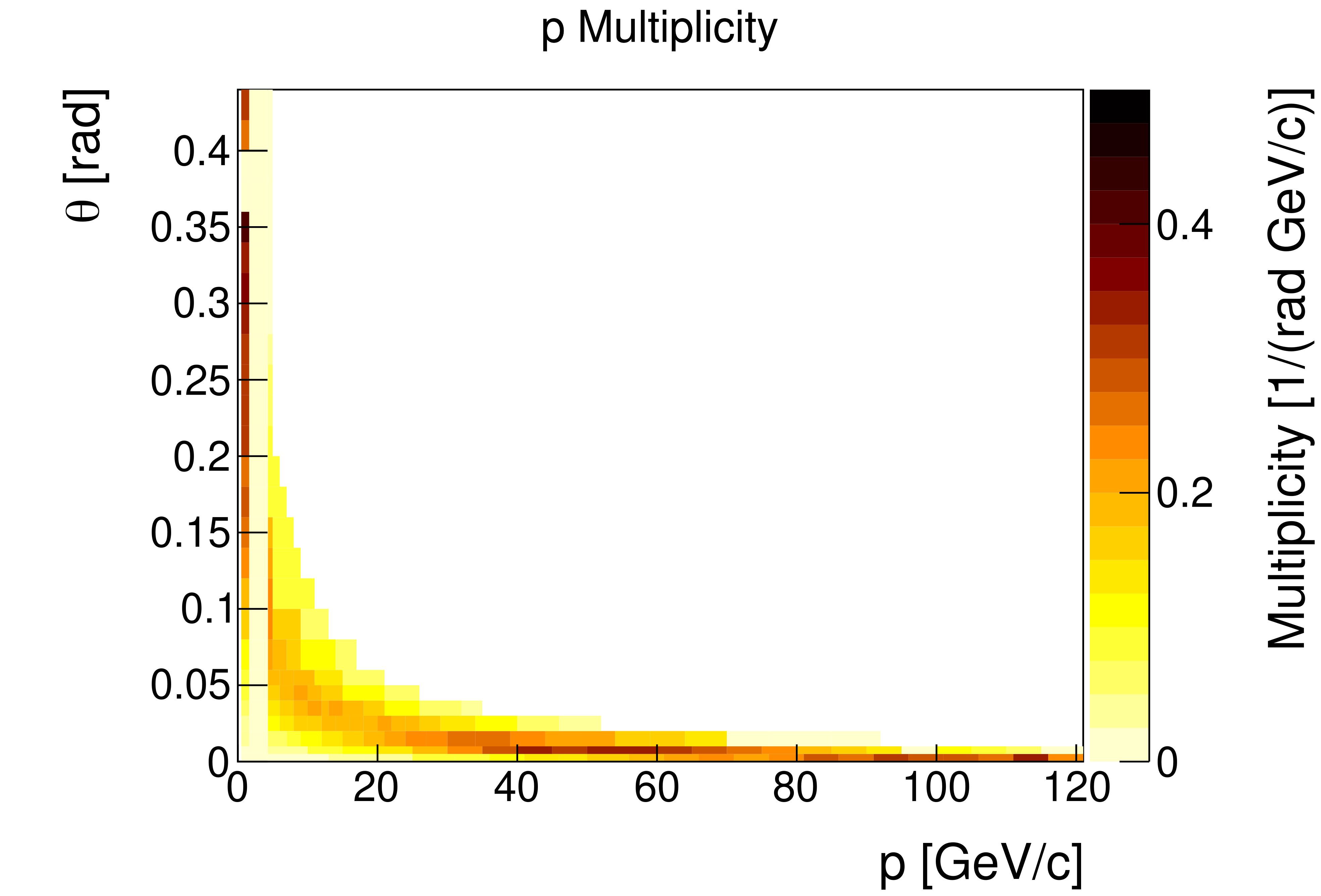}
  \includegraphics[width=0.49\textwidth]{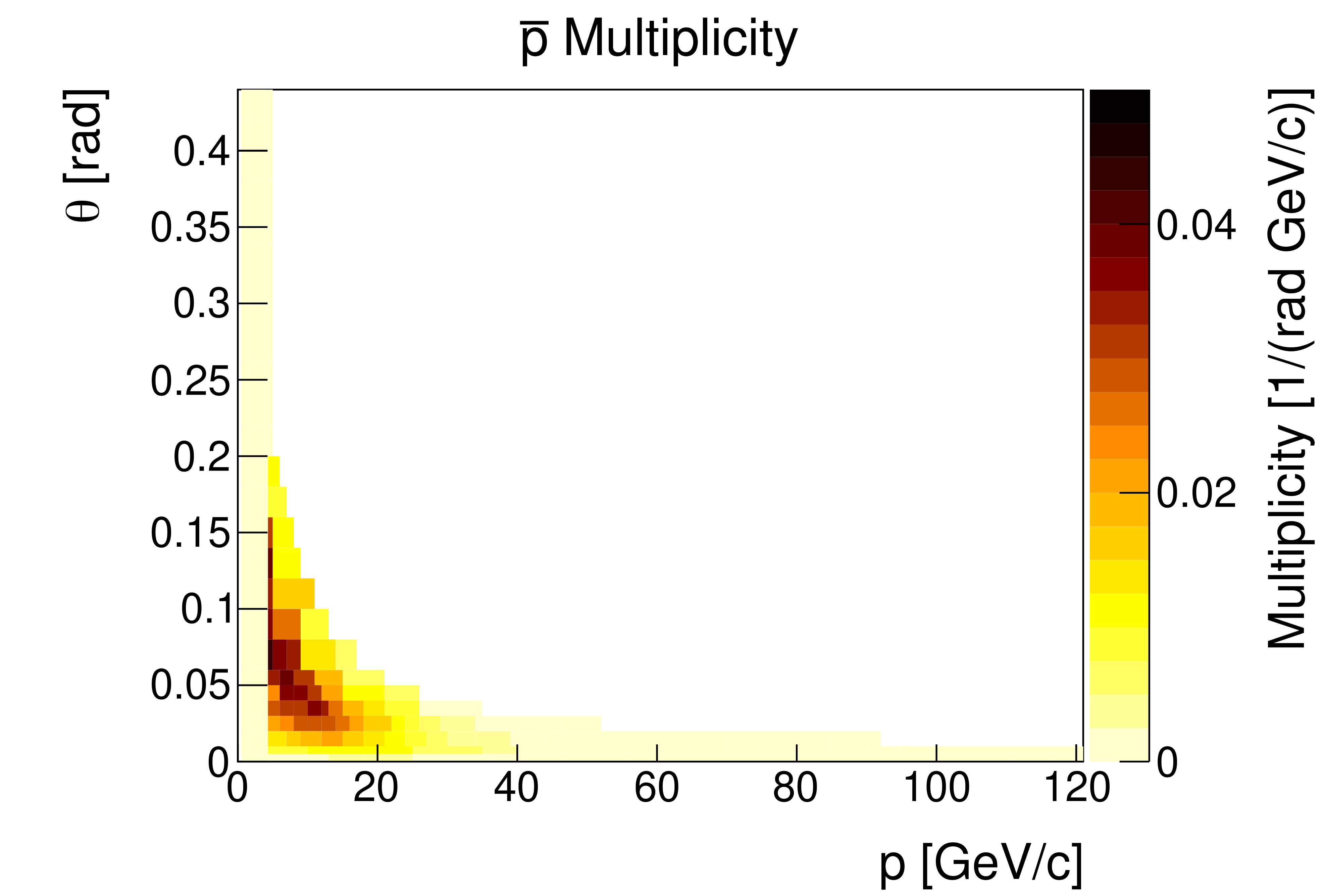}
\caption{Combined multiplicity measurements for proton and antiproton analyses. Numerical values can be found at~\cite{pC120EDMS}.}\label{fig:combinedResultP2D}
\end{figure*}

\begin{figure*}[!ht]
  \centering
  \includegraphics[width=0.49\textwidth]{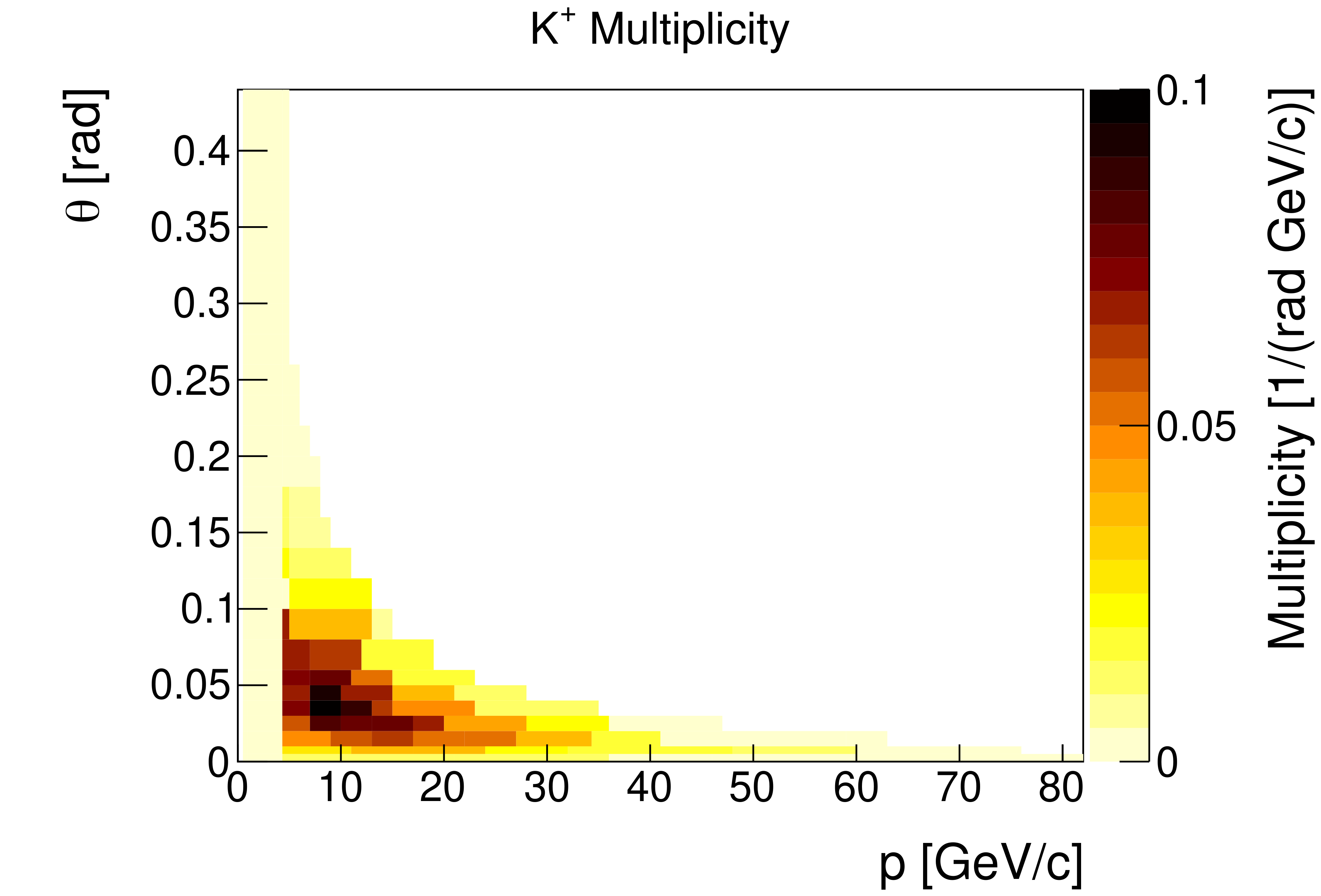}
  \includegraphics[width=0.49\textwidth]{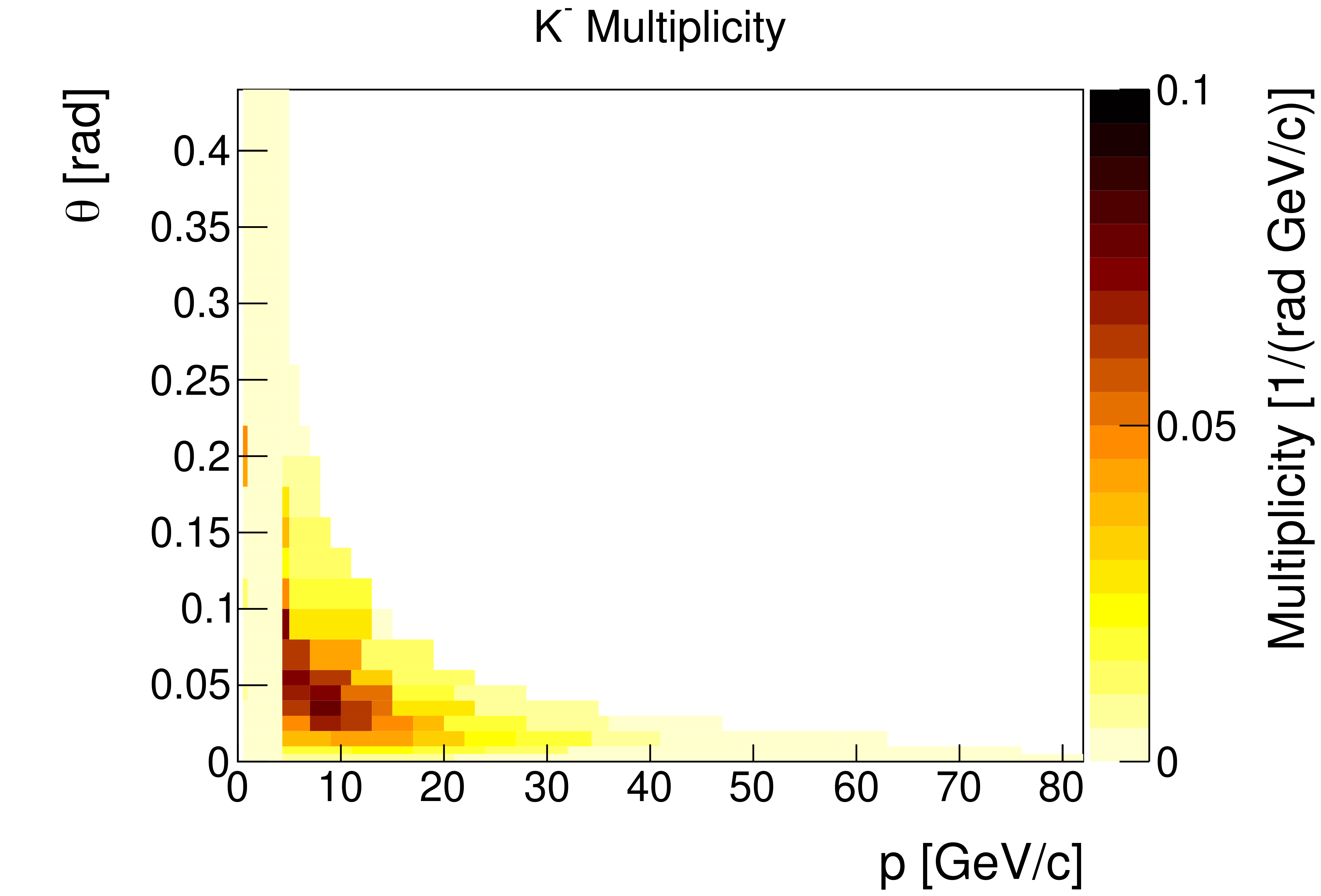}
\caption{Combined multiplicity measurements for \Kp and \Km analyses. Numerical values can be found at~\cite{pC120EDMS}.}\label{fig:combinedResultK2D}
\end{figure*}

\begin{figure*}[!ht]
  \centering
  \includegraphics[width=\textwidth]{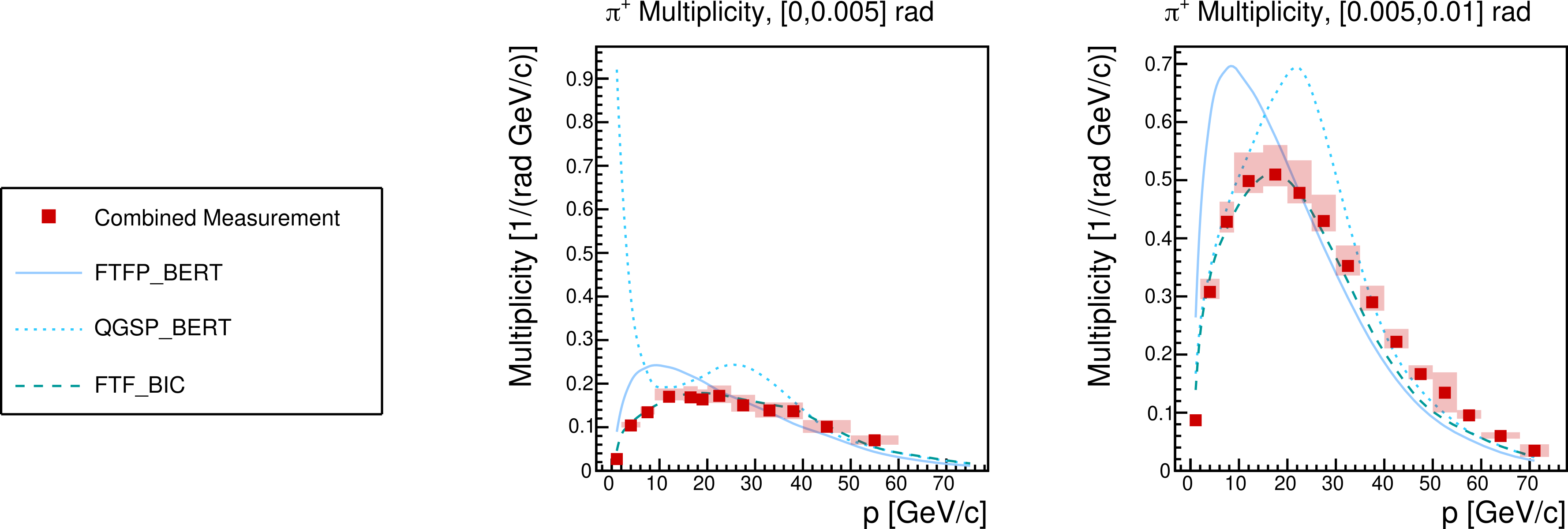}
\caption{Combined multiplicity measurements for \pip analysis. Error bars denote statistical uncertainty, and total systematic uncertainty is shown as a shaded band. Results are compared to three Monte Carlo models. Two representative angular bins are shown.}\label{fig:combinedResultPiPlus}
\end{figure*}

\begin{figure*}[!ht]
  \centering
  \includegraphics[width=\textwidth]{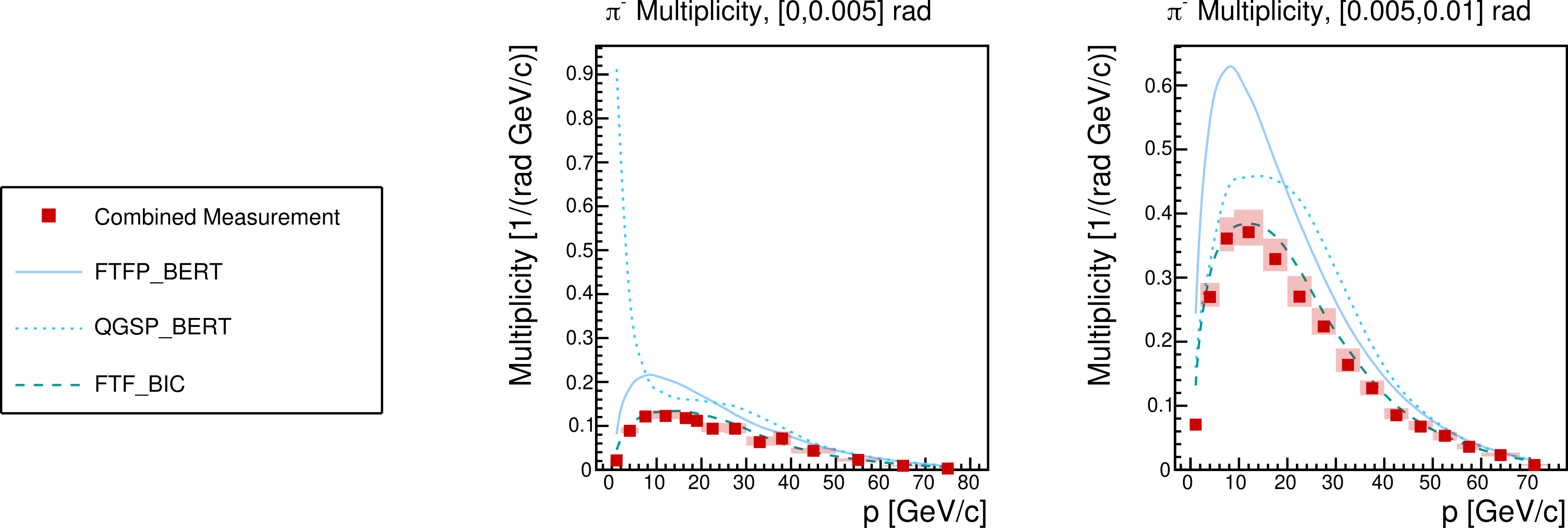}
\caption{Combined multiplicity measurements for \pim analysis. Error bars denote statistical uncertainty, and total systematic uncertainty is shown as a shaded band. Results are compared to three Monte Carlo models. Two representative angular bins are shown.}\label{fig:combinedResultPiMinus}
\end{figure*}

\begin{figure*}[!ht]
  \centering
  \includegraphics[width=\textwidth]{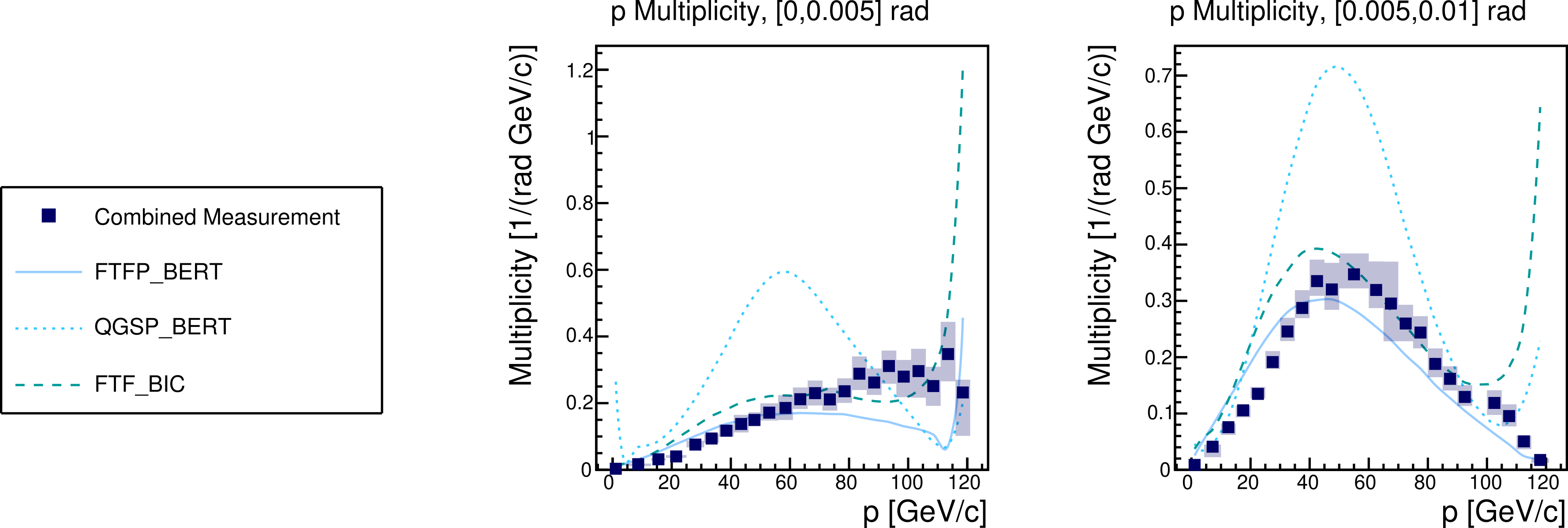}
\caption{Combined multiplicity measurements for $p$ analysis. Error bars denote statistical uncertainty, and total systematic uncertainty is shown as a shaded band. Results are compared to three Monte Carlo models. Two representative angular bins are shown.}\label{fig:combinedResultP}
\end{figure*}

\begin{figure*}[!ht]
  \centering
  \includegraphics[width=\textwidth]{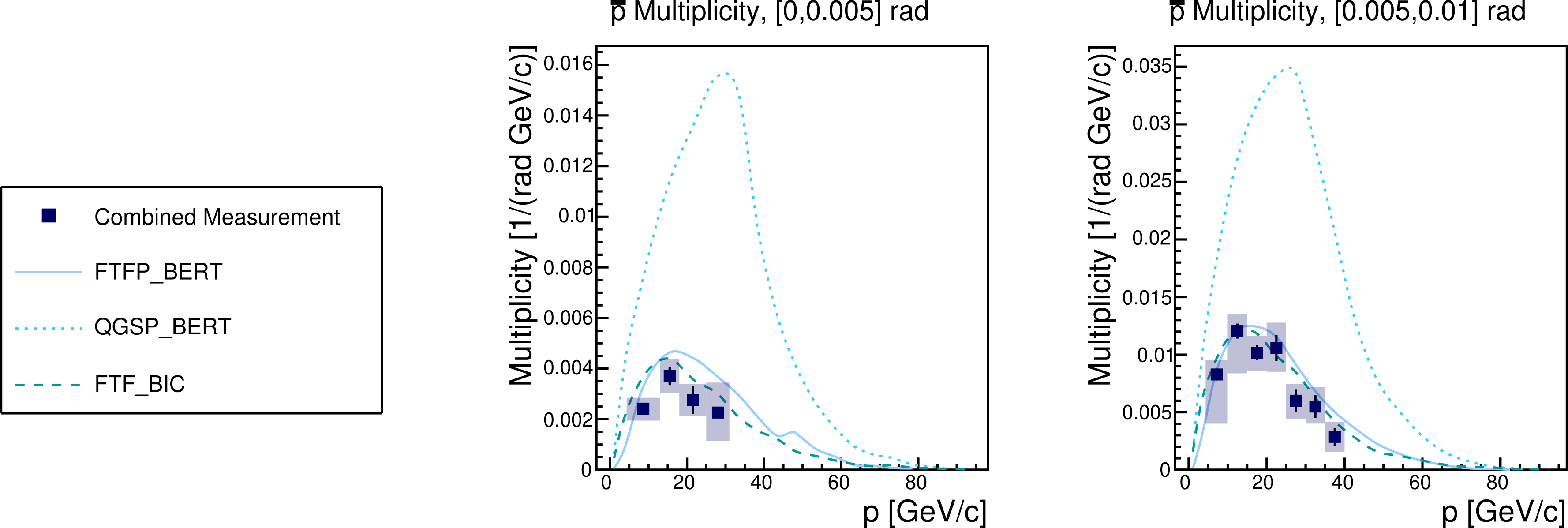}
\caption{Combined multiplicity measurements for $\bar{p}$ analysis. Error bars denote statistical uncertainty, and total systematic uncertainty is shown as a shaded band. Results are compared to three Monte Carlo models. Two representative angular bins are shown.}\label{fig:combinedResultPBar}
\end{figure*}

\begin{figure*}[!ht]
  \centering
  \includegraphics[width=\textwidth]{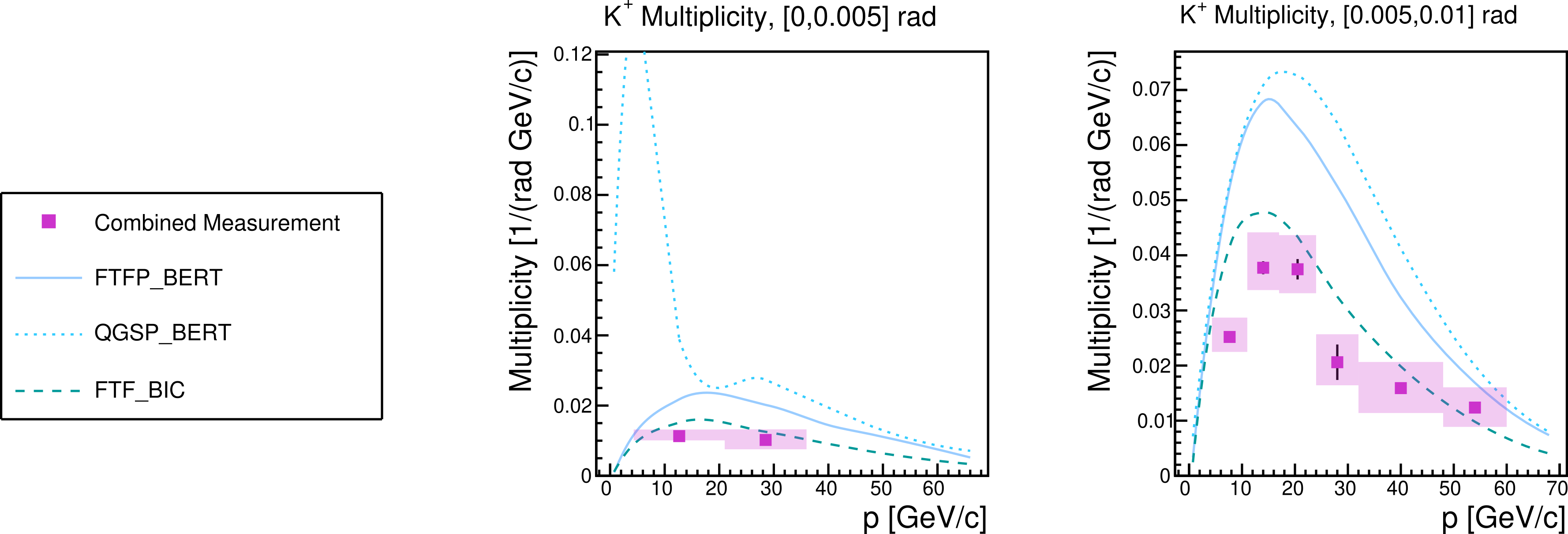}
\caption{Combined multiplicity measurements for \Kp analysis. Error bars denote statistical uncertainty, and total systematic uncertainty is shown as a shaded band. Results are compared to three Monte Carlo models. Two representative angular bins are shown.}\label{fig:combinedResultKPlus}
\end{figure*}

\begin{figure*}[!ht]
  \centering
  \includegraphics[width=\textwidth]{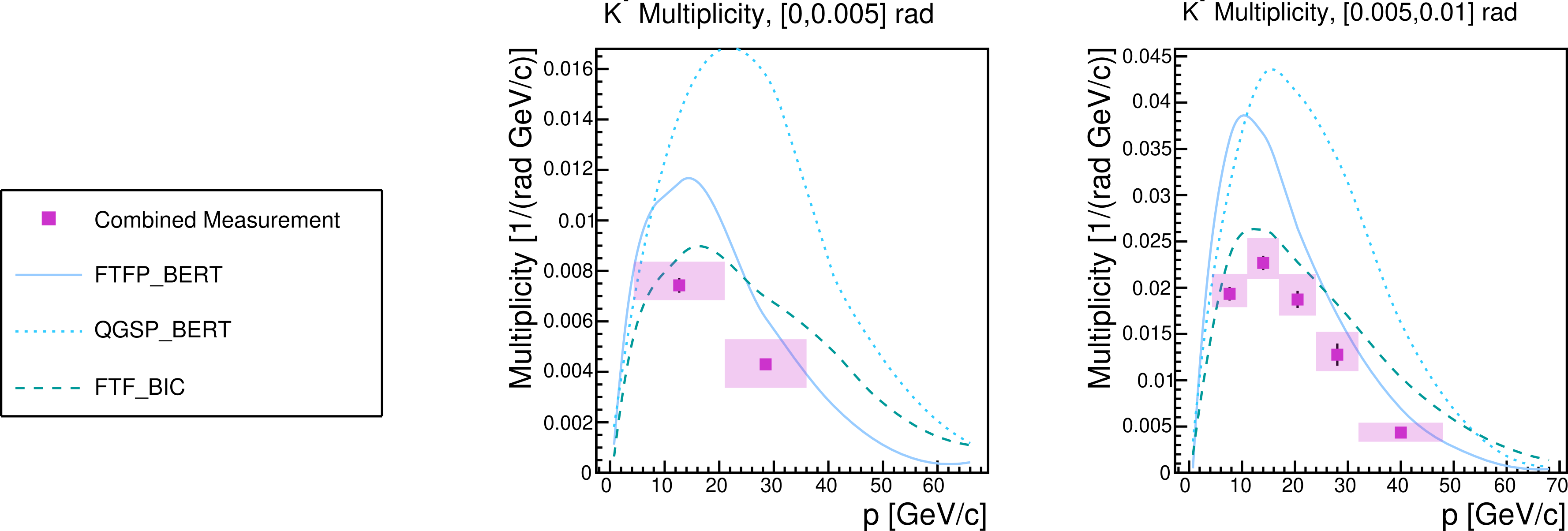}
\caption{Combined multiplicity measurements for \Km analysis. Error bars denote statistical uncertainty, and total systematic uncertainty is shown as a shaded band. Results are compared to three Monte Carlo models. Two representative angular bins are shown.}\label{fig:combinedResultKMinus}
\end{figure*}

\subsection{Combined Systematic Uncertainties}

The total systematic uncertainties on the combined multiplicities reflect both the uncorrelated uncertainties unique to each analysis and the correlated uncertainties common to both analyses. Uncorrelated uncertainties are added in quadrature in each kinematic bin. Fractional correlated uncertainties are treated differently, as they should not simply be added in quadrature. For each correlated uncertainty in each kinematic bin, fractional correlated uncertainty values were compared between the 2016 and 2017 analyses. The larger of the two was taken as the total contribution to the total uncertainty. The final values for the uncorrelated uncertainty and each correlated uncertainty were added in quadrature to obtain the total uncertainty.

\begin{figure*}[!ht]
  \centering
  \includegraphics[width=\textwidth]{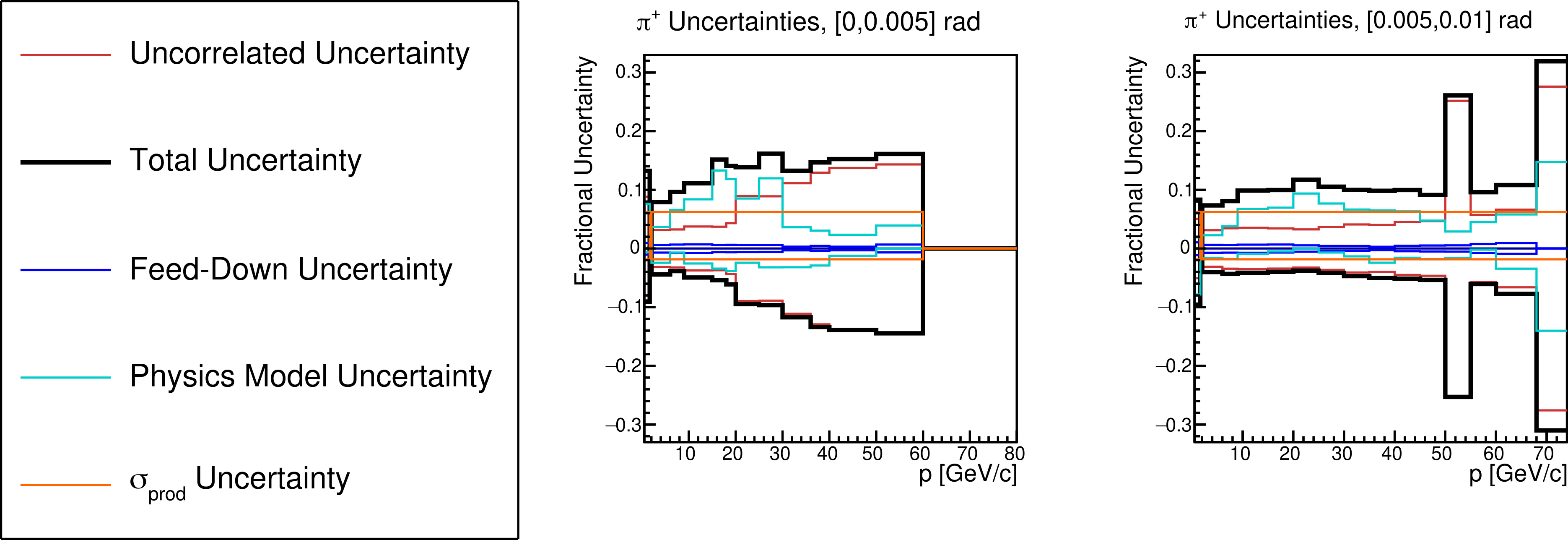}
\caption{Systematic uncertainty breakdown for the combined \pip analysis. Two representative angular bins are shown.}\label{fig:combinedUncertaintiesPiPlus}
\end{figure*}

\begin{figure*}[!ht]
  \centering
  \includegraphics[width=\textwidth]{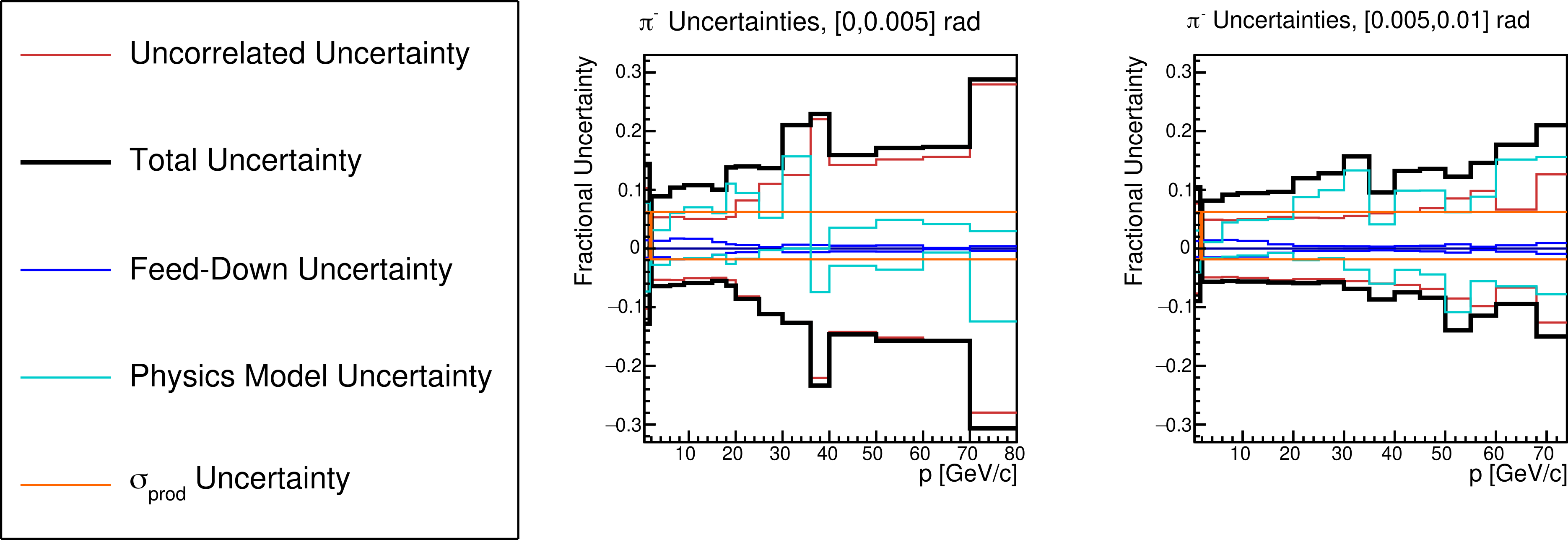}
\caption{Systematic uncertainty breakdown for the combined \pim analysis. Two representative angular bins are shown.}\label{fig:combinedUncertaintiesPiMinus}
\end{figure*}

\begin{figure*}[!ht]
  \centering
  \includegraphics[width=\textwidth]{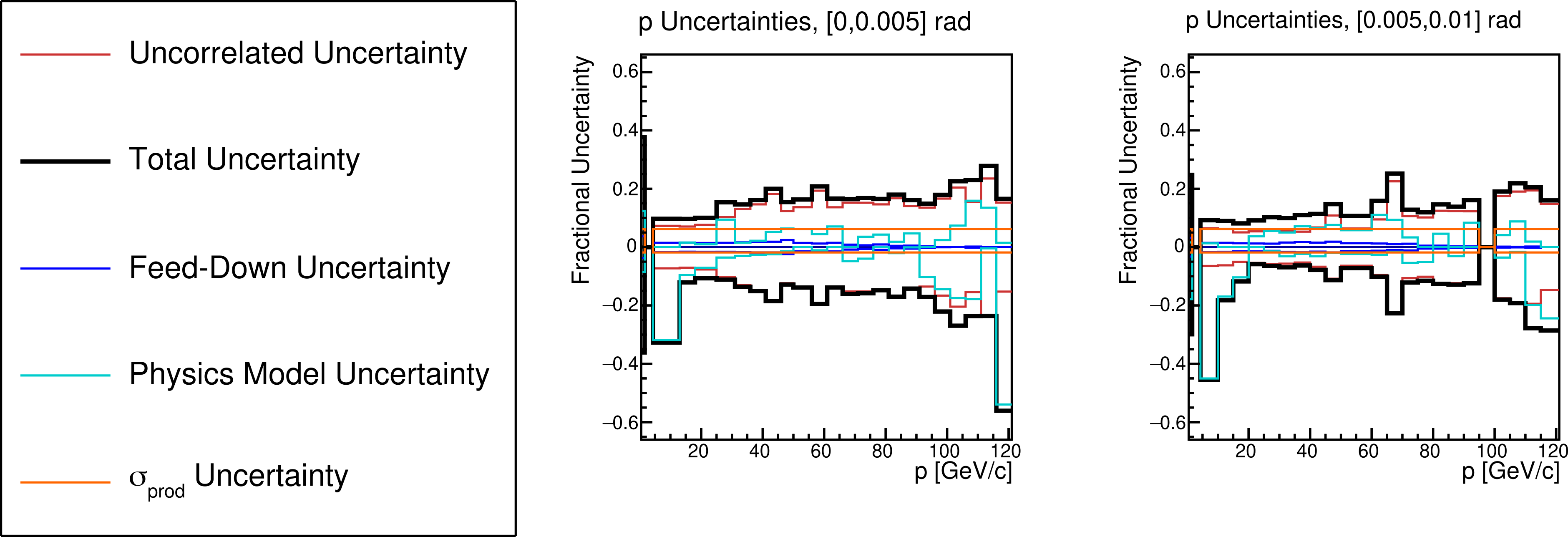}
\caption{Systematic uncertainty breakdown for the combined $p$ analysis. Two representative angular bins are shown.}\label{fig:combinedUncertaintiesP}
\end{figure*}

\begin{figure*}[!ht]
  \centering
  \includegraphics[width=\textwidth]{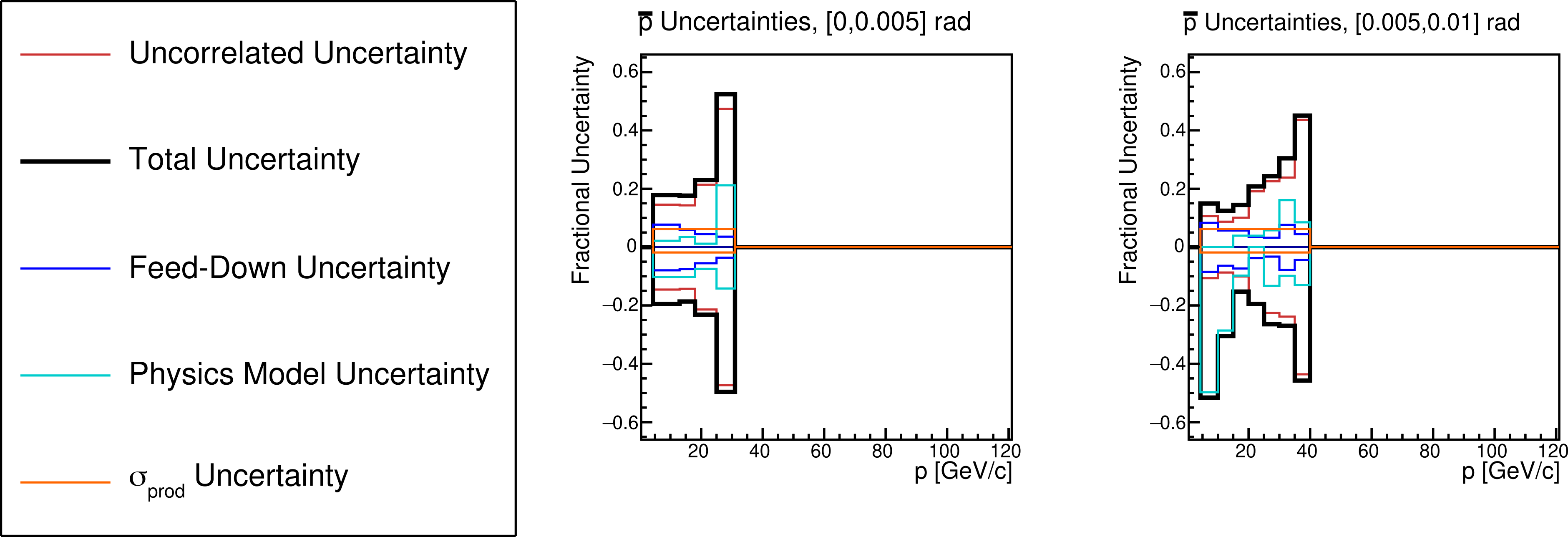}
\caption{Systematic uncertainty breakdown for the combined $\bar{p}$ analysis. Two representative angular bins are shown.}\label{fig:combinedUncertaintiesPBar}
\end{figure*}

\begin{figure*}[!ht]
  \centering
  \includegraphics[width=\textwidth]{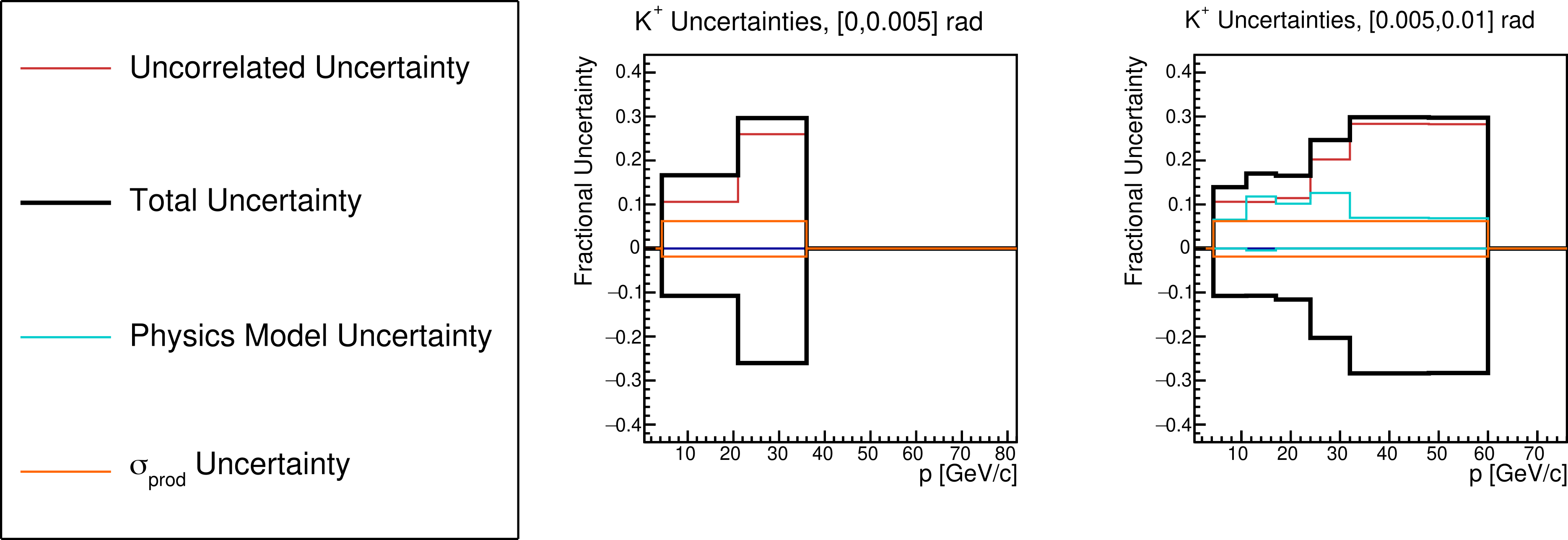}
\caption{Systematic uncertainty breakdown for the combined \Kp analysis. Two representative angular bins are shown.}\label{fig:combinedUncertaintiesKPlus}
\end{figure*}

\begin{figure*}[!ht]
  \centering
  \includegraphics[width=\textwidth]{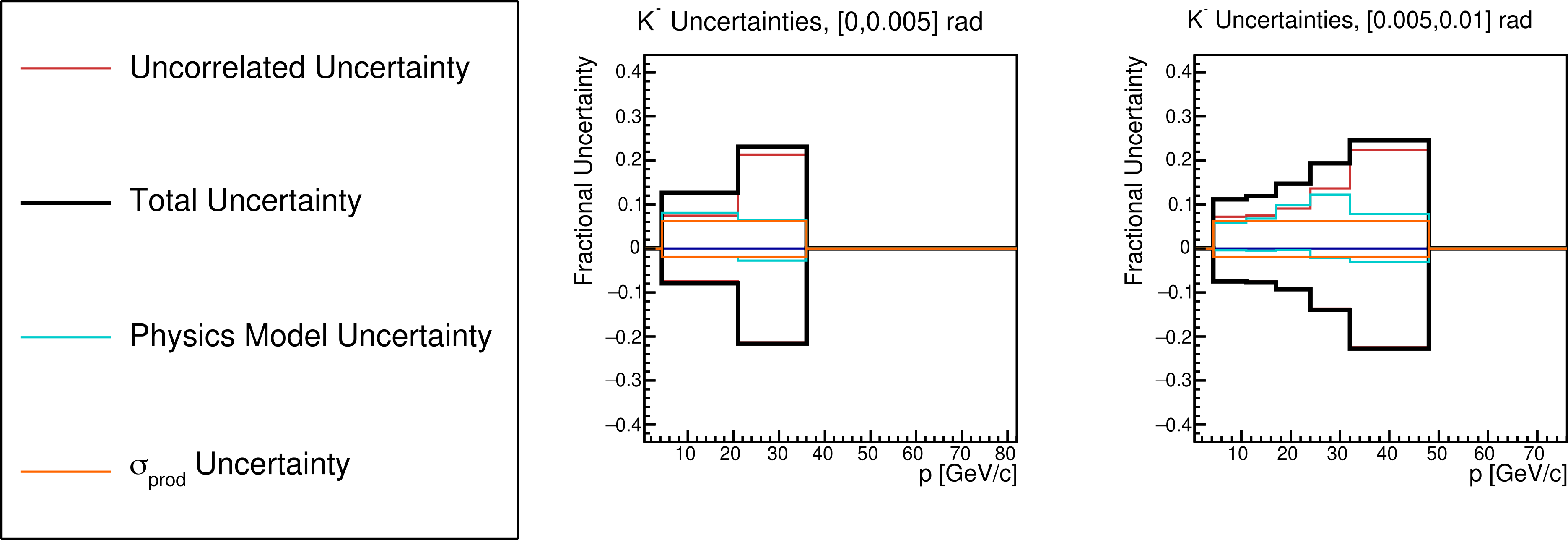}
\caption{Systematic uncertainty breakdown for the combined \Km analysis. Two representative angular bins are shown.}\label{fig:combinedUncertaintiesKMinus}
\end{figure*}

A breakdown of the systematic uncertainties for the combined measurements can be seen in Figs.~\ref{fig:combinedUncertaintiesPiPlus}--\ref{fig:combinedUncertaintiesKMinus}.

\section{Results and Data Release}\label{sec:results}

\subsection{Charged-Hadron Multiplicities}

Final multiplicity results for the charged-hadron analysis can be seen in Figs.~\ref{fig:combinedResultPi2D}--\ref{fig:combinedResultKMinus}. A two-dimensional overview of each particle species is shown in Figs.~\ref{fig:combinedResultPi2D}--\ref{fig:combinedResultK2D}. In addition, two representative angular bins are shown for each particle species. These two angular bins benefit significantly from the addition of the FTPCs. As can be seen in the two-dimensional overview plots, charged pion production dominates the majority of hadron production across phase space. In the forward region, proton production outweighs \pip production.   

Numerical results of the multiplicity measurements of \pipm, \p, \antip, and \Kpm are summarized in CERN EDMS~\cite{pC120EDMS} along with statistical, systematic and total uncertainties for each kinematic bin. Covariance matrices for each analysis are included.

\section{Summary}\label{sec:Summary}

Charged-hadron production measurements in 120 \gevc proton-carbon interactions were presented. The results are the combination of two complementary datasets recorded with significantly different detector configurations. Significant discrepancies between the measurements and popular Monte Carlo simulation models were highlighted. The results presented in this publication can be used to improve the accuracy of neutrino beam content estimation in existing and future experiments in which the neutrino beam is created using the 120 \gevc proton-carbon interaction. The results can also be used to improve Monte Carlo modeling of proton-nucleus interactions.

Dominant systematic uncertainties in the charged-hadron analysis originate from $dE/dx$ fits in the case of \Kpm and $p / \bar{p}$, uncertainties related to the production cross section in the case of \pipm, and, to a smaller extent, reconstruction uncertainty. The $dE/dx$ fit uncertainty is inherent to the stochastic nature of charged particle ionization and the finite number of $dE/dx$ samples collected in certain regions of phase space. The production cross-section uncertainty, on the other hand, can be significantly reduced if the quasielastic component of the interaction cross section is precisely measured. This would reduce the uncertainties on the \pipm spectra to just a few percent in the regions of phase space most pertinent to FNAL neutrino experiments.

The neutral-hadron multiplicity measurements previously reported by NA61/SHINE~\cite{neutralHadronArxivPaper} contributed significantly to reducing systematic uncertainties associated with modeling of \kos, \lam, and \alam decays and their contributions to charged-hadron multiplicities. Without these measurements, feed-down uncertainties associated with \lam production result in uncertainties up to 10\% on \p production multiplicities, and uncertainties associated with \alam production result in uncertainties up to 30\%. For \pip and \pim, the unconstrained uncertainties were as large as 6\% and 15\%, respectively. These uncertainties were all significantly reduced using the neutral hadron multiplicity measurements, as can be seen in Figs.~\ref{fig:feedDownComparisonPi}--\ref{fig:feedDownComparisonP}.

  \section*{Acknowledgments}



We would like to thank the CERN EP, BE, HSE and EN Departments for the
strong support of NA61/SHINE.

This work was supported by
the Hungarian Scientific Research Fund (grant NKFIH 138136\slash138152),
the Polish Ministry of Science and Higher Education
(DIR\slash WK\slash\-2016\slash 2017\slash\-10-1, WUT ID-UB), the National Science Centre Poland (grants
2014\slash 14\slash E\slash ST2\slash 00018, 
2016\slash 21\slash D\slash ST2\slash 01983, 
2017\slash 25\slash N\slash ST2\slash 02575, 
2018\slash 29\slash N\slash ST2\slash 02595, 
2018\slash 30\slash A\slash ST2\slash 00226, 
2018\slash 31\slash G\slash ST2\slash 03910, 
2019\slash 33\slash B\slash ST9\slash 03059, 
2020\slash 39\slash O\slash ST2\slash 00277), 
the Norwegian Financial Mechanism 2014--2021 (grant 2019\slash 34\slash H\slash ST2\slash 00585),
the Polish Minister of Education and Science (contract No. 2021\slash WK\slash 10),
the European Union's Horizon 2020 research and innovation programme under grant agreement No. 871072,
the Ministry of Education, Culture, Sports,
Science and Tech\-no\-lo\-gy, Japan, Grant-in-Aid for Sci\-en\-ti\-fic
Research (grants 18071005, 19034011, 19740162, 20740160 and 20039012),
the German Research Foundation DFG (grants GA\,1480\slash8-1 and project 426579465),
the Bulgarian Ministry of Education and Science within the National
Roadmap for Research Infrastructures 2020--2027, contract No. D01-374/18.12.2020,
Ministry of Education
and Science of the Republic of Serbia (grant OI171002), Swiss
Nationalfonds Foundation (grant 200020\-117913/1), ETH Research Grant
TH-01\,07-3,
the U.S. National Science Foundation grant PHY-2013228, the Fermi National Accelerator Laboratory (Fermilab), a U.S. Department of Energy, Office of Science, HEP User Facility managed by Fermi Research Alliance, LLC (FRA), acting under Contract No. DE-AC02-07CH11359, 
and the IN2P3-CNRS (France).\\

The data used in this paper were collected before February 2022.


\bibliography{main}

\clearpage

\end{document}